# On the Structural Origin of the Single-ion Magnetic Anisotropy in LuFeO$_3$

Shi Cao[1], Xiaozhe Zhang[2,1], Tula R. Paudel[1], Kishan Sinha[1], Xiao Wang[3], Xuanyuan Jiang[1], Wenbin Wang[4], Stuart Brutsche[1], Jian Wang[5], Philip J. Ryan[6], Jong-Woo Kim[6], Xuemei Cheng[3], Evgeny Y. Tsymbal[1], Peter A. Dowben[1] and Xiaoshan Xu[1]

[1]*Department of Physics and Astronomy & Nebraska Center for Materials and Nanoscience, University of Nebraska, Lincoln, Nebraska 68588, USA,*

[2]*Department of Physics, Xi'an Jiaotong University, Xi'an 710049, China*

[3]*Department of Physics, Bryn Mawr College, Bryn Mawr, Pennsylvania 19010, USA*

[4]*Department of Physics, Fudan University, Shanghai 200433, China*

[5]*Canadian Light Source, Saskatoon, SK S7N 2V3, Canada*

[6]*Advanced Photon Source, Argonne National Laboratory, Argonne, Illinois 60439, USA*

## Abstract
Electronic structures for the conduction bands of both hexagonal and orthorhombic LuFeO$_3$ thin films have been measured using x-ray absorption spectroscopy at oxygen K (O K) edge. Dramatic differences in both the spectra shape and the linear dichroism are observed. These differences in the spectra can be explained using the differences in crystal field splitting of the metal (Fe and Lu) electronic states and the differences in O 2p-Fe 3d and O 2p-Lu 5d hybridizations. While the oxidation states has not changed, the spectra are sensitive to the changes in the local environments of the Fe$^{3+}$ and Lu$^{3+}$ sites in the hexagonal and orthorhombic structures. Using the crystal-field splitting and the hybridizations that are extracted from the measured electronic structures and the structural distortion information, we derived the occupancies of the spin minority states in Fe$^{3+}$, which are non-zero and uneven. The single ion anisotropy on Fe$^{3+}$ sites is found to originate from these uneven occupancies of the spin minority states via spin-orbit coupling in LuFeO$_3$.

## Introduction
The structure of a crystalline material plays a determinant role in its physical properties. By fine-tuning the crystal structure, physical properties of a material may be modified and this offers great opportunities in engineering functional materials. [1–3] Particularly interesting is the effect of the crystal structure on the magnetic structure, including the relative alignment between the spins and the preferred overall orientation of the spins (magnetocrystalline anisotropy). While the exchange interactions determine the relative alignment of the spins, their effect on the magnetocrystalline anisotropy is indirect, because the exchange interactions are isotropic. Single-ion magnetic anisotropy is a critical factor for the overall magnetocrystalline anisotropy, although the latter is also affected by the topological arrangements of the spins. [4] The crystal structure, particularly the local environments of the magnetic ions, is expected to decide the single-ion magnetic anisotropy, by changing their orbital states and affecting spin orientations via the spin-orbit coupling. [5–7]

Here we are concerned with the effect of the crystal structure on the magnetic anisotropy in antiferromagnetic LuFeO$_3$. LuFeO$_3$ is a rare example of a material that exists in both orthorhombic and stabilized hexagonal structures [3,8–15], which are different both in symmetry of the lattice and in the symmetry of the local environment of the metal (Fe and Lu) sites [Fig. 1 (a) and (b)]. [16] These differences in structure, give rise to the dramatic differences in properties such as ferroelectricity and magnetism. [8,10,11,13,14] In hexagonal LuFeO$_3$ (h-LuFeO$_3$), the inversion symmetry of the lattice structure is broken by the rotation of the FeO$_5$ trigonal bipyramids, generating ferroelectricity below 1050 K with a polarization on the order of 10 μC/cm$^2$. [3,9,10,13,17] The spins on the Fe sites in h-LuFeO$_3$ order in a 120-degree antiferromagnetic fashion in the $a − b$ plane [Fig. 1(a)]; a canting of the spins out of the $a − b$ plane results in a weak ferromagnetism below 130 K. [8,9,11,13,18] In orthorhombic LuFeO$_3$ (o-LuFeO$_3$), ferroelectricity is unexpected due to the symmetric arrangement of the atoms. The spins on the Fe sites in o-LuFeO$_3$ order antiferromagnetically in a chain-like fashion in the $a − b$ plane below 620 K [Fig. 1(b)]; a canting toward the $c$ axis generates a weak ferromagnetism. [19] The single-ion magnetic anisotropy is critical for the magnetic orders in LuFeO$_3$: In o-LuFeO$_3$, it is the prerequisite for the magnetocrystalline anisotropy; in h-LuFeO$_3$, the weak ferromagnetism is not allowed unless the spins are along the $a$ axis. [13,14,20] Therefore, elucidating the origin of the single-ion magnetic anisotropy is important to understanding and tuning the magnetism in LuFeO$_3$; this is especially true in h-LuFeO$_3$, a promising magnetoelectric material that exhibits ferromagnetism and ferroelectricity simultaneously [21].

In this work, we attempt to understand the single-ion magnetic anisotropy in LuFeO$_3$, by studying the effect of the crystal structure on the orbital states, and the consequential effect on the spin states according to the spin-orbit coupling. To investigate the orbital states of Fe, we measured the electronic structures of LuFeO$_3$ using x-ray absorption spectroscopy; the results are consistent with the D$_{3h}$ and O$_h$ local symmetry of Fe sites in the hexagonal and orthorhombic LuFeO$_3$ respectively. More details of the orbital states are calculated according to the low temperature structure of LuFeO$_3$ (C$_S$ and D$_{2h}$ local symmetry for Fe sites in the hexagonal and orthorhombic structures respectively) using the multiplets theory. [22] The low temperature structure of h-LuFeO$_3$ was measured in this work using single-crystal x-ray diffractions, since it has not been reported. We found that the low local symmetry split the orbital states, generating preferred spin orientations of these states via spin-orbit coupling. The magnetic anisotropy for the whole Fe$^{3+}$ ion is then caused by the non-zero and uneven occupancies of the spin-minority states due to the uneven hybridizations of these states to O 2p states. For orthorhombic LuFeO$_3$, the predicted easy axis for the spins is the shortest axis ($a$ axis) after the D$_{2h}$ distortion. For hexagonal LuFeO$_3$, the preferred spin orientation are in the intersection between the basal plane and the mirror plane of the C$_S$ symmetry. Both predictions are consistent with the experimental observations. [13,18,19]

## Methods

### Experimental

Hexagonal and orthorhombic LuFeO$_3$ films (~50 nm) have been grown on Al$_2$O$_3$ (0001) and SrTiO$_3$ (001) substrates respectively using pulsed laser deposition at 750 ºC with 5 mtorr oxygen environment. [13] The surfaces of the film samples are (0001) for h-LuFeO$_3$ and (001) for o-LuFeO$_3$. X-ray absorption spectroscopy (XAS) studies via the X-ray Photoemission Electron Microscope (X-PEEM) have been carried out at the SM beamline of the Canadian Light Source with a linearly polarized x-rays at room temperature in ultrahigh vacuum; the incident angle is 16 degree (See S1) [23]. Structural refinement using x-ray diffraction has been carried out at 6-ID-B beam line of the Advanced Photon Source by measuring 43 diffraction peaks at 7 temperatures.

### Theoretical Methods

Theoretical modeling of the bulk h-LuFeO$_3$ and o-LuFeO$_3$ was performed using density functional theory, the projected augmented wave method [21], and Perdew-Burke-Ernzerhof pseudopotentials [22], as implemented in Vienna *ab initio* simulation package [22]. We fully relaxed the structure with the force convergence limit of 0.01 eV/atom. Correlation effects beyond generalized gradient approximation (GGA) were treated at a semi-empirical GGA+U level within a rotationally invariant formalism [24] with a U = 5 eV chosen for the Fe 3d-orbitals [15]. In k-edge XAS spectroscopy, an electron is excited from the O 1s core level to the conduction band. In the semiconductor like LuFeO$_3$, the resulting core hole is only partially screened affecting the orbital energy. In order to take into account the effect of the core hole on measured XAS spectra, we introduce a frozen ½ hole [25] in apex (6c Wyckoff's position, see S9 for definition) [23] oxygen site in a 2×2×1 supercell for h-LuFeO$_3$, and a 2×2×2 supercell for o-LuFeO$_3$. The supercell is necessary to minimize the interaction between the core holes introduced to the periodic boundary conditions.

# Results and discussion

## Energy and spatial distribution of the orbital states measured using XAS

As the first step, we investigate the effect of the crystal structure on the orbital states of the metal ions (Fe and Lu), because the crystal structure affect the spin states of the magnetic ions by first changing their orbital states. X-ray absorption spectroscopy was employed to study the unoccupied orbital states (conduction band). The measured energy distribution (spectra shape) and spatial distribution (linear dichroism) of these orbital states are compared with the crystal field splitting and hybridization (with O 2p states) analyzed according to the crystal structure.

As shown in Fig. 2, absorption spectra as functions of x-ray energy with linearly polarized x-ray have been collected in the energy range 525 eV to 560 eV; the energy range corresponds to the excitation of O 1s orbital to O 2p orbital (O K edge). The fact that the O 1s → O 2p excitations are clearly observed indicates significant hybridization between the metal (Fe and Lu) states and the oxygen states, making the effective occupation of the O 2p orbital different from the full 2p$^6$ occupation. The presence of the O 1s → O 2p excitations also means that the electronic occupancy for the metal (Fe and Lu) sites is more complex than suggested by their nominal valence. Hence, the energies of the unoccupied oxygen orbitals actually correspond to the energies of the metal (Fe and Lu) states, as illustrated in Fig. 1(c). Therefore, using the O K edge absorption spectra, one can infer the properties of the states that include metal (Fe and Lu) atomic contributions through hybridization. [24,26] In the case of LuFeO$_3$, the conduction (unoccupied) states include Fe 3d, Fe 4s, Lu 6s and Lu 5d. Among these states, Fe 3d and Lu 5d are expected to be more localized and the energy distributions are narrow enough to be resolved in the x-ray absorption spectra.

By comparing the observed spectra in this work to the previous studies on YMnO$_3$ and LuFe$_2$O$_4$ (see S2) [23,24,27] one can divide the absorption spectra into two parts that correspond to the contribution from Fe 3d and Lu 5d respectively, as shown in Fig. 2.

For the spectra related to Fe 3d unoccupied states (conduction band), the differences between h-LuFeO$_3$ and o-LuFeO$_3$ are huge, not only in the spectra shape, but also in the dichroism. These differences appear to be correlated with the local environments of the Fe centers. As shown in Fig. 1 (b), in o-LuFeO$_3$, the local environment of the Fe centers are the FeO$_6$ octahedra; no strong anisotropy or optical dichroism is expected due to the O$_h$ local symmetry of the Fe-site. In contrast, as shown in Fig. 1(a), the local environment of the Fe centers in h-LuFeO$_3$ is the FeO$_5$ trigonal bipyramid; the D$_{3h}$ local symmetry suggests strong anisotropy and optical dichroism between the $a - b$ plane and the $c$ axis. Below, we try to understand the spectra shape (energy distribution) in terms of the crystal field splitting, and to understand the dichroism (spatial distribution) in terms of the hybridization between the Fe 3d and O 2p states.

In h-LuFeO$_3$, the degeneracy of the Fe 3d states are broken by the crystal field from the 5 neighboring oxygen sites in the trigonal bipyramid FeO$_5$. Applying group theory analysis (see S5), [23,28,29] the D$_{3h}$ local symmetry splits the 5 Fe 3d states into $a_1'$ ($2z^2 - x^2 - y^2$), $e'(x^2 - y^2, xy)$, and $e''(xz, yz)$; the $z$ direction is approximately parallel to the three fold rotational axis of FeO$_5$ and the $c$ axis of the h-LuFeO$_3$ unit cell. Analysis using the multiplets model (see S6) [22,23] provides the order of these states in energy as $E_{a_1'} > E_{e'} > E_{e''}$ (see S6). [12,23] One can roughly understand this lifting of degeneracies, using the electrostatic energy between the oxygen sites and the Fe 3d electrons in the FeO$_5$ bipyramid: The $a_1'$ state is very close to the apex oxygen sites, while the $e''$ state is far away from all the oxygen sites. However, our first-principles calculations show that $e'$ states to lie below $e''$ states. The calculated density of states that are resolved in cubic harmonics (Fig 3(a)), shows that bottom of conduction band is dominated by the Fe- $e'$ states. This is in part supported by the shorter Fe-O bond length (~1.95 Å) along apex ($z$) direction compared to slightly longer bond (~1.99 Å) along equator direction. In additions we found slight variations on in-plane bonds as well indicating symmetry lowering during relaxations, leading to somewhat different density of states than that calculated using full potential approximation [12]. An additional calculation performed for an FeO$_5$ cluster with the in-plane ∠O-Fe-O =120° also shows that in the conduction band $e'$ to lies below $e''$(see S10) [23]. However, when the angle is rotated to 135°, $e''$ states becomes lower in energy than $e'$ states (see S10) [23]. We argue that in XAS spectra, the core hole that is created by the incoming x-ray beam may not be completely screened in semiconductor or near dielectric like LuFeO$_3$. When such a hole is present in oxygen at the apex site, the $e''$ and $a_1'$ state are affected by larger electrostatic attraction compared to other orbitals that lowers their energy. As a result, these states may appear below the $e'$ states. To test this hypothesis, we constructed an 80 atom supercell, including the core hole in the apex site, and calculated the orbital-dependent density of electronic states. Fig 3(b) shows the density of states projected on the Fe site bonded to the oxygen at the apex site, indicating that the $e''$ states lie lower in energy than the $e'$ states. On the other hand, as shown below, the energy ordering of the $e''$ and $e'$ states does not play important role for the single-ion magnetic anisotropy in h-LuFeO$_3$. Instead, the significant hybridization of the $a_1'$ states and the splitting of the $e''$ and $e'$ states due to the lattice distortion are the key.

Due to the different spatial distribution of these crystal field states, their hybridizations with O 2p orbitals are different, which is schematically shown in Fig. 4. For h-LuFeO$_3$, one needs to consider two inequivalent O sites: the apex O and the equator O, as shown in Fig. 1(a). The O 2p states are divided into $p$ (along the $c$-axis) and $s$ (in the $a - b$ plane) to match the linearly polarized x-ray. By calculating the hybridization using Harrison's method (see S8), [23,30] five non-zero scenarios can be identified, as depicted in Fig. 4 using boxes. The hybridization can be appreciated by looking at the overlap of the wave function between the Fe 3d and O 2p orbitals.

With linearly $s$ (in plane) and $p$ (out of plane) polarized x-ray, the excitation from O 1s to $s$ and $p$ branches of the O 2p states can be chosen respectively using their spatial distribution according to the optical selection rules [29]. As shown in Fig. 4, one expects two peaks in the XAS of $p$ polarization and three peaks in the XAS of $s$ polarization, which matches the experimental observation in Fig. 2 (a) closely. The calculated hybridization strength using the Harrison's method is also displayed in Fig. 2(a), which qualitatively agrees with the observed spectra intensity (see S8). [12,23,26] From the experimentally obtained XAS peak positions, one can extract the energy separations between the Fe 3d $a_1'$, $e'$, and $e''$ states, as well as the energy difference $\delta_{1s}$ between the O 1s states in the apex and equator sites. The results are $E_{a_1'} - E_{e'} = 1.1$ eV, $E_{a_1'} - E_{e''} = 1.4$ eV, and $\delta_{1s} = 0.6$ eV, in fair agreement with the values we found in our previous work. [12]

The analysis is more straightforward in o-LuFeO$_3$. The Fe 3d states are split into the well-known $t_{2g}$ $(xz, yz, xy)$ and $e_g$ $(2z^2 - x^2 - y^2, x^2 - y^2)$ states [31], in which the $t_{2g}$ states have lower energy consistent with calculated density of state as shown in Fig 3(c). From the spectra in Fig. 2(b), one finds that $E_{e_g} - E_{t_{2g}} = 1.4$ eV. Again, no dichroism is expected due to the O$_h$ local symmetry. Additionally, we found that presence of core hole does not change the ordering of the states.

The electronic structure of the Lu 5d states may also be inferred from the corresponding spectra. Figure 2(c) and (d) display the XAS related to the Lu 5d states in h-LuFeO$_3$ and o-LuFeO$_3$. The local environments of Lu correspond to C$_{3v}$ symmetry in both h-LuFeO$_3$ and o-LuFeO$_3$. According to the group theory analysis (see S5), [23,28,29] the five Lu 5d states are split into two doubly degenerate $e$ states and one $a_1(2z^2 - x^2 - y^2)$ state. Using multiplet model (see S6) [22,23], one can gain more insight of the symmetry of the two $e$ states: they are $e^\sigma = [(x^2 - y^2) + 2\lambda xz, xy + \lambda yz]$ and $e^\pi = [(x^2 - y^2) - 2\lambda xz, xy - \lambda yz]$, where $\lambda \sim 1$ is a mixing factor. The order of these states in energy is $E_{e^\sigma} > E_{a_1} > E_{e^\pi}$.

In h-LuFeO$_3$, the three-fold rotational axis of the LuO$_7$ local environment is aligned with the crystalline $c$ axis, which is also the out-of-plane direction for the film samples. This definitive alignment between the high symmetric axis and the polarization of the x-ray generates dichroism, as observed in Fig. 2(c). For the $a_1$ state, since the probability density of the wave function is mostly along the z axis, its hybridization with the equator O 2p$_z$ is expected to be the largest, which corresponds to an enhancement with the $p$ polarization in the XAS. For the $e^\sigma$ and $e^\pi$ states, the hybridization are mostly with the 2p of the apex oxygen sites, resulting in higher intensity of XAS in the $s$ polarization. The calculated hybridization strength is also plotted in Fig. 2(c) using the Harrison's method as a comparison (see S8). [23,26] We extract the energy separation as approximately $E_{e^\sigma} - E_{a_1} = 2.7$ eV and $E_{e^\sigma} - E_{e^\pi} = 4.9$ eV. The crystal field splitting is larger in Lu 5d than that in Fe 3d, suggesting that the Lu 5d is more exposed to the surrounding oxygen sites.

In contrast, there is no overall alignment between the rotation axis of the LuO$_6$ moieties and the crystal axis, which greatly reduces the dichroism effects, because of the averaging over various orientations. Nevertheless, the crystal field splitting feature does not vanish, as observed in the spectra in Fig. 2(d). We extract the energy separations of $e^\sigma$ to $a_1$ and $e^\sigma$ to $e^\pi$ as approximately $E_{e^\sigma} - E_{a_1} = 2.6$ eV, and $E_{e^\sigma} - E_{e^\pi} = 4.9$ eV, quite similar to the results from the h-LuFeO$_3$. The similarities in these energy separations of unoccupied states is consistent with the fact that the Lu-O bond length and local symmetry are similar in o-LuFeO$_3$ and h-LuFeO$_3$.

Therefore, the energy and spatial distributions of the metal states (Fe and Lu) measured using XAS are consistent with the crystal field splitting and hybridization analyzed according to the crystal structural. Another key result from the XAS study is the significant Fe 3d-O 2p hybridizations. In LuFeO$_3$, the Fe 3d is nominally half-filled, corresponding full spin majority states and empty spin minority states. On the other hand, significant Fe 3d-O 2p hybridizations make the effective occupancy of the spin minority states non-zero and uneven; this turns out to be critical for the single-ion magnetic anisotropy in LuFeO$_3$.

## Splitting of orbital states in low symmetry structure and single-ion anisotropy

As the second step, we study the single-ion anisotropy of Fe based on the 3d orbital states measured in the XAS study. We calculate spin anisotropy of the individual Fe 3d states according to spin orbit coupling; from the spin anisotropy of these individual states, we calculate the spin anisotropy of the whole Fe (single-ion magnetic anisotropy) by considering the uneven occupancies of these states, according to the Fe 3d-O 2p hybridizations found in the XAS measurements.

The following one-electron Hamiltonian is used to model the effect of crystal structure on spin anisotropy of the Fe 3d individual states:

$$H_\alpha = V_{CF}(d) + \xi \vec{L} \cdot \vec{S} - J\vec{S} \cdot \hat{\alpha}, \quad (1)$$

where the basis are the 10 Fe 3d states considering both orbital and spin degrees of freedom.

The first term $V_{CF}(d)$ is a matrix that takes into account the crystal field on the Fe orbital states. The crystal field splitting (about 1 eV) of Fe 3d measured using XAS (D$_{3h}$ symmetry and O$_h$ symmetry for hexagonal and orthorhombic structures respectively) are included in the matrix as constants (see S6 and S7). In addition, we need to consider more details of crystal field splitting that cannot be revealed by XAS measurements (because of the experimental uncertainty), i.e. the splitting due to the C$_S$ and D$_{2h}$ local symmetry of the Fe in hexagonal and orthorhombic LuFeO$_3$ respectively. Figure 5(c) and (d) display these local displacement in hexagonal and orthorhombic LuFeO$_3$ respectively. In the ferroelectric phase of h-LuFeO$_3$ (below 1050 K), [13] the Fe shift from the center of the FeO$_5$ toward (or way from) one of the equator oxygens by $\delta_{Fe}$, which removes the 3-fold rotation symmetry as well as the 2-fold rotation symmetry; the corresponding symmetry of the FeO$_5$ local environment is reduced from D$_{3h}$ to C$_S$, which contains one vertical mirror plane that is parallel to the $c$ axis and passes the O3 site [see Fig. 5(c)]. For orthorhombic LuFeO$_3$, the FeO6 octahedra are distorted so that the Fe-O bond length along the $x$, $y$ and $z$ directions are all different. Taking the $x - y$ plane for example, the distortion can be viewed as the displacement of oxygen atoms by $\delta_O$, as shown in Fig. 5(d). We represent the distortions in the crystal field potentials $V_{CF}$ using a perturbation parameter $d$ that are proportional to the displacement of atoms [$d \propto -\delta_{Fe}$ and $d \propto \delta_O$, see Fig. 5 (c) and (d)]. The orbital states in these low symmetry structures ($C_S$ and $D_{2h}$) can be calculated as functions of the distortion parameter $d$.

The second term $\xi \vec{L} \cdot \vec{S}$ represents the spin orbit coupling, where $\vec{L}$ and $\vec{S}$ are the orbital and spin angular momentum respectively, and $\xi$ is the spin-orbit coupling strength (taken as 0.05 eV).

The third term $J\vec{S} \cdot \hat{\alpha}$ represents the exchange interaction between the Fe sites. The net effect of the spins from all the other Fe sites is treated as a molecular field $J\hat{\alpha}$, where $J > 0$ (taken as 4 eV) represents the molecular field strength from the exchange interaction and $\alpha = x, y$, or $z$ is the orientation of the molecular field. The molecular field splits the individual states on Fe into spin majority and minority states, in which the spins are parallel and antiparallel to the molecular field respectively.

Diagonalizing the Hamiltonian in Eq. (1), one can find the energy of every individual states, modified by the structural distortion, as well as the molecular field. By varying the direction of the molecular field $\hat{\alpha}$, one can find the spin anisotropy energy of individual states. The total energy of $Fe^{3+}$ ion can be calculated using $E_\alpha = \sum n_i E_{i\alpha}$, where $n_i$ and $E_{i\alpha}$ are the occupancy and energy of the one-electron state $i$ respectively. The single-ion magnetic anisotropy energy can be found from the dependence of total energy on the direction $\hat{\alpha}$. In $LuFeO_3$, the Fe 3d is nominally half-full, corresponding to fully occupied spin majority states and empty minority states; the single-ion magnetic anisotropy is then expected to be zero. On the other hand, as observed in the XAS, significant hybridization between Fe 3d and O 2p makes the effective occupancy of the spin minority states non-zero and uneven. A non-zero single-ion magnetic anisotropy of $Fe^{3+}$ is then expected. Below, we discuss the single-ion magnetic anisotropy of $Fe^{3+}$ in hexagonal and orthorhombic $LuFeO_3$ and the dependence on the distortion parameter $d$.

To discuss the spin anisotropy between the $z$ axis and the $x - y$ plane in h-$LuFeO_3$, as a good approximation, the $D_{3h}$ point group can be taken as the symmetry of the $FeO_5$ local environment. In this case, the hybridization of the $2z^2 - x^2 - y^2$ states ($a_1'$ of Fe 3d) with the O 2p states is the largest [see Fig. 2(a)]. So the effect of spin orientation on the energy of the $2z^2 - x^2 - y^2$ state is the most important. When the spins are along the $z$ axis, the $2z^2 - x^2 - y^2$ state only interacts with the spin majority states (see S7) [23], causing a smaller increase of its energy. When the spins are along the $x$ axis, the $2z^2 - x^2 - y^2$ state interacts with both spin minority and majority states (see S7) [23], causing a larger increase of its energy. So in $D_{3h}$ local environment, one has $E_x > E_z$, as shown in Fig. 5(a). Similarly, $E_y > E_z$ is found. The anisotropy energy $E_x - E_z$ is on the order of $\frac{\xi^2}{E_{EXCF}} \langle n_i \rangle$, where $E_{EXCF}$ is the energy scale of crystal field and exchange interactions (a few eV), and $\langle n_i \rangle$ is the average occupancies of the minority states (see S7) [23].

To discuss the spin anisotropy in the $x - y$ plane in h-$LuFeO_3$, the $D_{3d} \rightarrow C_S$ distortion of the local environment needs to be considered. As shown in Fig. 5(a), when $d = 0$, $E_x - E_y = 0$, indicating no anisotropy in the $x - y$ plane when there is no distortion. When $d < 0$, $E_x - E_y <$ 0, corresponding to an easy $x$ axis within the $x - y$ plane. The anisotropy energy $E_x - E_y$ is on the order of $\frac{\xi^2}{E_{EXCF}} \frac{d}{E_{EXCF}} \langle n_i \rangle$.

In o-LuFeO3, we consider $O_h \rightarrow D_{2h}$ distortion that breaks the symmetry of $x$, $y$, and $z$, again using the parameter $d$ (distortion energy) that has the same sign as the displacement $\delta_O$ (see S7) [23]. As shown in Fig. 5(b), when $d > 0$, $E_x < E_z < E_y$; the $x$ axis is the easy axis. The anisotropy energy $E_z - E_x$ is also on the order of $\frac{\xi^2}{E_{EXCF}} \frac{d}{E_{EXCF}} \langle n_i \rangle$.

To verify the calculated relations between the structure and the single-ion magnetic anisotropy, we compare the above predictions with the experimental observations.

In o-LuFeO3, the observed lattice distortion is displayed in Fig. 5(d) [32], where $\delta_O > 0$ and $d > 0$, suggesting $E_x < E_z < E_y$ according to Fig 5(b). This means that the shortest axis is the easy axis, which is consistent with the observed single-ion anisotropy [see S7] [19].

In h-LuFeO3, at low temperature, the spins on the Fe sites prefer lying in the intersection between the basal plane and the mirror plane of the $C_S$ symmetry, or the $x$ direction displayed in Fig. 5(c). [13,14,33] According to Fig. 5(a), this corresponds to a negative distortion parameter ($d < 0$) and a positive displacement of Fe ($\delta_{Fe} > 0$). On the other hand, the details of the $C_S$ distortion at low temperature, e.g. the sign and magnitude of $\delta_{Fe}$, have not been reported. We need to measure the structural distortion pattern of h-LuFeO3 at low temperature to verify the predicted effect of crystal structure on the single-ion anisotropy.

In order to clarify the lattice distortion in h-LuFeO3, we measured the low temperature (7 temperatures, down to 6 K) single-crystal x-ray diffractions (43 peaks) and carried out structure refinements (see S9) [23]; the results are shown in Table I. The room temperature distortion agrees with the previous work. [10] At low temperature, Fe moves away from the O3 site (see site definition in S9) ($\delta_{Fe} > 0, d < 0$), suggesting that the single-ion magnetic anisotropy energy $E_x - E_y < 0$, according to the analysis above, which is observed experimentally.

While the single-ion magnetic anisotropy of h-LuFeO3 in the $x - y$ plane is correctly predicted by the model, the model also predicts that the $z$ axis is the overall easy axis. This controversy may have to do with the geometric frustration on a triangular lattice when the spins are pointing out of plane and the spins are coupled antiferromagnetically. In other words, the spins cannot be along the $z$ axis and form the 120 degree order at the same time. [4] Therefore, whether the spins are out of the $x - y$ plane is not solely determined by the single-ion magnetic anisotropy. On the other hand, the rotation within the $x - y$ plane does not change the 120 degree order and has no effect on the total energies from the exchange interactions. So in the $x - y$ plane, the single-ion magnetic anisotropy play a dominant role in determining the preferred spin orientation, as verified by the correct prediction of the relation between the crystal structure and the preferred spin orientation in h-LuFeO3.

In general, quantitative comparison between the predicted single-ion magnetic anisotropy and the observation is difficult, because the occupancy $\{n_i\}$ of the spin minority states, the magnitude of the distortion energy, and the single-ion magnetic anisotropy are hard to estimate. Nevertheless, in h-LuFeO$_3$, the reversal of the weak ferromagnetic moments corresponds to the rotation of the spins by 180 degree within the $x - y$ plane; so we can estimate the single-ion magnetic anisotropy energy using the coercivity. In addition, the distortion energy can be estimated using the onset temperature of the structural distortion. Assuming that the model is correct, one can estimate the order of magnitude of $\{n_i\}$ in h-LuFeO$_3$. Taking 2 tesla as the coercive field, and 0.025 $\mu_B/Fe$ as the canted moment [11,13], the anisotropy energy can be estimated as approximately 3 $\mu eV$. The order of magnitude of the distortion energy is estimates as 10 meV from the onset temperature (~150 K) of the structural distortion observed in Table 1. Using the results in Fig. 5(a), we found the occupancies of $n_{e''} = 0.05$, $n_{e'} = 0.08$, and $n_{a_1'} = 0.2$ for spin minority states. The large $n_{a_1'}$ is consistent with the recent optical spectroscopy results. [12,34,35]

## Conclusion

In conclusion, we have studied the effect of the crystal structure on the single-ion anisotropy of the Fe in hexagonal and orthorhombic LuFeO$_3$. We found that the low structural symmetry of the local environment splits the Fe 3d orbital states by the crystal fields; the spin anisotropy of these one-electron states is then generated via spin orbit coupling. In addition, the electronic configurations of Fe$^{3+}$ is found more complex than nominal valency arguments, i.e. the spin minority states are also partly occupied due to the Fe 3d-O 2p hybridization. These occupancies of the different spin minority states are uneven because of their different spatial distributions. The single-ion magnetic anisotropy of Fe$^{3+}$ is then caused by the spin anisotropy of the one-electron 3d states and the uneven occupancies of these states. For h-LuFeO$_3$, the D$_{3h}$ symmetry of the FeO$_5$ is responsible for the anisotropy between the $x - y$ plane and the $z$ axis, while the distortion from the D$_{3h}$ to C$_S$ symmetry generates the anisotropy within the $x - y$ plane; 2) for o-LuFeO$_3$, the distortion from O$_h$ to D$_{2h}$ symmetry of the FeO$_6$ generates the anisotropy between the $x$, $y$ and $z$ directions. The key role of the local structural symmetry suggests a route in tuning the magnetism in LuFeO$_3$ by fine adjustment of the crystal structures, as well as a route in coupling the electric and magnetic degrees of freedom via structural distortions.

## Acknowledgement

This project was primarily supported by the National Science Foundation through the Nebraska Materials Research Science and Engineering Center (Grant No. DMR-1420645). Additional support was provided by the Semiconductor Research Corporation through the Center for Nanoferroic Devices and the SRC-NRI Center under Task ID 2398.001. The Canadian Light Source is funded by the Canada Foundation for Innovation, the Natural Sciences and Engineering Research Council of Canada, the National Research Council Canada, the Canadian Institutes of Health Research, the Government of Saskatchewan, Western Economic Diversification Canada, and the University of Saskatchewan. Use of the Advanced Photon Source was supported by the U.S. Department of Energy, Office of Science, Office of Basic Energy Sciences, under Contract No. DE-AC02-06CH11357. Computations were performed utilizing the Holland Computing Center of the University of Nebraska. X.M. Cheng acknowledges support from the National Science Foundation under Grant No. 1053854.

# References

[1]  J. H. Lee, L. Fang, E. Vlahos, X. Ke, Y. W. Jung, L. F. Kourkoutis, J.-W. Kim, P. J. Ryan, T. Heeg, M. Roeckerath, V. Goian, M. Bernhagen, R. Uecker, P. C. Hammel, K. M. Rabe, S. Kamba, J. Schubert, J. W. Freeland, D. a Muller, C. J. Fennie, P. Schiffer, V. Gopalan, E. Johnston-Halperin, and D. G. Schlom, Nature **466**, 954 (2010).

[2]  M. Nakamura, Y. Tokunaga, M. Kawasaki, and Y. Tokura, Appl. Phys. Lett. **98**, 082902 (2011).

[3]  A. A. Bossak, I. E. Graboy, O. Y. Gorbenko, A. R. Kaul, M. S. Kartavtseva, V. L. Svetchnikov, and H. W. Zandbergen, Chem. Mater. **16**, 1751 (2004).

[4]  A. P. Ramirez, Annu. Rev. Mater. Sci. **24**, 453 (1994).

[5]  J. Stohr and H. C. Siegmann, *Magnetism from Fundamentals to Nanoscale Dynamics* (Springer, Berlin, 2006).

[6]  J. Stöhr, J. Magn. Magn. Mater. **200**, 470 (1999).

[7]  M. T. J. and P. J. H. B. and F. J. A. den B. and J. J. de Vries, Reports Prog. Phys. **59**, 1409 (1996).

[8]  A. R. Akbashev, A. S. Semisalova, N. S. Perov, and A. R. Kaul, Appl. Phys. Lett. **99**, 122502 (2011).

[9]  Y. K. Jeong, J. Lee, S. Ahn, S.-W. Song, H. M. Jang, H. Choi, J. F. Scott, H. Wang, I. V Solovyev, W. Wang, X. Wang, P. Ryan, D. J. Keavney, T. Z. Ward, L. Zhu, X. M. Cheng, J. Shen, L. He, X. Xu, and X. Wu, J. Am. Chem. Soc. **134**, 1450 (2012).

[10]  E. Magome, C. Moriyoshi, Y. Kuroiwa, A. Masuno, and H. Inoue, Jpn. J. Appl. Phys. **49**, 09ME06 (2010).

[11]  J. a. Moyer, R. Misra, J. a. Mundy, C. M. Brooks, J. T. Heron, D. a. Muller, D. G. Schlom, and P. Schiffer, APL Mater. **2**, 012106 (2014).

[12]  W. Wang, H. Wang, X. Xu, L. Zhu, L. He, E. Wills, X. Cheng, D. J. Keavney, J. Shen, X. Wu, and X. Xu, Appl. Phys. Lett. **101**, 241907 (2012).

[13]  W. Wang, J. Zhao, W. Wang, Z. Gai, N. Balke, M. Chi, H. N. Lee, W. Tian, L. Zhu, X. Cheng, D. J. Keavney, J. Yi, T. Z. Ward, P. C. Snijders, H. M. Christen, W. Wu, J. Shen, and X. Xu, Phys. Rev. Lett. **110**, 237601 (2013).

[14]  X. Xu and W. Wang, Mod. Phys. Lett. B **28**, 1430008 (2014).

[15]  S. Cao, T. R. Paudel, K. Sinha, X. Jiang, W. Wang, E. Y. Tsymbal, X. Xu, and P. a Dowben, J. Phys. Condens. Matter **27**, 175004 (2015).

[16]  *A Small Barrier and Energy Difference between the Face Centered Cubic (fcc) and Hexagonal Close-Packed (hcp) Structures Is Much More Common.*

[17]  H. Wang, I. V. Solovyev, W. Wang, X. Wang, P. P. J. Ryan, D. J. Keavney, J.-W. Kim, T. Z. Ward, L. Zhu, J. Shen, X. M. Cheng, L. He, X. Xu, X. Wu, J. Shen, L. He, X. Xu, and X. Wu, Phys. Rev. B **90**, 014436 (2014).

[18] S. M. M. Disseler, J. A. A. Borchers, C. M. M. Brooks, J. A. A. Mundy, J. A. A. Moyer, D. A. A. Hillsberry, E. L. L. Thies, D. A. A. Tenne, J. Heron, M. E. E. Holtz, J. D. D. Clarkson, G. M. M. Stiehl, P. Schiffer, D. A. A. Muller, D. G. G. Schlom, and W. D. D. Ratcliff, Phys. Rev. Lett. **114**, 217602 (2015).

[19] R. L. White, J. Appl. Phys. **40**, 1061 (1969).

[20] A. Munoz, J. A. Alonso, M. J. Martinez-Lope, M. T. Casais, J. L. Martinez, M. T. Fernandez-Diaz, M. J. Martínez-Lope, M. T. Casáis, J. L. Martínez, M. T. Fernández-Díaz, M. J. Martinez-Lope, M. T. Casais, J. L. Martinez, and M. T. Fernandez-Diaz, Phys. Rev. B **62**, 9498 (2000).

[21] H. Das, A. L. Wysocki, Y. Geng, W. Wu, and C. J. Fennie, Nat Commun **5**, 2998 (2014).

[22] E. Pavarini, E. Koch, F. Anders, M. Jarrell, I. for A. Simulation, and A. S. C. Electrons, *Correlated Electrons: From Models to Materials Lecture Notes of the Autumn School Correlated Electrons 2012 at Forschungszentrum Jülich, 3 - 7 September 2012* (Forschungszentrum Jülich, Zentralbibliothek, Verl., Jülich, 2012).

[23] *See Supplementary Material at for More Detailed Information on X-Ray Absorption Spectroscopy, Crystal Field Splitting, Hybridization, and Single-Ion Anisotropy*.

[24] D.-Y. Cho, J.-Y. Kim, B.-G. Park, K.-J. Rho, J.-H. Park, H.-J. Noh, B. J. Kim, S.-J. Oh, H.-M. Park, J.-S. Ahn, H. Ishibashi, S.-W. Cheong, J. H. Lee, P. Murugavel, T. W. Noh, A. Tanaka, and T. Jo, Phys. Rev. Lett. **98**, 217601 (2007).

[25] L. Köhler and G. Kresse, Phys. Rev. B **70**, 165405 (2004).

[26] D. Y. Cho, S. J. Oh, D. G. Kim, A. Tanaka, and J. H. Park, Phys. Rev. B **79**, 1 (2009).

[27] K.-T. Ko, H.-J. Noh, J.-Y. Kim, B.-G. Park, J.-H. Park, A. Tanaka, S. B. Kim, C. L. Zhang, and S.-W. Cheong, Phys. Rev. Lett. **103**, 207202 (2009).

[28] M. S. Dresselhaus, G. Dresselhaus, and A. Jorio, *Group Theory Application to the Physics of Condensed Matter* (Springer-Verlag, Berlin, 2007).

[29] F. A. Cotton, *Chemical Applications of Group Theory* (Wiley, New York, 1990).

[30] W. A. Harrison, *Electronic Structure and the Properties of Solids : The Physics of the Chemical Bond* (Freeman, San Francisco, 1980).

[31] S. Sugano, Y. Tanabe, and H. Kamimura, *Multiplets of Transition-Metal Ions in Crystals* (Academic Press, New York, 1970).

[32] M. Marezio, J. P. Remeika, and P. D. Dernier, Acta Crystallogr. Sect. B **26**, 2008 (1970).

[33] S. M. Disseler, J. A. Borchers, C. M. Brooks, J. A. Mundy, J. A. Moyer, D. A. Hillsberry, E. L. Thies, D. A. Tenne, J. Heron, M. E. Holtz, J. D. Clarkson, G. M. Stiehl, P. Schiffer, D. A. Muller, D. G. Schlom, and W. D. Ratcliff, Phys. Rev. Lett. **114**, 217602 (2015).

[34] B. S. Holinsworth, D. Mazumdar, C. M. Brooks, J. A. Mundy, H. Das, J. G. Cherian, S. A. McGill, C. J. Fennie, D. G. Schlom, and J. L. Musfeldt, Appl. Phys. Lett. **106**, 082902 (2015).

[35] V. V. Pavlov, a. R. Akbashev, a. M. Kalashnikova, V. a. Rusakov, a. R. Kaul, M.

Bayer, and R. V. Pisarev, J. Appl. Phys. **111**, 56103 (2012).

# Figures and captions

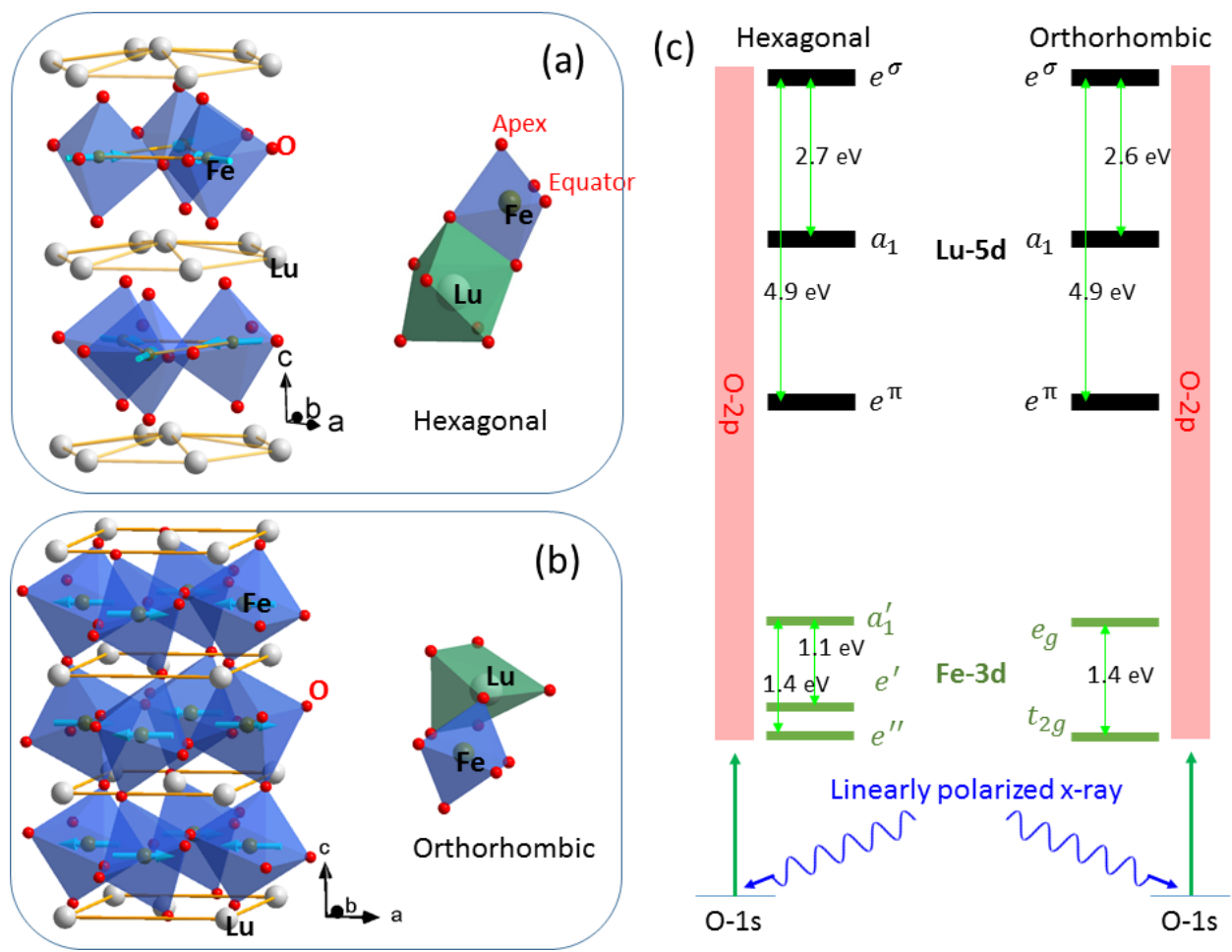

**Figure 1**. Lattice structures of hexagonal (a) and orthorhombic (b) LuFeO$_3$ as well as the local environments of the Lu and Fe sites. The thick arrows in (a) and (b) indicate the orientations of the spins. (c) Schematics of the O K edge excitation in LuFeO$_3$. The crystal-field-splitting energies are measured from the XAS spectra (see text).

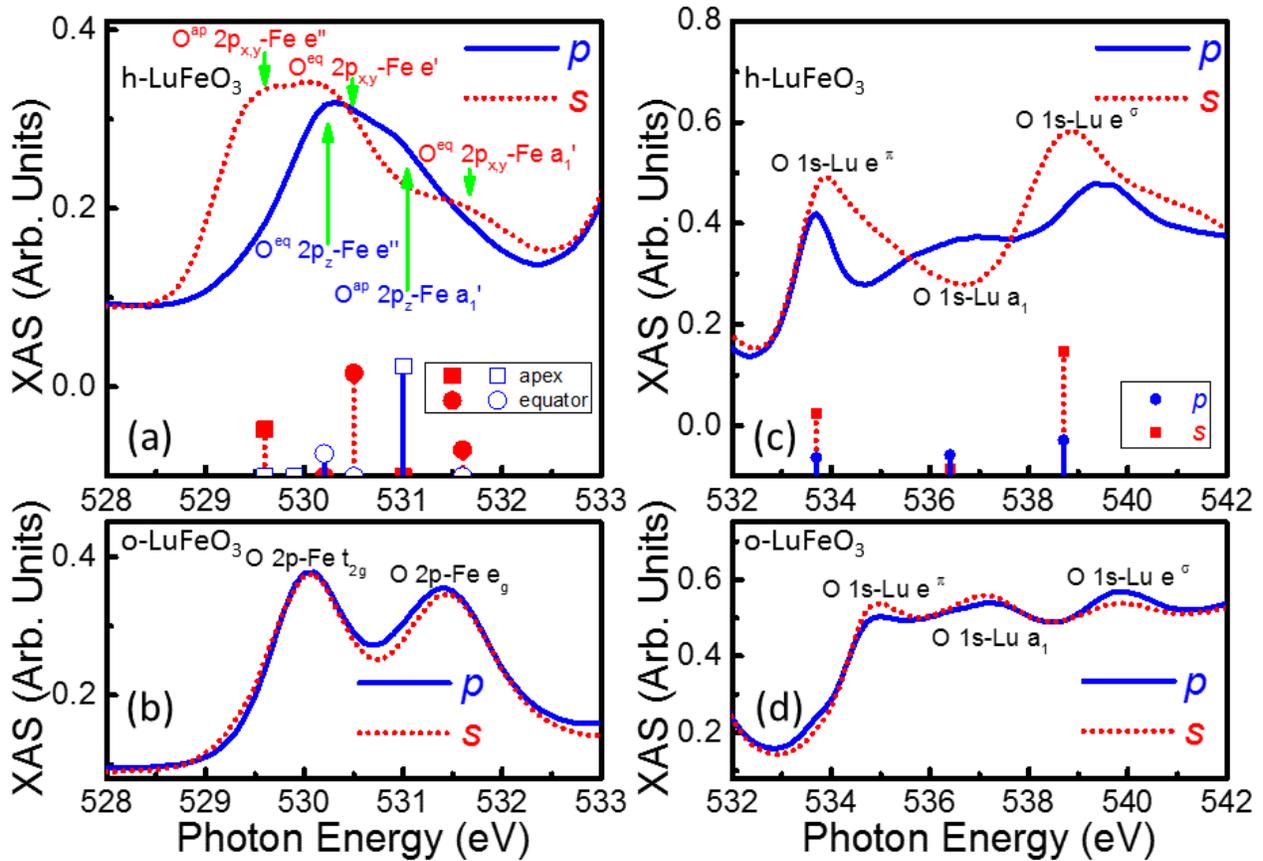

**Figure 2**. Absorption spectra corresponding to the O K edge with linearly polarized ($s$: in plane, $p$: out of plane, see S1 [23]) x-ray in LuFO$_3$. The spectra corresponding to the Fe 3d-O 2p hybridizations are displayed for hexagonal (a) and orthorhombic (b) LuFeO$_3$. The spectra corresponding to the Lu 5d-O 2p hybridizations are displayed for hexagonal (c) and orthorhombic (d) LuFeO$_3$. The vertical lines (solid: $p$ polarization, dashed: $s$ polarization) in (a) and (c) are the results from the calculation of hybridization using the Harrison's method (see text). The arrows in (a) point to the energies of the excitations corresponding to the 5 hybridized states in Fig. 4. In (c)-(e), the hybridizations corresponding to the excitations peaks are labelled.

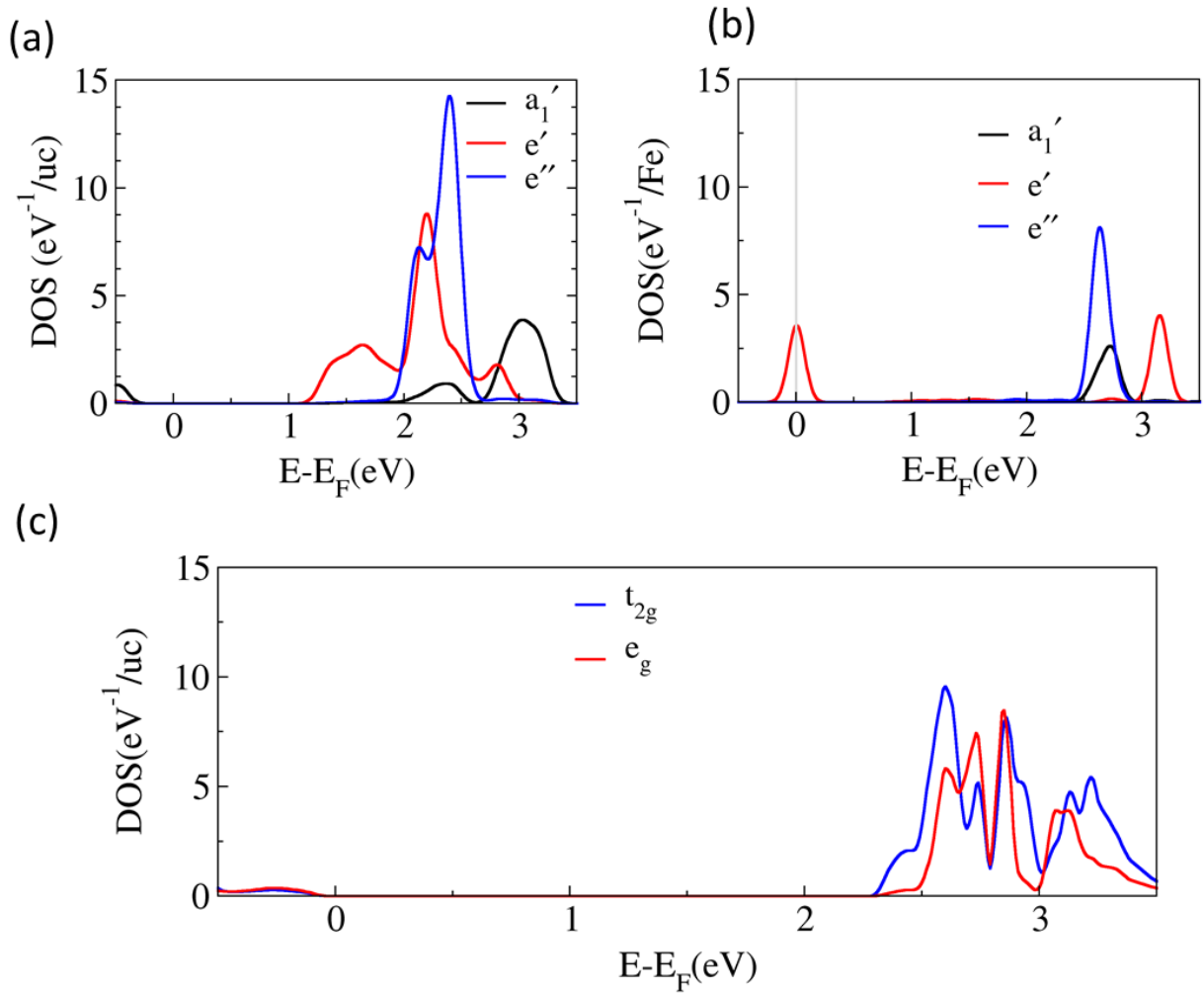

**Figure 3**. Calculated density of states (DOS) of h-LuFeO$_3$ resolved into $e'$ and $e''$ according to the D$_{3d}$ symmetry of h-LuFeO$_3$ (a), DOS projected at the Fe atom bonded with oxygen at apex site containing ½ hole in core 1s states (b), and $t_{2g}$ and $e_g$ resolved DOS of o-LuFeO$_3$ projected at Fe atom according to the O$_h$ symmetry of the crystal (c).

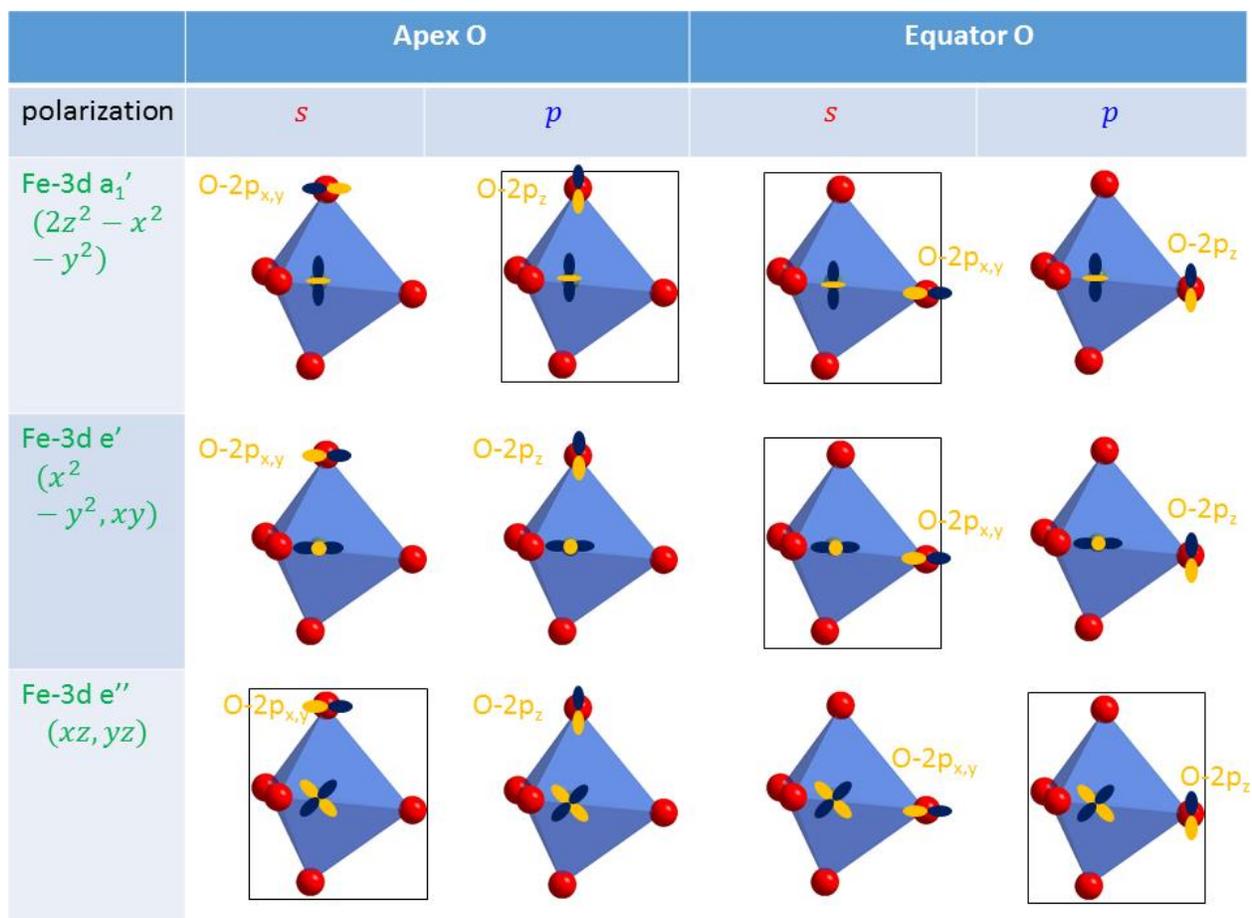

**Figure 4**. Model of hybridization between Fe 3d and O 2p illustrated using the relative position between the wave functions at different configurations. The configurations that correspond to significant hybridizations are boxed.

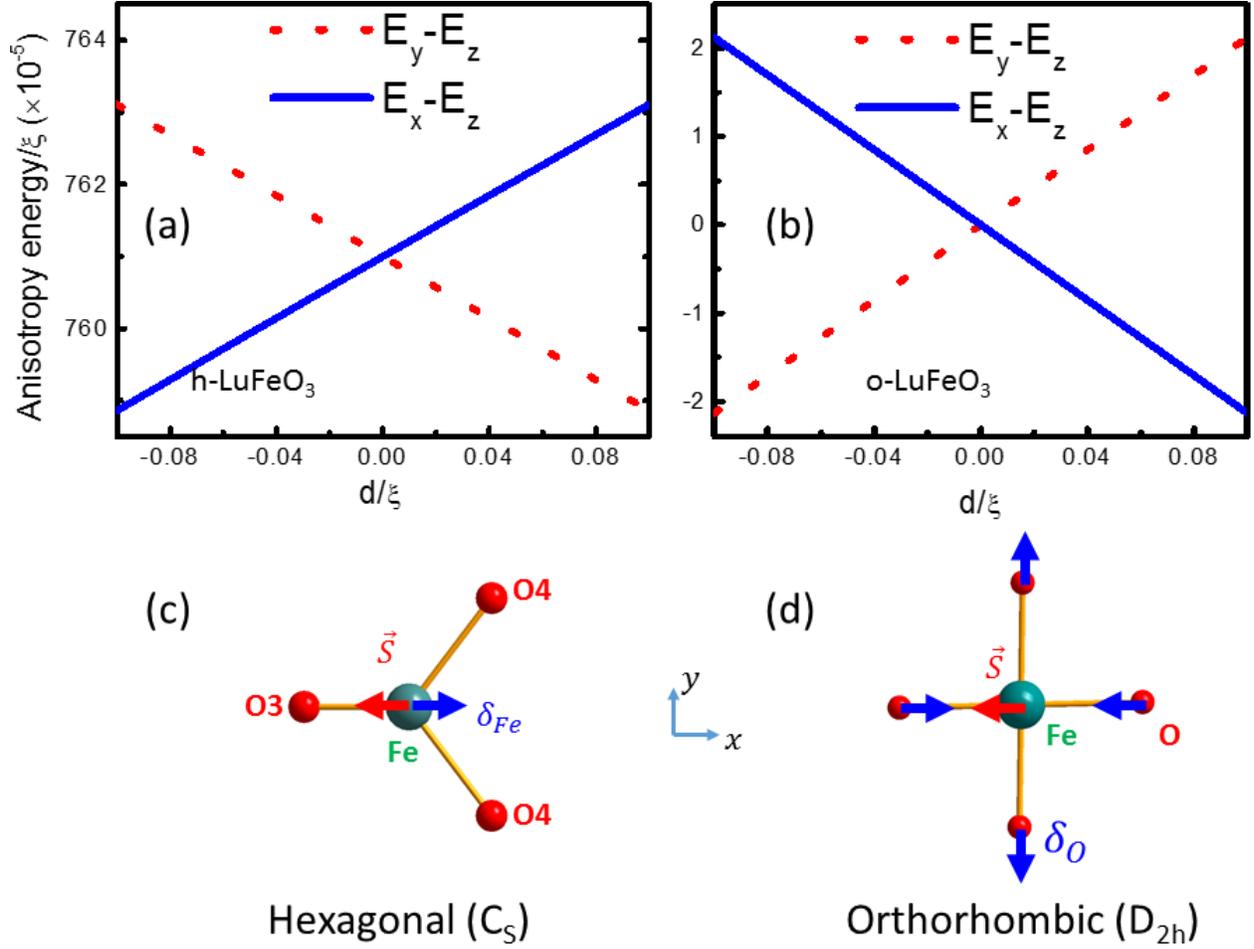

**Figure 5**. Calculated anisotropy energy as a function of lattice distortion energy parameter $d$ for h-LuFeO$_3$ (a) and o-LuFeO$_3$ (b). Observed lattice distortion pattern and preferred spin orientation from single-ion anisotropy at low temperature in h-LuFeO$_3$ (c) and o-LuFeO$_3$ (d) (See S7) [23]. The structural models are viewed along the $c$ axis. The O3 and O4 are the equator oxygen atoms (see Fig. 1 and S9 for definition). The exchange interaction and spin-orbit coupling parameters are assumed as $J = 4$ eV and $\xi = 0.05$ eV. For the crystal field interaction, the experimentally observed parameters are used. The occupancy of the spin majority states are set as one. The occupancy for the spin minority states are set as small numbers proportional to the calculated hybridizations according to the Harrison's method. For h-LuFeO$_3$, $n_{e''} = 0.046$, $n_{e'} = 0.068$, and $n_{a_1'} = 0.18$; for o-LuFeO$_3$, $n_{t_{2g}} = 0.02$ and $n_{e_g} = 0.071$. Notice that $d \propto -\delta_{Fe}$ and $d \propto \delta_O$.
19

## Table and Caption

| Displacement ($\times 10^{-3}$) | 6 K | 100 K | 110 K | 130 K | 150 K | 200 K | 300 K | Error | 300 K [10] |
|---|---|---|---|---|---|---|---|---|---|
| $\delta_{Fe}/a$ | 1.9 | 1.6 | 1.5 | 1.4 | 1.4 | 1.0 | 0.1 | 0.9 | 0 |

**Table 1**. Displacements of the Fe sites ($\delta_{Fe}$), defined as the displacement of Fe site away from the nearest equator oxygen site (see Fig. 5(c)) [23], where $a$ is the lattice constants of the basal plane in h-LuFeO$_3$.

# On the Structural Origin of the Single-ion Magnetic Anisotropy in LuFeO$_3$: supplementary material

## S1. X-ray absorption at the oxygen K-edge and the role of metal-oxygen hybridization

Figure S1.1 (a) depicts the experimental configuration of the x-ray absorption experiments. The x-ray has an incident angle of 16 degree above the plane of the sample. For the *p* polarization, the electric vector of the photon is in the sample surface plane. For the *s* polarization, the electric vector of the photon is almost perpendicular to the sample surface, since $\cos(16°) = 0.96$.

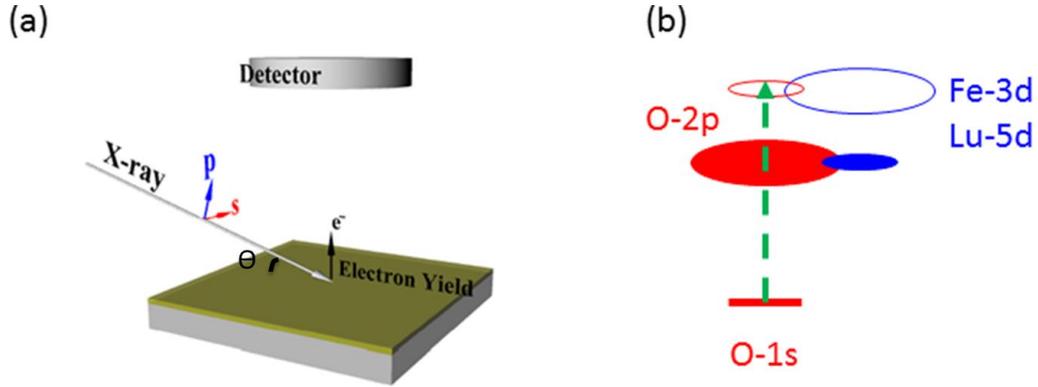

**Figure S1.1** Illustration of the x-ray absorption experiment setup (a) and the physical process for excitations at the oxygen K edge (b). The filled (empty) shapes in (b) indicate the occupied (unoccupied) orbitals.

The physical process of the excitation from O-1s orbital (K edge) is illustrated in the Fig. S2(b). The intensity of the optical transition is determined by the transition matrix $<\phi_{O-1s}|\vec{r}|\phi_f>$, where $\phi_{O-1s}$ is the O-1s orbital and $\phi_f = \phi_{Me} + a\phi_{O-2p}$, where $\phi_{Me}$ is the metal (Fe and Lu) valence orbital and the $\phi_{O-2p}$ is the O-2p orbital. The matrix element $<\phi_{O-1s}|\vec{r}|\phi_{O-2p}>$ is significant because the O-1s and O-2p are on the same atom and the two wave functions have different symmetries with respect to inversion. In contrast, the matrix element $<\phi_{O-1s}|\vec{r}|\phi_{Me}>$ is much less significant because the overlap between the O-1s and valence orbital of metal (Fe and Lu) atoms are much less than that between O-1s and O-2p. Therefore, although the coefficient $a$ may be small, the observed intensity of the O-K edge x-ray absorption are coming mainly from the transition of partially unoccupied O-2p orbital. The partially unoccupied O-2p excitation is coming from the hybridization described in Fig. S1.1

## S2. O-K edge of various compounds

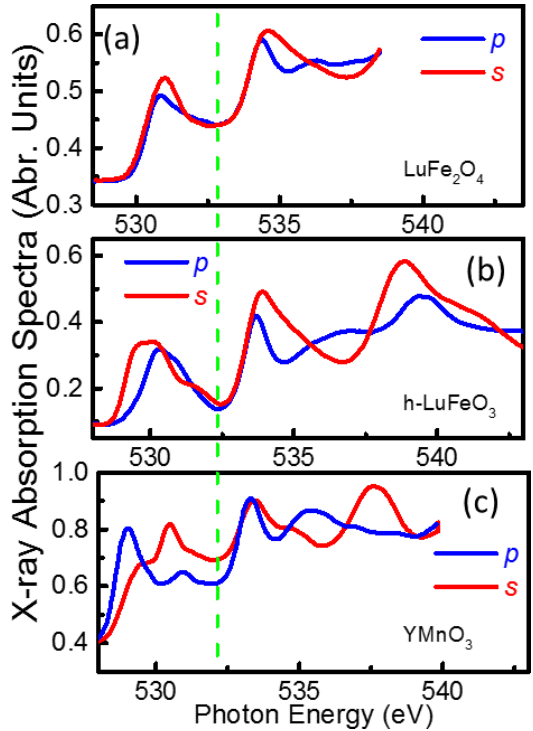

**Figure S2.1** Comparison between the XAS spectra of O-K edge in $LuFe_2O_4$, h-$LuFeO_3$, and $YMnO_3$. The data of the $LuFe_2O_4$ and $YMnO_3$ are digitized from Ref. [1] and [2] respectively.

As shown in Fig. S2.1, XAS spectra of the K-edge of $LuFe_2O_4$, h-$LuFeO_3$, and $YMnO_3$ are displayed. [1,2] The spectra below 532 eV are very different, while the part above 532 eV are more similar. This has to do with the structure and electronic structure. For $LuFe_2O_4$ and $LuFeO_3$, the local environment of Lu are similar except for the difference Lu-O bond length caused in the ferroelectric distortion, which explains the similarity for energy above 532 eV. Although the local environment of the Fe in $LuFe_2O_4$ and h-$LuFeO_3$ are both trigonal bipyramid, the valence difference (2.5+ in $LuFe_2O_4$ and 3+ in $LuFeO_3$) causes the difference in the electronic structure, which is the origin of the spectra difference below 523 eV. The h-$LuFeO_3$ and $YMnO_3$ are isomorphic, i.e. the local environment of the Y and Lu have the same symmetry and the local environment of the Mn and Fe have the same symmetry. The spectra above 532 eV corresponds to Lu-5d and Y-4d respectively. Because of the similar orbital symmetry and local environment, the spectra look similar. The spectra of Y-4d is more like a shrunk version of the Lu-5d, suggesting that the Lu-5d orbitals are more extended and sensitive to the crystal field generated by the local environment.

## S3. Final states of the O K-edge excitations

In LuFeO$_3$, an electron can be excited from O-1s orbital to O-2p orbital, which is not totally filled because of the hybridization with metal (Fe and Lu) orbitals. Effectively, the electron is excited from an O-1s orbital to a metal (Fe and Lu) orbital. The observed spectra may span a sizable range of energy around the O-1s to O-2p energy difference, depending on the metal (Fe anf Lu) orbital that the O-2p orbitals hybridize with. The structure of the spectra also reflects the density of states of the metal (Fe and Lu) orbitals and the selection rules of the excitations.

### S3.1 O-K edge excitation related to O-2p Fe-3d hybridization

In the ground state of LuFeO$_3$, O-1s is fully filled and Fe$^{3+}$ is expected to take the high spin ($S_z = \frac{5}{2}$) state. Using the language of second quantization, a many body wave function can be written as $\phi = \prod_{v_i} C^\dagger_{v_i} |vac>$, where $|vac>$ represent the vacuum state and $\{v_j\}$ are the one-electron wave functions. So we can write down the wave function of the ground state $\phi_g = C^\dagger_{1s,\uparrow} C^\dagger_{1s,\downarrow} C^\dagger_{xz,\uparrow} C^\dagger_{yz,\uparrow} C^\dagger_{xx-yy,\uparrow} C^\dagger_{xz,\uparrow} C^\dagger_{yz,\uparrow} |0>$. When an electron is excited from an O-1s orbital to an Fe-3d orbital, there are multiple final states of Fe-3d$^6$ with $S_z = 2$. The energies of these states are determined by the Coulomb energy (electron-electron interaction) and the crystal field energies (one-electron energy), if the spin-orbit coupling is small. The possible energy distribution of these final state will determine the corresponding x-ray absorption spectra. Below, we will show that the Coulomb energy of different O-1s$^1$ Fe-3d$^6$ states (for $S_z = 2$) are the same, as long as the atomic orbitals are treated as hydrogen-like orbitals as an approximation. Therefore, the total energy will mostly be determined by the crystal-field energies, which can be treated using one-electron picture.

In order to calculate the energy distribution of these final states, we select the following basis:

$$\phi_{-2} = C^\dagger_{O1s,\uparrow} C^\dagger_{Fe3d-2,\downarrow} C^\dagger_{Fe3d-2,\uparrow} C^\dagger_{Fe3d-1,\uparrow} C^\dagger_{Fe3d0,\uparrow} C^\dagger_{Fe3d1,\uparrow} C^\dagger_{Fe3d2,\uparrow} |vac\rangle,$$

$$\phi_{-1} = C^\dagger_{O1s,\uparrow} C^\dagger_{Fe3d-1,\downarrow} C^\dagger_{Fe3d-2,\uparrow} C^\dagger_{Fe3d-1,\uparrow} C^\dagger_{Fe3d0,\uparrow} C^\dagger_{Fe3d1,\uparrow} C^\dagger_{Fe3d2,\uparrow} |vac\rangle,$$

$$\phi_0 = C^\dagger_{O1s,\uparrow} C^\dagger_{Fe3d0,\downarrow} C^\dagger_{Fe3d-2,\uparrow} C^\dagger_{Fe3d-1,\uparrow} C^\dagger_{Fe3d0,\uparrow} C^\dagger_{Fe3d1,\uparrow} C^\dagger_{Fe3d2,\uparrow} |vac\rangle,$$

$$\phi_1 = C^\dagger_{O1s,\uparrow} C^\dagger_{Fe3d1,\downarrow} C^\dagger_{Fe3d-2,\uparrow} C^\dagger_{Fe3d-1,\uparrow} C^\dagger_{Fe3d0,\uparrow} C^\dagger_{Fe3d1,\uparrow} C^\dagger_{Fe3d2,\uparrow} |vac\rangle,$$

$$\phi_2 = C^\dagger_{O1s,\uparrow} C^\dagger_{Fe3d2,\downarrow} C^\dagger_{Fe3d-2,\uparrow} C^\dagger_{Fe3d-1,\uparrow} C^\dagger_{Fe3d0,\uparrow} C^\dagger_{Fe3d1,\uparrow} C^\dagger_{Fe3d2,\uparrow} |vac\rangle,$$

where the index from -2 to 2 is the magnetic quantum number $m$ of the 3d orbital, and $|vac\rangle$ is the vacuum state. Below, we will show that the calculation of the Coulomb interaction can be simplified due to many special conditions

1) The Hamiltonian of the Coulomb interaction in the basis above is diagonal

Since the total $L_z$ is different for all these states, they do not mix due to the Coulomb interaction, i.e. the Hamiltonian that represents Coulomb interaction will be diagonal. In other words, only the diagonal terms $\langle \phi_i | \hat{V}_c | \phi_i \rangle$ needs to be calculated, where $\hat{V}_c = \frac{1}{2} \sum_{v_1, v_2, v_3, v_4} C^\dagger_{v_1} C^\dagger_{v_2} C_{v_3} C_{v_4}$ is the Coulomb interaction, where $\{v_j\}$ are atomic orbitals that can be specified by quantum numbers $n_j, l_j, m_j, \sigma_j$ ($\sigma_j$ is the spin states).

For the diagonal terms

$$\langle \phi_i | \hat{V}_c | \phi_i \rangle = \left\langle \phi_i \left| \frac{1}{2} \sum_{v_1, v_2} V_{v_1, v_2, v_2, v_1} C^\dagger_{v_1} C^\dagger_{v_2} C_{v_2} C_{v_1} \right| \phi_i \right\rangle - \left\langle \phi_i \left| \frac{1}{2} \sum_{v_1, v_2} V_{v_1, v_2, v_1, v_2} C^\dagger_{v_1} C^\dagger_{v_2} C_{v_1} C_{v_2} \right| \phi_i \right\rangle,$$

and

$$V_{\nu_1,\nu_2,\nu_2,\nu_1} = \sum_{k=|l_1-l_2|}^{k=l_1+l_2} c^k(l_1m_1;l_1m_1)c^k(l_2m_2;l_2m_2)R^k(n_1l_1,n_2l_2)$$

where $c^k(l_1m_1;l_1m_1)$ and $c^k(l_2m_2;l_2m_2)$ are the Gaunt coefficients and $R^k(n_1l_1,n_2l_2)$ are the integral of the radial part of the wave function. [3]

The definition of the Gaunt coefficients is

$$c^k(l_1m_1;l_2m_2) = \sqrt{\frac{4\pi}{2k+1}} \int_0^{2\pi} d\phi \int_0^{2\pi} d\cos(\theta)\, Y_{l_1,m_1}^*(\theta,\phi) Y_{k,m_1-m_2}^*(\theta,\phi) Y_{l_2,m_2}^*(\theta,\phi).$$

2) The Coulomb interaction between the wave function $v_{O1s}$ and a wave function $v_{Fe3dm}$ does not depend on the quantum number $m$.

To show this condition, one just has to calculate $c^k(l_1m_1;l_1m_1)$ and $c^k(l_2m_2;l_2m_2)$, where $l_1 = m_1 = 0$, and $l_2 = 2$. The results are

$c^k(l_1m_1;l_1m_1) = c^0(0,0;0,0)$, and $c^k(l_2m_2;l_2m_2) = c^k(2,m_2;2,m_2)$.

Using the definition of the Gaunt coefficient, one can see that $c^k(2,m_2;2,m_2)$ is independent of $m_2$.

Therefore, the Coulomb interaction between the O-1s orbital and the Fe-3d orbitals are independent of the quantum number $m$. In addition, because the O-1s orbitals and the Fe-3d orbitals are on different atomic sites (the distance is approximately 2.0 Å), the Coulomb interactions between the O-1s orbitals and the Fe-3d orbitals are smaller than that between the Fe-3d orbitals.

3) Due to the particle-hole symmetry, the $3d^6$ problem can be treated as the $3d^4$ problem.

Therefore, we can rewrite the basis of the excited states as

$$\phi_{-2} = C_{Fe3d-1\uparrow}^\dagger C_{Fe3d0\uparrow}^\dagger C_{Fe3d1\uparrow}^\dagger C_{Fe3d2\uparrow}^\dagger |vac\rangle,$$

$$\phi_{-1} = C_{Fe3d-2\uparrow}^\dagger C_{Fe3d0\uparrow}^\dagger C_{Fe3d1\uparrow}^\dagger C_{Fe3d2\uparrow}^\dagger |vac\rangle,$$

$$\phi_0 = C_{Fe3d-2\uparrow}^\dagger C_{Fe3d-1\uparrow}^\dagger C_{Fe3d1\uparrow}^\dagger C_{Fe3d2\uparrow}^\dagger |vac\rangle,$$

$$\phi_1 = C_{Fe3d-2\uparrow}^\dagger C_{Fe3d-1\uparrow}^\dagger C_{Fe3d0\uparrow}^\dagger C_{Fe3d2\uparrow}^\dagger |vac\rangle,$$

$$\phi_2 = C_{Fe3d-2\uparrow}^\dagger C_{Fe3d-1\uparrow}^\dagger C_{Fe3d0\uparrow}^\dagger C_{Fe3d1\uparrow}^\dagger |vac\rangle.$$

Here we ignore the Coulomb interactions between the O 1s electrons and Fe 3d electrons because they are much smaller and independent on $m$, as discussed above. To calculate the Coulomb interaction $\langle\phi_i|\hat{V}_c|\phi_i\rangle$, one needs to calculate the interaction between the electron pairs, which is listed as the following (here we omit the notation Fe 3d and the spin for simplicity):

$$\langle vac|C_{-2}^\dagger C_{-1}^\dagger \hat{V}_c C_{-1} C_{-2}|vac\rangle = R^0 - \frac{8}{49}R^2 - \frac{9}{441}R^4,$$

$$\langle vac|C_{-2}^\dagger C_0^\dagger \hat{V}_c C_0 C_{-2}|vac\rangle = R^0 - \frac{8}{49}R^2 - \frac{9}{441}R^4,$$

$$\langle vac|C_{-2}^\dagger C_1^\dagger \hat{V}_c C_1 C_{-2}|vac\rangle = R^0 - \frac{2}{49}R^2 - \frac{39}{441}R^4,$$

$$\langle vac|C_{-2}^\dagger C_2^\dagger \hat{V}_c C_2 C_{-2}|vac\rangle = R^0 + \frac{4}{49}R^2 - \frac{69}{441}R^4,$$

$$\langle vac|C_{-1}^\dagger C_0^\dagger \hat{V}_c C_0 C_{-1}|vac\rangle = R^0 + \frac{1}{49}R^2 - \frac{54}{441}R^4,$$

$$\langle vac|C_{-1}^\dagger C_1^\dagger \hat{V}_c C_1 C_{-1}|vac\rangle = R^0 - \frac{5}{49}R^2 - \frac{24}{441}R^4,$$

$$\langle vac|C_{-1}^\dagger C_2^\dagger \hat{V}_c C_2 C_{-1}|vac\rangle = R^0 - \frac{2}{49}R^2 - \frac{39}{441}R^4,$$

$$\langle vac|C_0^\dagger C_1^\dagger \hat{V}_c C_1 C_0|vac\rangle = R^0 + \frac{1}{49}R^2 - \frac{54}{441}R^4,$$

$$\langle vac|C_0^\dagger C_2^\dagger \hat{V}_c C_2 C_0|vac\rangle = R^0 - \frac{8}{49}R^2 - \frac{9}{441}R^4,$$

$$\langle vac|C_1^\dagger C_2^\dagger \hat{V}_c C_2 C_1|vac\rangle = R^0 - \frac{8}{49}R^2 - \frac{9}{441}R^4,$$

where $R^0, R^2$, and $R^4$ are abbreviation of $R^0(3,2,3,2), R^2(3,2,3,2)$, and $R^4(3,2,3,2)$.

In addition, $\langle vac|C_{m1}^\dagger C_{m2}^\dagger \hat{V}_c C_{m2} C_{m1}|vac\rangle = \langle vac|C_{m2}^\dagger C_{m1}^\dagger \hat{V}_c C_{m1} C_{m2}|vac\rangle$, and $\langle vac|C_m^\dagger C_m^\dagger \hat{V}_c C_m C_m|vac\rangle = 0$.

Using these results, one gets $<\phi_m|\hat{V}_c|\phi_m> = R^0 - \frac{21}{49}R^2 - \frac{189}{441}R^4$, for all m=-2 to 2.

Hence, the Coulomb interaction provides a constant energy for the 3d$^6$ configuration, meaning the total energy is determined by the one-electron (e.g. crystal field) energies.

### S3.2 O-K edge excitation related to O-2p Lu-5d hybridization

This initial states of the this excitation spectroscopy at the O 1s core (the K edge) is the ground state which corresponds to $\phi_g = C_{1s,\uparrow}^\dagger C_{1s,\downarrow}^\dagger|0>$. For the excited state, there will be one electron in the O-1s orbital and one electron in the Lu-5d orbital. The only many-body energy will be the Coulomb interaction between one Lu-5d electron with one O-1s electron, which is shown small and independent of quantum number m. Therefore, the energy of the final states $\phi_f = C_{1s\uparrow}^\dagger C_{Lu-5dm\downarrow}^\dagger|0>$ can be calculated according to the one-electron energy, which is mainly determined by the crystal field splitting.

## S4. Crystal field splitting from x-ray absorption spectra

Since the x-ray absorption spectra of O K edge can be approximately treated using one electron picture (see section 3), it is possible to extract the crystal field splitting from the spectra. We show here an example of analyzing crystal field splitting of Fe-3d using the O K edge x-ray absorption spectra.

Table S4.1

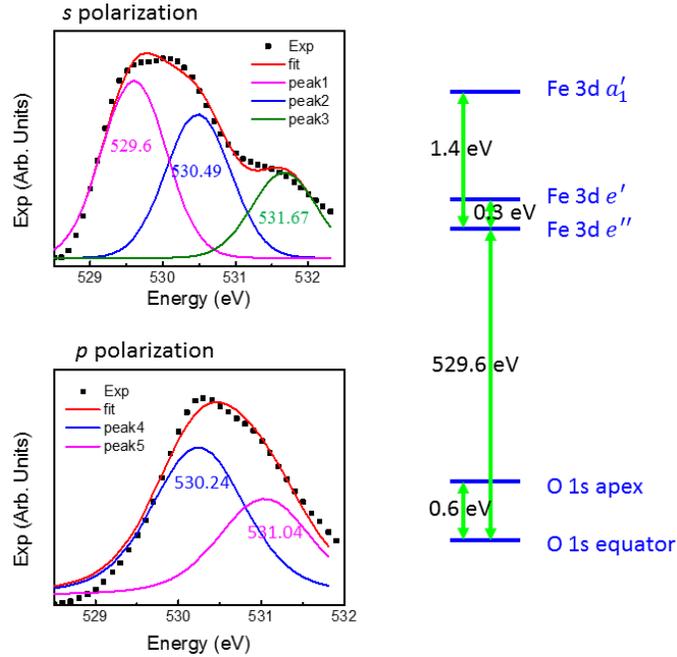

**Figure S4.1** Fit to the x-ray absorption spectra using Voigt peaks and the extracted values for the energy levels.

| Peak index | Peak positions (eV) | Assignment | Peak positions from fit (eV) |
|---|---|---|---|
| 1 | 529.6 | $O^{ap} 1s \rightarrow Fe\ 3d\ e''$ | 529.6 |
| 2 | 530.49 | $O^{eq} 1s \rightarrow Fe\ 3d\ e'$ | 530.49 |
| 3 | 531.67 | $O^{eq} 1s \rightarrow Fe\ 3d\ a_1'$ | 531.67 |
| 4 | 530.24 | $O^{eq} 1s \rightarrow Fe\ 3d\ e''$ | 530.23 |
| 5 | 531.04 | $O^{ap} 1s \rightarrow Fe\ 3d\ a_1'$ | 531.04 |

First, we fit the x-ray absorption spectra using Voigt peaks, as shown in Fig S4.1. The resulting peak positions are listed in the Table S4.1.

Second, we calculate the crystal field energies using these peak positions. There are totally 4 unknown variables here. The first three are the three crystal field energies of Fe 3d. The fourth one has to do with the two oxygen positions (apex and equator). Assuming that the crystal field energies are $E_{e''}$, $E_{e'}$, $E_{a_1'}$, and the energy difference between the 1s of apex and equator oxygen atoms is $\delta$, one can solve these four variables using the peak positions extracted from the fit.

Note that there are five peak positions which give five equations, but there are only 4 unknown variables. So we use the least square fit to find the 4 variables. The results are $E_{e''} = 529.6$ eV, $E_{e'} - E_{e''} = 0.25$ eV, $E_{a_1'} - E_{e''} = 1.43$ eV, and $\delta = 0.64$ eV. As shown in Table S4.1, the calculated energies (from the least square fit) match the observed energies almost exactly, indicating the validity of the model.

## S5. Crystal field splitting in Fe 3$d$ and Lu 5$d$: analysis using group theory

In crystalline LuFeO$_3$, metal (Fe and Lu) atoms sites are surrounded by oxygen sites, forming so-called local environments. For the atomic orbitals of metal (Fe and Lu) atoms (e.g. $d$ orbital) that are not very delocalized but exposed to neighboring oxygen atoms, the change of the one-electron energy can be analyzed using the crystal field model. In this case, the oxygen atoms are treated as point charges at their atomic positions; the degeneracy of the one-electron energy of the atomic orbital (e.g. $d$ orbitals) of the metals in general splits in the electric field generated by the point charges; this is called crystal field splitting.

Here we show the two-step analysis. The first step is to qualitatively analyze the crystal field splitting according to the symmetry of the local environment. The second step is to analyze the crystal field splitting semi-quantitatively using multipole expansion of the field.

In group theory analysis, the first step is to find the symmetry of the local environment of the metal (Fe anf Lu) atoms; the symmetry is described using a point group. In many cases (e.g. in LuFeO$_3$), the symmetry of the local environment is actually very low. For example, the symmetry of FeO$_5$ in h-LuFeO$_3$ should be described by C$_S$, since there is only one symmetry operation (mirror plane) in addition to identity. But the deviation from the high symmetry D$_{3h}$ is small, so we can approximately treat the symmetry of FeO$_5$ using the D$_{3h}$ symmetry, unless this small energy splitting is important, which is the case when the orbital angular momentum is discussed. For the LuO$_7$ local environment in h-LuFeO$_3$, we choose C$_{3v}$ symmetry as the approximation. In o-LuFeO$_3$, the local symmetry of the FeO$_6$ and LuO$_6$ are approximately O$_h$ and C$_{3v}$ respectively.

Once the symmetry of the local environment is determined in terms of point group, the crystal field splitting can be analyzed using the representations of the point group. In principle, any sequence of functions form a representation of the point group, which can be reduced to a few irreducible representations. For the atomic orbitals of certain angular quantum number ($l$=2 for $d$ orbital), the calculation of the representation (more precisely, the characters of the representation) can be found in Dressohouse's book. [4] We list below the calculation for the $d$ orbitals in D$_{3h}$, O$_h$, and C$_{3v}$ orbitals. The decomposition is done by carrying out the dot product between the representation and irreducible representations.

In the tables below, the first line shows the group elements in classes. The following lines show the character of the irreducible representations. The last line shows the representation of the interested atomic orbital. The last two columns show typical linear and quadratic representation basis. The third column from the right shows the decomposition of the representation at the last line in terms of the irreproducible representations.

Table S5.1 $d$ orbital in D$_{3h}$ symmetry (Fe-3$d$ in h-LuFeO$_3$ approximation)

| D$_{3h}$ | E | 2C$_3$ | 3C$_2$ | σ$_h$ | 2S$_3$ | 3σ$_v$ | | | |
|---|---|---|---|---|---|---|---|---|---|
| $a_1'$ | 1 | 1 | 1 | 1 | 1 | 1 | 1 | | $z^2$ |
| $e'$ | 2 | -1 | 0 | 2 | -1 | 0 | 1 | $(x, y)$ | $(x^2 - y^2, xy)$ |
| $a_2''$ | 1 | 1 | -1 | -1 | -1 | 1 | 0 | $z$ | |
| $e''$ | 2 | -1 | 0 | -2 | 1 | 0 | 1 | | $(xz, yz)$ |

| | | | | | |
|---|---|---|---|---|---|
| d (l=2) | 5 | -1 | 1 | 1 | 1 |

According to the table above, the d orbitals in a D$_{3h}$ symmetry are decomposed into $a_1' + e' + e''$. By looking at the typical quadratic basis, one can associate these irreproducible representations with different components of the d orbitals: $a_1'\,(z^2) + e'\,(x^2 - y^2) + e''\,(xz, yz)$. Experimentally, one then expects to see that the degeneracy of the d orbital is broken; three one-electron energies are supposed to be observed.

Table S5.2 d orbital in C$_s$ symmetry (Fe-3d in h-LuFeO$_3$)

| C$_s$ | E | σ$_v$ | | | |
|---|---|---|---|---|---|
| a' | 1 | 1 | 3 | x, y | $x^2, y^2, z^2, xy$ |
| a'' | 1 | -1 | 2 | z | $xz, yz$ |
| d (l=2) | 5 | 1 | | | |

The real local environment of the Fe in h-LuFeO$_3$ has three distortions: $\Gamma_2^-$, $K_1$, and $K_3$. The $K_3$ distortion is a rigid rotation that does not change the local environment. The $K_1$ distortion destroys the 3-fold rotational symmetry and the $\Gamma_2^-$ distortion combined with the $K_3$ mode removes the 2-fold rotational symmetry as well as the horizontal mirror plane. What's left is only the vertical mirror plane. Hence the point group becomes $C_S$. From the last two columns of the character table above, one can infer that the e' representation in D$_{3h}$ is decomposed into two a' in C$_s$; the e'' representation in D$_{3h}$ is decomposed into two a'' in C$_s$; the $a_1'$ representation in D$_{3h}$ becomes a' in C$_S$.

Table S5.3 d orbital in C$_{3v}$ symmetry (Lu-5d in h-LuFeO$_3$ and o-LuFeO$_3$ approximately)

| C$_{3v}$ | E | 2C$_3$ | 3σ$_v$ | | | |
|---|---|---|---|---|---|---|
| $a_1$ | 1 | 1 | 1 | 1 | z | $z^2$ |
| $a_2$ | 1 | 1 | -1 | 0 | | |
| e | 2 | -1 | 0 | 2 | (x, y) | $(x^2 - y^2, xy)\,(xz, yz)$ |
| d (l=2) | 5 | -1 | 1 | | | |

According to the table above, the d orbitals in a C$_{3v}$ local environment are decomposed into $a_1 + 2e$.

Table S5.4 d orbital in O$_h$ symmetry (Fe-3d in o-LuFeO$_3$ approximately)

| O$_h$ | E | 8C$_3$ | 6C$_2$ | 6C$_4$ | 3C$_2$ | i | 6S$_4$ | 8S$_6$ | 3σ$_h$ | 6σ$_d$ | | |
|---|---|---|---|---|---|---|---|---|---|---|---|---|
| $a_{1g}$ | 1 | 1 | 1 | 1 | 1 | 1 | 1 | 1 | 1 | 0 | | $x^2 + y^2 + z^2$ |
| $a_{2g}$ | 1 | 1 | -1 | -1 | 1 | 1 | -1 | 1 | 1 | -1 | 0 | |
| $e_g$ | 2 | -1 | 0 | 0 | 2 | 2 | 0 | -1 | 2 | 0 | 1 | $(z^2, x^2 - y^2)$ |
| $t_{1g}$ | 3 | 0 | -1 | 1 | -1 | 3 | 1 | 0 | -1 | -1 | 0 | |
| $t_{2g}$ | 3 | 0 | 1 | -1 | -1 | 3 | -1 | 0 | -1 | 1 | 1 | (xz, yz, xy) |
| $a_{1u}$ | 1 | 1 | 1 | 1 | 1 | -1 | -1 | -1 | -1 | -1 | 0 | |
| $a_{2u}$ | 1 | 1 | -1 | -1 | 1 | -1 | 1 | -1 | -1 | 1 | 0 | |
| $e_u$ | 2 | -1 | 0 | 0 | 2 | -2 | 0 | 1 | -2 | 0 | 0 | |
| $t_{1u}$ | 3 | 0 | -1 | 1 | -1 | -3 | -1 | 0 | 1 | 1 | 0 | (x, y, z) |
| $t_{2u}$ | 3 | 0 | 1 | -1 | -1 | -3 | 1 | 0 | 1 | -1 | 0 | |
| d (l=2) | 5 | -1 | 1 | -1 | 1 | 5 | -1 | -1 | 1 | 1 | | |

Table S5.5 $d$ orbital in D$_{2h}$ symmetry (Fe-3d in o-LuFeO$_3$)

| D$_{2h}$ | E | C$_2$(z) | C$_2$(y) | C$_2$(x) | i | σ (xy) | σ (xz) | σ (yz) | | | |
|---|---|---|---|---|---|---|---|---|---|---|---|
| $a_{1g}$ | 1 | 1 | 1 | 1 | 1 | 1 | 1 | 1 | 2 | | $x^2, y^2, z^2$ |
| $b_{1g}$ | 1 | 1 | -1 | -1 | 1 | 1 | -1 | -1 | 1 | | xy |
| $b_{2g}$ | 1 | -1 | 1 | -1 | 1 | -1 | 1 | -1 | 1 | | xz |
| $b_{3g}$ | 1 | -1 | -1 | 1 | 1 | -1 | -1 | 1 | 1 | | yz |
| $a_u$ | 1 | 1 | 1 | 1 | -1 | -1 | -1 | -1 | 0 | | |
| $b_{1u}$ | 1 | 1 | -1 | -1 | -1 | -1 | 1 | 1 | 0 | z | |
| $b_{2u}$ | 1 | -1 | 1 | -1 | -1 | 1 | -1 | 1 | 0 | x | |
| $b_{3u}$ | 1 | -1 | -1 | 1 | -1 | 1 | 1 | -1 | 0 | y | |
| d (l=2) | 5 | 1 | 1 | 1 | 5 | 1 | 1 | 1 | | | |

According to the table above, the $d$ orbitals in an D$_{2h}$ local environment are decomposed into $2a_{1g} + b_{1g} + b_{2g} + b_{3g}$. From the quadratic form of the irreducible representations in the character table above, one can infer that the $t_{2g}$ in O$_h$ symmetry is decomposed into $b_{1g} + b_{2g} + b_{3g}$ in the D$_{2h}$ symmetry; the $e_g$ in O$_h$ symmetry is decomposed into $2a_{1g}$.

The group theory analysis provides quick and qualitative analysis without having to know much detail about the local environment other than the symmetry. In addition, the symmetry of the split orbitals can also be found in the analysis, which is very useful. The main limitation is that it does not directly provide information about the energy relations between the spilt levels. This becomes inconvenient when a certain irreducible representation appears more than once in the decomposition, for example when a $d$ orbital is put in a C$_{3v}$ local environment.

To resolve this problem, we carry out semi-quantitative analysis using the multipole expansion of the crystal field.

## S6. Crystal field splitting in Fe $3d$ and Lu $5d$: analysis using multiplets model

In principle, the splitting of atomic levels (with angular momentum $l$) in a local environment can be found by diagonalizing the corresponding matrix of the crystal field potential $V_{CF}(\vec{r})$:

$V_{CF}^{m_1,m_2} = \int R_{n,l}^2(r) Y_{l,m_1}^*(\theta,\phi) V_{CF}(\vec{r}) Y_{l,m_2}(\theta,\phi) r^2 \sin(\theta)\, d\theta d\phi\, dr$, where

$Y_{l,m_1}(\theta,\phi)$ and $Y_{l,m_2}(\theta,\phi)$ are spherical harmonic functions, and $R_{n,l}(r)$ are the radial function of atomic orbital with main quantum number $n$, angular quantum number $l$, and magnetic quantum number $m$.

Using the spherical harmonic expansion,

$V_{CF}(\vec{r}) = -\frac{Z_o e^2}{4\pi\epsilon R} \sum_{k=0}^{\infty} \sum_{m=-k}^{k} \gamma_{k,m} \left(\frac{r}{R}\right)^k \sqrt{\frac{4\pi}{2k+1}} Y_{k,m}(\theta,\phi)$, where

$\gamma_{k,m} = \sqrt{\frac{4\pi}{2k+1}} \sum_{i=1}^{n_o} \left(\frac{r}{R_i}\right)^{k+1} Y_{k,m}^*(\theta_i,\phi_i)$ is called the structural factor, $Z_o$, $R$, and $i$ are point charge of the crystal field (from oxygen), the average distance between the charges and the metal (Fe and Lu) atom, and $i$ is the index of the point charges, respectively. [3]

Using the above definition, one can calculate the matrix elements using

$V_{CF}^{m_1,m_2} = \sum_{k \in \{0,2,\ldots,2l\}} \gamma_{k,m_1-m_2} c^k(l,m_1,l,m_2) U_{n,l,k}$, where

$U_{n,l,k} = -\frac{Z_o e^2}{4\pi\epsilon R^{k+1}} \int_0^\infty R_{n,l}^2(r) r^{k+2} dr = -\frac{Z_o e^2}{4\pi\epsilon R} \frac{\langle r^k \rangle_{n,l}}{R^k}$ is the integral that has the dimension of energy,

$\langle r^k \rangle_{n,l} \equiv \int_0^\infty R_{n,l}^2(r) r^{k+2} dr$,

and $c^k(l,m_1,l,m_2) = \int Y_{l,m_1}^*(\theta,\phi) Y_{k,m_1-m_2}(\theta,\phi) Y_{l,m_2}(\theta,\phi) \sin(\theta)\, d\theta d\phi$ are called the Gaunt coefficients which have been calculated and tabulated by Slater. [3]

Therefore, the problem of finding the matrix elements is reduced to find the structure factor $\gamma_{k,m}$ and energy integral $U_{n,l,k}$.

Before discussing any specific crystal field, we can examine a few points:

1) For the sum in $V_{CF}^{m_1,m_2}$, the term $k=0$ corresponds to $\gamma_{0,m_1-m_2} c^k(l,m_1,l,m_2) U_{n,l,k}$. For non-zero $\gamma_{0,m_1-m_2}$, $m_1 - m_2 = 0$. According to the definition of the Gaunt coefficients, the value of $c^k(l,m,l,-m)$ is independent of $m$. So $k=0$ only contributes an identity matrix multiplied by a factor $\gamma_{0,0} c^k(l,m,l,-m) U_{n,l,k}$. Therefore, in real calculations, one can ignore the $k=0$ term if only the relative energies are interested.

2) $\langle r^k \rangle$: $R_{n,l}(r)$ can be written as $a^{-\frac{3}{2}} \bar{R}_{n,l}\left(\frac{r}{a}\right)$, where $a = \frac{a_0}{Z_{eff}}$, $a_0 = 0.53 \times 10^{-10}$ meter and $Z_{eff}$ is the effective charge of the nucleus for the atomic orbital.

$\langle r^k \rangle_{n,l} \equiv \int_0^\infty a^{-3} \bar{R}_{n,l}^2\left(\frac{r}{a}\right) r^{k+2} dr = a^k \int_0^\infty \bar{R}_{n,l}^2\left(\frac{r}{a}\right) \left(\frac{r}{a}\right)^{k+2} d\left(\frac{r}{a}\right) = a^k \int_0^\infty \bar{R}_{n,l}^2(\rho) \rho^{k+2} d\rho$,

where the integral $I_{n,l,k} = \int_0^\infty \bar{R}_{n,l}^2(\rho) \rho^{k+2} d\rho$ depends only on $n$, $l$, and $k$.

Therefore $U_{n,l,k} = -\frac{Z_o e^2}{4\pi\epsilon R} \frac{a^k}{R^k} I_{n,l,k} = -\frac{Z_o e^2}{4\pi\epsilon R} \left(\frac{a_0}{R Z_{eff}}\right)^k I_{n,l,k}$.

3) Here we will be calculating $d$ ($l = 2$) orbit exclusively, so the possible values for $k$ are 0, 2, and 4.

4) Note that $\frac{I_{3,2,4}}{I_{3,2,2}} = \frac{9}{4}\frac{\Gamma(11)}{\Gamma(9)} = \frac{405}{2}$. For Fe atoms, if we treat the 1s, 2s, 2p, 3s and 3p electrons as part of the ionic core, we have $Z_{eff} = 8$. In this case, $\frac{U_{3,2,4}}{U_{3,2,2}} = 0.231 \approx \frac{1}{4}$.

## S6.1 Fe-3$d$ orbitals in an O$_h$ local environment

Here $l=2$, $R_i = R$.

Since this local environment is totally symmetric, $\gamma_{k,m}$ vanishes for all the odd $m$.

The six oxygen atoms will be located at $(R, 0, 0)$, $(-R, 0, 0)$, $(0, R, 0)$, $(0, -R, 0)$, $(0, 0, R)$, and $(0, 0, -R)$. In addition, $\gamma_{2,m} = 0$.

Only $\gamma_{4,4} = \gamma_{4,-4} = \sqrt{\frac{35}{8}}$, $\gamma_{4,0} = \frac{7}{2}$ are non-zero.

The resulting matrix for $V_{CF} = \begin{bmatrix} C^4_{-2,-2}\gamma_{4,0} & 0 & 0 & 0 & C^4_{-2,2}\gamma_{4,-4} \\ 0 & C^4_{-1,-1}\gamma_{4,0} & 0 & 0 & 0 \\ 0 & 0 & C^4_{0,0}\gamma_{4,0} & 0 & 0 \\ 0 & 0 & 0 & C^4_{1,1}\gamma_{4,0} & 0 \\ C^4_{2,-2}\gamma_{4,4} & 0 & 0 & 0 & C^4_{2,2}\gamma_{4,0} \end{bmatrix} U_{3,2,4}$, where $l_1 = l_2 = 2$ are omitted for the Gaunt coefficients.

Plugging the numbers $C^4_{-2,-2} = C^4_{2,2} = \frac{1}{21}$, $C^4_{-1,-1} = C^4_{1,1} = -\frac{4}{21}$, $C^4_{0,0} = \frac{6}{21}$, $C^4_{-2,2} = C^4_{2,-2} = \frac{\sqrt{70}}{21}$

$$V_{CF} = \frac{I_{3,2,4}}{6}\begin{bmatrix} 1 & 0 & 0 & 0 & 5 \\ 0 & -4 & 0 & 0 & 0 \\ 0 & 0 & 6 & 0 & 0 \\ 0 & 0 & 0 & -4 & 0 \\ 5 & 0 & 0 & 0 & 1 \end{bmatrix}.$$

One can diagonalize $V_{CF}$, the 3d orbitals are split into two levels in energy:

$-\frac{2}{3}U_{n,2,4}$ for $(xz, xy, yz)$

$U_{n,2,4}$ for $(z^2, x^2 - y^2)$.

This is consistent with the group theory analysis. In addition to the splitting, the relative energy difference is also revealed.

## S6.2 Fe 3$d$ orbitals in an D$_{2h}$ local environment

In Pbnm o-LuFeO$_3$, the octahedrons are actually distorted according to the orthorhombic symmetry. Then the six oxygen atoms will be located at $(R + a, 0, 0)$, $(-R - a, 0, 0)$, $(0, R + b, 0)$, $(0, -R - b, 0)$, $(0, 0, R + c)$, and $(0, 0, -R - c)$. Note that here the a and b are not lattice constants.

We can now calculate the structural factors. Notice that the structure factors $\gamma_{k,m}$ are zero when $m$ is odd.

$$\gamma_{2,0} = 2\left(\frac{R}{R+c}\right)^3 - \left(\frac{R}{R+b}\right)^3 - \left(\frac{R}{R+a}\right)^3 \approx \frac{3a + 3b - 6c}{R}$$

$$\gamma_{2,2} = -\frac{\sqrt{6}}{4}\left[\left(\frac{R}{R+a}\right)^3 - \left(\frac{R}{R+b}\right)^3\right] \approx \frac{3\sqrt{6}}{4}\frac{a-b}{R}$$

$$\gamma_{4,0} = \frac{1}{4}\left[3\left(\frac{R}{R+a}\right)^5 + 3\left(\frac{R}{R+b}\right)^5 + 8\left(\frac{R}{R+c}\right)^5\right] \approx \frac{7}{2}$$

$$\gamma_{4,2} = -\frac{\sqrt{10}}{2}\left[\left(\frac{R}{R+a}\right)^5 - \left(\frac{R}{R+b}\right)^5\right] \approx \frac{5\sqrt{10}}{2}\frac{a-b}{R}$$

$$\gamma_{4,4} = -\frac{\sqrt{70}}{8}\left[\left(\frac{R}{R+a}\right)^5 + \left(\frac{R}{R+b}\right)^5\right] \approx \sqrt{\frac{35}{8}}.$$

There are two types of distortions here.

The first type of distortion is represented by the non-zero $\gamma_{2,0}$; it modifies the on-diagonal terms in $V_{CF}$. Using $C^2_{-2,-2} = C^2_{2,2} = -\frac{2}{7}, C^2_{-1,-1} = C^2_{1,1} = \frac{1}{7}, C^2_{0,0} = \frac{2}{7}$, this additional term is:

$$H_1 = \frac{U_{3,2,2}}{7}\frac{3a+3b-6c}{R}\begin{bmatrix} -2 & 0 & 0 & 0 & 0 \\ 0 & 1 & 0 & 0 & 0 \\ 0 & 0 & 2 & 0 & 0 \\ 0 & 0 & 0 & 1 & 0 \\ 0 & 0 & 0 & 0 & -2 \end{bmatrix}.$$

Diagonalizing $V_{CF} + H_1$ results in splitting of the $(xz, yz, xy)$ into $(xz, yz)$ and $xy$; $(z^2, x^2 - y^2)$ is split into $z^2$ and $x^2 - y^2$. This is actually a symmetry transformation from $O_h$ to $D_{4h}$. This splitting is proportional to $\frac{a+b-2c}{2R}$. If we assume $a > c > b$, then $\frac{a+b-2c}{2R} = 0.1\%$ in orthorhombic LuFeO$_3$.

The second type of distortion is represented by $\gamma_{2,2}$ and $\gamma_{4,2}$.

Using $C^2_{-2,0} = C^2_{2,0} = C^2_{0,-2} = C^2_{0,2} = -\frac{2}{7}, C^2_{-1,1} = C^2_{1,-1} = -\frac{\sqrt{6}}{7}, C^4_{-2,0} = C^4_{2,0} = C^4_{0,-2} = C^4_{0,2} = \frac{\sqrt{15}}{21}, C^4_{-1,1} = C^4_{1,-1} = -\frac{\sqrt{40}}{21}$, the additional term in the Hamiltonian is

$$H_2 = \frac{U_{3,2,4}\gamma_{4,2}}{21}\begin{bmatrix} 0 & 0 & \sqrt{15} & 0 & 0 \\ 0 & 0 & 0 & -\sqrt{40} & 0 \\ \sqrt{15} & 0 & 0 & 0 & \sqrt{15} \\ 0 & -\sqrt{40} & 0 & 0 & 0 \\ 0 & 0 & \sqrt{15} & 0 & 0 \end{bmatrix} + \frac{U_{3,2,2}\gamma_{2,2}}{7}\begin{bmatrix} 0 & 0 & -2 & 0 & 0 \\ 0 & 0 & 0 & -\sqrt{6} & 0 \\ -2 & 0 & 0 & 0 & -2 \\ 0 & -\sqrt{6} & 0 & 0 & 0 \\ 0 & 0 & -2 & 0 & 0 \end{bmatrix}.$$

This will couple the $m = \pm 2$ and $m = \pm 1$ states. By diagonalizing the Hamiltonian including the $H_2$ term, the $t_{2g}$ states will split into $xy, yz, xz$; the $e_g$ states split into $2z^2 + (\frac{\sqrt{6}}{2} - 1)x^2 - (\frac{\sqrt{6}}{2} + 1)y^2, 2z^2 - (\frac{\sqrt{6}}{2} + 1)x^2 + (\frac{\sqrt{6}}{2} - 1)y^2$. This is consistent with the group theory analysis of the $O_h$ to $D_{2h}$ distortion. The splitting is proportional to $\frac{a-b}{R}$ (about 1.3% in orthorhombic LuFeO$_3$). This distortion is significantly larger than the $O_h$ to $D_{4h}$ distortion.

From the form of $H_2$ above, we can assume

$$H_2 = \begin{bmatrix} 0 & 0 & \alpha' & 0 & 0 \\ 0 & 0 & 0 & \beta' & 0 \\ \alpha' & 0 & 0 & 0 & \alpha' \\ 0 & \beta' & 0 & 0 & 0 \\ 0 & 0 & \alpha' & 0 & 0 \end{bmatrix},$$

where

$$\alpha' = \frac{\sqrt{15}}{21}U_{3,2,4}Y_{4,2} - \frac{2}{7}U_{3,2,2}Y_{2,2} = \frac{5\sqrt{30}}{42}\frac{a-b}{R}U_{3,2,4} - \frac{6\sqrt{6}}{28}\frac{a-b}{R}U_{3,2,2},$$

$$\beta' = \frac{-\sqrt{40}}{21}U_{3,2,4}Y_{4,2} - \frac{\sqrt{6}}{7}U_{3,2,2}Y_{2,2} = \frac{-50}{21}\frac{a-b}{R}U_{3,2,4} - \frac{9}{14}\frac{a-b}{R}U_{3,2,2}.$$

If we assume $\frac{U_{3,2,4}}{U_{3,2,2}} \approx \frac{1}{4}$, then $\alpha' \approx -0.36\frac{a-b}{R}U_{3,2,2}$, $\beta' \approx -1.2\frac{a-b}{R}U_{3,2,2}$.

To find the energy change due to $H_2$, one can transform $H_2$ to the basis $\{z^2, x^2 - y^2, xz, xy, yz\}$, which means $H'_2 = T^\dagger H_2 T$, where the transformation matrix is:

$$T = \begin{bmatrix} 0 & \frac{1}{\sqrt{2}} & 0 & \frac{i}{\sqrt{2}} & 0 \\ 0 & 0 & \frac{1}{\sqrt{2}} & 0 & \frac{i}{\sqrt{2}} \\ 1 & 0 & 0 & 0 & 0 \\ 0 & 0 & \frac{1}{\sqrt{2}} & 0 & \frac{-i}{\sqrt{2}} \\ 0 & \frac{1}{\sqrt{2}} & 0 & \frac{-i}{\sqrt{2}} & 0 \end{bmatrix}.$$

The result is

$$H'_2 = \begin{bmatrix} 0 & \sqrt{2}\alpha' & 0 & 0 & 0 \\ \sqrt{2}\alpha' & 0 & 0 & 0 & 0 \\ 0 & 0 & \beta' & 0 & 0 \\ 0 & 0 & 0 & 0 & 0 \\ 0 & 0 & 0 & 0 & -\beta' \end{bmatrix}.$$

The results correspond to energies for the states:

$U_{n,2,4} + \sqrt{2}\alpha'$ for $\sqrt{2}|m=0\rangle + |m=2\rangle + |m=-2\rangle$ or $2z^2 + (\frac{\sqrt{6}}{2} - 1)x^2 - (\frac{\sqrt{6}}{2} + 1)y^2$

$U_{n,2,4} - \sqrt{2}\alpha'$ for $\sqrt{2}|m=0\rangle - |m=2\rangle - |m=-2\rangle$ or $2z^2 - (\frac{\sqrt{6}}{2} + 1)x^2 + (\frac{\sqrt{6}}{2} - 1)y^2$

$-\frac{2}{3}U_{n,2,4} + \beta'$ for $xz$

$-\frac{2}{3}U_{n,2,4}$ for $xy$

$-\frac{2}{3}U_{n,2,4} - \beta'$ for $yz$.

If we assume $\frac{U_{3,2,4}}{U_{3,2,2}} \approx \frac{1}{4}$, the splitting of the $e_g$ level is about half of that in $t_{2g}$ level.

Therefore, whether a state goes up or down in energy is decided by the sign of $\alpha'$ and $\beta'$. Here we assume $\alpha$ and $\beta$ are both greater than zero (meaning $a < b$). Then these states, sorted in ascending energy are:

$$yz, xy, xz, 2z^2 - \left(\frac{\sqrt{6}}{2} + 1\right)x^2 + (\frac{\sqrt{6}}{2} - 1)y^2, 2z^2 + (\frac{\sqrt{6}}{2} - 1)x^2 - (\frac{\sqrt{6}}{2} + 1)y^2.$$

### S6.3 Fe-3$d$ orbitals in an D$_{3h}$ local environment

Here $l=2$, $R = R_a$ for the equator oxygen atoms and $R = R_c$ for the apex oxygen atoms.

The angles for the oxygen atoms are listed in Table S5.1

Table S6.1

| index | $\theta$ | $\phi$ | R |
|---|---|---|---|
| 1 | 0 | 0 | $R_c$ |
| 2 | $\pi$ | 0 | $R_c$ |
| 3 | $\frac{\pi}{2}$ | 0 | $R_a$ |
| 4 | $\frac{\pi}{2}$ | $\frac{2\pi}{3}$ | $R_a$ |
| 5 | $\frac{\pi}{2}$ | $\frac{4\pi}{3}$ | $R_a$ |

If we define $\alpha = \frac{R_c}{R_a}$, we get $\gamma_{4,0} = \frac{1}{8}\left(\frac{5}{2\alpha+3}\right)^5 (16\alpha^5 + 9)$, $\gamma_{2,0} = \frac{1}{2}\left(\frac{5}{2\alpha+3}\right)^3 (4\alpha^3 - 3)$.

All the other structural factors are zero.

Note that for h-LuFeO$_3$, $\alpha \approx 0.94$. So $\gamma_{4,0} \approx 2.92$ and $\gamma_{2,0} \approx 0.17$.

Therefore, the matrix of crystal field is diagonal.

$$V_{CF} = \begin{bmatrix} C^4_{-2,-2}\gamma_{4,0} & 0 & 0 & 0 & 0 \\ 0 & C^4_{-1,-1}\gamma_{4,0} & 0 & 0 & 0 \\ 0 & 0 & C^4_{0,0}\gamma_{4,0} & 0 & 0 \\ 0 & 0 & 0 & C^4_{1,1}\gamma_{4,0} & 0 \\ 0 & 0 & 0 & 0 & C^4_{2,2}\gamma_{4,0} \end{bmatrix} U_{3,2,4} +$$

$$\begin{bmatrix} C^2_{-2,-2}\gamma_{2,0} & 0 & 0 & 0 & 0 \\ 0 & C^2_{-1,-1}\gamma_{2,0} & 0 & 0 & 0 \\ 0 & 0 & C^2_{0,0}\gamma_{2,0} & 0 & 0 \\ 0 & 0 & 0 & C^2_{1,1}\gamma_{2,0} & 0 \\ 0 & 0 & 0 & 0 & C^2_{2,2}\gamma_{2,0} \end{bmatrix} U_{3,2,2}.$$

Using $C^2_{-2,-2} = C^2_{2,2} = -\frac{2}{7}$, $C^2_{-1,-1} = C^2_{1,1} = \frac{1}{7}$, $C^2_{0,0} = \frac{2}{7}$, one finds

$$V_{CF} = \frac{U_{3,2,4}\gamma_{4,0}}{21}\begin{bmatrix} 1 & 0 & 0 & 0 & 0 \\ 0 & -4 & 0 & 0 & 0 \\ 0 & 0 & 6 & 0 & 0 \\ 0 & 0 & 0 & -4 & 0 \\ 0 & 0 & 0 & 0 & 1 \end{bmatrix} + \frac{U_{3,2,2}\gamma_{2,0}}{7}\begin{bmatrix} -2 & 0 & 0 & 0 & 0 \\ 0 & 1 & 0 & 0 & 0 \\ 0 & 0 & 2 & 0 & 0 \\ 0 & 0 & 0 & 1 & 0 \\ 0 & 0 & 0 & 0 & -2 \end{bmatrix}.$$

To further study the energy splitting, we look at the ratio

$$\beta' = \frac{7U_{3,2,4}\gamma_{4,0}}{21U_{3,2,2}\gamma_{2,0}} = \frac{1}{3}\frac{\gamma_{4,0}}{\gamma_{2,0}}\left(\frac{a_0}{RZ_{eff}}\right)^2 \frac{I_{3,2,4}}{I_{3,2,2}}.$$

Note that $\left(\frac{a_0}{R}\right)^2 \approx \left(\frac{0.53}{1.95}\right)^2 \approx 0.074$. In addition, $\frac{\gamma_{4,0}}{\gamma_{2,0}} \approx 17.2$.

So $\beta' = 1.325$ ($\frac{U_{3,2,4}}{U_{3,2,2}} = 0.231$) for $Z_{eff} = 8$.

Therefore, the 3d orbitals are split into three levels:

$(1 - 4\beta')I_{3,2,2} = -4.3I_{3,2,2}$ for $(xz, yz)$ or $e''$,

$(\beta' - 2)I_{3,2,2} = -0.67I_{3,2,2}$ for $(xx - yy, xy)$ or $e'$,

$(6\beta' + 2)I_{3,2,2} = 10.0I_{3,2,2}$ for $(zz)$ or $a_1'$.

This is consistent with the group theory analysis. The additional information is the energy difference and the order of the levels in energy.

### S6.4 Fe-3$d$ orbitals in an $C_S$ local environment

In P6$_3$cm h-LuFeO$_3$, the trigonal bipyramid is actually distorted due to the displacement of the apex oxygen atoms in the $\Gamma_2^-$ and $K_1$ mode (Note that the $K_3$ mode corresponds to a rigid rotation that does not distort the FeO$_5$ trigonal bipyramid itself).

#### $\Gamma_2^-$ mode

The effect of the $\Gamma_2^-$ mode is to displace the two apex atoms toward the same direction. This effect makes the two apex oxygen atoms (1 and 2, see Fig. S5.1) inequivalent. In other words, one has $R_c^1 = R_c - \delta, R_c^2 = R_c + \delta$.

The result of this distortion is a small modification of $\gamma_{4,0}$ and $\gamma_{2,0}$; there is no additional structural factor generated.

In addition, this displacement of apex oxygen atoms, combined with the $K_3$ rotation, changes the O$_{ap}$-Fe-O$_{ap}$ angle from 180 degree. Therefore, one can parameterize the effect using a small angle $\theta_{g2}$. The corresponding O$_{ap}$-Fe-O$_{ap}$ angle is $\pi - \theta_{g2}$. In this case, the $\sin(\theta_{g2})$ is a small quantity that can be treated as the order of magnitude of the perturbation.

One can show that with this distortion, all the structure factor $\gamma_{k,m}$ are non-zero. On the other hand, one can show that $\gamma_{k,m}$ is proportional to $[\sin(\theta_{g2})]^m$. Therefore, we only keep the lowest order, i.e. $\gamma_{2,1}$ and $\gamma_{4,1}$.

Again, here $l=2$, $R = R_a$ for the equator oxygen atoms and $R = R_{c1}, R_{c2} \approx R_c$ for the apex oxygen atoms. To keep the lowest order, the perturbation Hamiltonian can be written as

$$H_1 = \frac{U_{3,2,4}\gamma_{4,1}}{21} \begin{bmatrix} 0 & -\sqrt{5} & 0 & 0 & 0 \\ -\sqrt{5} & 0 & \sqrt{30} & 0 & 0 \\ 0 & \sqrt{30} & 0 & \sqrt{30} & 0 \\ 0 & 0 & \sqrt{30} & 0 & -\sqrt{5} \\ 0 & 0 & 0 & -\sqrt{5} & 0 \end{bmatrix} + \frac{U_{3,2,2}\gamma_{2,1}}{7} \begin{bmatrix} 0 & \sqrt{6} & 0 & 0 & 0 \\ \sqrt{6} & 0 & 1 & 0 & 0 \\ 0 & 1 & 0 & 1 & 0 \\ 0 & 0 & 1 & 0 & \sqrt{6} \\ 0 & 0 & 0 & \sqrt{6} & 0 \end{bmatrix}.$$

where $\gamma_{4,1} = 10\sqrt{5}\left(\frac{3+2\alpha}{5\alpha}\right)^5 \sin^2(\theta_{g2}) = \gamma_{4,-1}$, and $\gamma_{2,1} = 3\sqrt{6}\left(\frac{3+2\alpha}{5\alpha}\right)^3 \sin^2(\theta_{g2}) = \gamma_{2,-1}$.

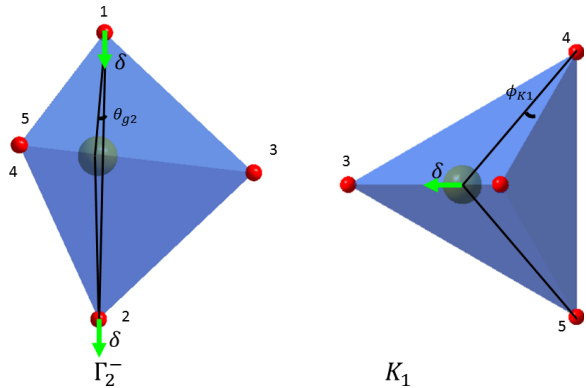

**Figure S6.1** The (exaggerated) angles in the two distortions.

If we define $\alpha' = \frac{-\sqrt{5}}{21}U_{3,2,4}\gamma_{4,1} + \frac{\sqrt{6}}{7}U_{3,2,2}\gamma_{2,1}$ and $\beta' = \frac{\sqrt{30}}{21}U_{3,2,4}\gamma_{4,1} + \frac{1}{7}U_{3,2,2}\gamma_{2,1}$, the perturbation Hamiltonian can be rewritten as

$$H_1 = \begin{bmatrix} 0 & \alpha' & 0 & 0 & 0 \\ \alpha' & 0 & \beta' & 0 & 0 \\ 0 & \beta' & 0 & \beta' & 0 \\ 0 & 0 & \beta' & 0 & \alpha' \\ 0 & 0 & 0 & \alpha' & 0 \end{bmatrix},$$

where

$$\alpha' = \frac{-50}{21}U_{3,2,4}\left(\frac{3+2\alpha}{5\alpha}\right)^5 \sin^2(\theta_{g2}) + \frac{18}{7}U_{3,2,2}\left(\frac{3+2\alpha}{5\alpha}\right)^3 \sin^2(\theta_{g2}) \text{ and}$$

$$\beta' = \frac{50\sqrt{6}}{21}U_{3,2,4}\left(\frac{3+2\alpha}{5\alpha}\right)^5 \sin^2(\theta_{g2}) + \frac{3\sqrt{6}}{7}U_{3,2,2}\left(\frac{3+2\alpha}{5\alpha}\right)^3 \sin^2(\theta_{g2}).$$

Transforming to the basis $\{z^2, x^2 - y^2, xy, xz, yz\}$, one gets

$$H_1' = \begin{bmatrix} 0 & 0 & 0 & \sqrt{2}\beta' & 0 \\ 0 & 0 & 0 & \alpha' & 0 \\ 0 & 0 & 0 & 0 & \alpha' \\ \sqrt{2}\beta'\alpha' & 0 & 0 & 0 & 0 \\ 0 & 0 & \alpha' & 0 & 0 \end{bmatrix}.$$

Assuming $\frac{U_{3,2,4}}{U_{3,2,2}} = 0.25$, one can estimate $\alpha' = 2.2U_{3,2,2}\sin^2(\theta_{g2})$ and $\beta' = 2.5U_{3,2,2}\sin^2(\theta_{g2})$.

Adding this perturbation results in coupling between the individual states. On the other hand, since the off diagonal terms in $H_1'$ are zero and the degenerate states in D$_{3h}$ (i.e. $\{xy, x^2 - y^2\}$,$\{xz, yz\}$) are not coupled, the resulting perturbation are all second order in energy, which means that the modification of levels caused by $\Gamma_2^-$ mode is actually proportional to $[\sin(\theta_{g2})]^4$, which is small enough to be ignored.

In principle, $\gamma_{2,2}, \gamma_{4,2}, \gamma_{4,3}$, and $\gamma_{4,4}$ are also nonzero. Since the effect of $\gamma_{4,1}$ and $\gamma_{2,1}$ are actually fourth order in terms of $\sin(\theta_{g2})$, one needs to consider $\gamma_{2,2}$ and $\gamma_{4,2}$, where

$$\gamma_{2,2} = \gamma_{2,-2} = \frac{\sqrt{6}}{2}\left(\frac{3+2\alpha}{5\alpha}\right)^3 \sin^2(\theta_{g2})$$

$$\gamma_{4,2} = \gamma_{4,-2} = \frac{3\sqrt{10}}{2}\left(\frac{3+2\alpha}{5\alpha}\right)^5 \sin^2(\theta_{g2}).$$

This additional Hamiltonian is

$$H_2 = \frac{U_{3,2,4}\gamma_{4,2}}{21}\begin{bmatrix} 0 & 0 & \sqrt{15} & 0 & 0 \\ 0 & 0 & 0 & -\sqrt{40} & 0 \\ \sqrt{15} & 0 & 0 & 0 & \sqrt{15} \\ 0 & -\sqrt{40} & 0 & 0 & 0 \\ 0 & 0 & \sqrt{15} & 0 & 0 \end{bmatrix} + \frac{U_{3,2,2}\gamma_{2,2}}{7}\begin{bmatrix} 0 & 0 & -2 & 0 & 0 \\ 0 & 0 & 0 & -\sqrt{6} & 0 \\ -2 & 0 & 0 & 0 & -2 \\ 0 & -\sqrt{6} & 0 & 0 & 0 \\ 0 & 0 & -2 & 0 & 0 \end{bmatrix}.$$

This may be transformed into the basis $\{z^2, x^2 - y^2, xy, xz, yz\}$ as

$$H_2' = \begin{bmatrix} 0 & \sqrt{2}\alpha' & 0 & 0 & 0 \\ \sqrt{2}\alpha' & 0 & 0 & 0 & 0 \\ 0 & 0 & 0 & 0 & 0 \\ 0 & 0 & 0 & \beta' & 0 \\ 0 & 0 & 0 & 0 & -\beta' \end{bmatrix}.$$

where

$$\alpha' = \frac{\sqrt{15}}{21}U_{3,2,4}\gamma_{4,2} - \frac{2}{7}U_{3,2,2}\gamma_{2,2} = \left[\frac{5\sqrt{6}}{14}U_{3,2,4}\left(\frac{3+2\alpha}{5\alpha}\right)^5 - \frac{\sqrt{6}}{7}U_{3,2,2}\left(\frac{3+2\alpha}{5\alpha}\right)^3\right]\sin^2(\theta_{g2}),$$

$$\beta' = -\frac{\sqrt{40}}{21}U_{3,2,4}\gamma_{4,2} - \frac{\sqrt{6}}{7}U_{3,2,2}\gamma_{2,2} = \left[-\frac{10}{7}U_{3,2,4}\left(\frac{3+2\alpha}{5\alpha}\right)^5 - \frac{3}{7}U_{3,2,2}\left(\frac{3+2\alpha}{5\alpha}\right)^3\right]\sin^2(\theta_{g2}).$$

Assuming $\frac{U_{3,2,4}}{U_{3,2,2}} = 0.25$, one can estimate

$\alpha' = -0.13 U_{3,2,2} \sin^2(\theta_{g2})$ and $\beta' = -2.2\, U_{3,2,2} \sin^2(\theta_{g2})$.

In h-LuFeO$_3$, $\sin(\theta_{g2}) \approx 0.044$, meaning $\sin^2(\theta_{g2}) \approx 0.19\%$.

In addition, since this is a second order effect, no matter which direction the distortion occurs the splitting of the crystal field levels always follow the similar pattern. Sorted in descending energy, the five states are $z^2, xy, x^2 - y^2,\ yz,\ xz$.

### $K_1$ mode

The effect of the $K_1$ mode can be divided into two types. The first type is the horizontal displacement of the apex oxygen atoms; this generates a O$_{ap}$-Fe-O$_{ap}$ angle that is similar to the effect of $\Gamma_2^-$, which is the second order. The second type comes from the displacement of the Fe atoms ($\delta$) in the basal plan. This can also be describe by the deviation of the O$_{eq}$-Fe-O$_{eq}$ angle from 120 degree, which can be called $\phi_{k1}$ (see Fig. S5.1) Next, we investigate the effect of this angle.

If only the displacement of Fe atoms are concerned, all the equator oxygen atoms still have $\theta = \frac{\pi}{2}$. Therefore, in addition to $\gamma_{2,0}$, $\gamma_{4,0}$, the other non-zero structural factors are $\gamma_{2,2}$, $\gamma_{4,2}$, and $\gamma_{4,4}$.

We can calculate these factors. The results are:

$$\gamma_{2,2} = -\frac{15\sqrt{2}}{4}\left(\frac{3+2\alpha}{5}\right)^3 \sin(\phi_{k1})$$

$$\gamma_{4,2} = \frac{7\sqrt{30}}{8}\left(\frac{3+2\alpha}{5}\right)^5 \sin(\phi_{k1})$$

$$\gamma_{4,4} = -\frac{\sqrt{210}}{16}\left(\frac{3+2\alpha}{5}\right)^5 \sin(\phi_{k1}).$$

The perturbation Hamiltonian can be written as

$$H_2 = \frac{U_{3,2,4}\gamma_{4,4}}{21}\begin{bmatrix} 0 & 0 & 0 & 0 & 1 \\ 0 & 0 & 0 & 0 & 0 \\ 0 & 0 & 0 & 0 & 0 \\ 0 & 0 & 0 & 0 & 0 \\ 1 & 0 & 0 & 0 & 0 \end{bmatrix} + \frac{U_{3,2,4}\gamma_{4,2}}{21}\begin{bmatrix} 0 & 0 & \sqrt{15} & 0 & 0 \\ 0 & 0 & 0 & -\sqrt{40} & 0 \\ \sqrt{15} & 0 & 0 & 0 & \sqrt{15} \\ 0 & -\sqrt{40} & 0 & 0 & 0 \\ 0 & 0 & \sqrt{15} & 0 & 0 \end{bmatrix} + \frac{U_{3,2,2}\gamma_{2,2}}{7}\begin{bmatrix} 0 & 0 & -2 & 0 & 0 \\ 0 & 0 & 0 & -\sqrt{6} & 0 \\ -2 & 0 & 0 & 0 & -2 \\ 0 & -\sqrt{6} & 0 & 0 & 0 \\ 0 & 0 & -2 & 0 & 0 \end{bmatrix}.$$

To find the energy change due to $H_2$, one can transform $H_2$ to the basis $\{z^2, x^2-y^2, xy, xz, yz\}$, which is

$$H_2' = \begin{bmatrix} 0 & \sqrt{2}\alpha' & 0 & 0 & 0 \\ \sqrt{2}\alpha' & -\delta' & 0 & 0 & 0 \\ 0 & 0 & -\delta' & 0 & 0 \\ 0 & 0 & 0 & \beta' & 0 \\ 0 & 0 & 0 & 0 & -\beta' \end{bmatrix}.$$

where

$$\alpha' = \frac{\sqrt{15}}{21}U_{3,2,4}\gamma_{4,2} - \frac{2}{7}U_{3,2,2}\gamma_{2,2} = \left[\frac{5\sqrt{2}}{8}U_{3,2,4}\left(\frac{3+2\alpha}{5}\right)^5 + \frac{15\sqrt{2}}{14}U_{3,2,2}\left(\frac{3+2\alpha}{5}\right)^3\right]\sin(\phi_{k1}),$$

$$\beta' = -\frac{\sqrt{40}}{21}U_{3,2,4}\gamma_{4,2} - \frac{\sqrt{6}}{7}U_{3,2,2}\gamma_{2,2} = \left[-\frac{5\sqrt{3}}{3}U_{3,2,4}\left(\frac{3+2\alpha}{5}\right)^5 + \frac{15\sqrt{3}}{14}U_{3,2,2}\left(\frac{3+2\alpha}{5}\right)^3\right]\sin(\phi_{k1}),$$

$$\delta' = -\frac{U_{3,2,4}\gamma_{4,4}}{21} = \frac{\sqrt{210}}{16}U_{3,2,4}\left(\frac{3+2\alpha}{5}\right)^5\sin(\phi_{k1}).$$

If we assume $\frac{U_{3,2,4}}{U_{3,2,2}} \approx \frac{1}{4}$, and $\alpha = 0.94$, these parameters are

$$\alpha' \approx 1.6\,U_{3,2,2}\sin(\phi_{k1}), \qquad \beta' \approx 1.1\,U_{3,2,2}\sin(\phi_{k1}), \qquad \delta' \approx 0.2\,U_{3,2,2}\sin(\phi_{k1}).$$

Therefore, the result of splitting of the levels relative to those in the $D_{3h}$ symmetry is on the order of $\sin(\phi_{k1})$.

The sign of the coupling terms in the perturbation Hamiltonian determines whether a level goes up or down in energy. If we assume $\sin(\phi_{k1}) > 0$ ($\delta > 0$), the 3d orbitals are split into five levels. They are approximately $yz, xz, x^2-y^2, xy, z^2$, in the order of ascending energy. In additional, the splitting between the $x^2-y^2, xy$ is second order since it comes from the off-diagonal term $\sqrt{2}\alpha'$. So the splitting between the $yz$ and $xz$ states are most important since $\beta'$ is much larger than $\delta'$.

### S6.5 Lu 5d orbitals in an C$_{3v}$ local environment

Here $l=2$, $R = R_a$ for the apex oxygen atoms (six of them) and $R = R_c$ for the equator oxygen atom (only one). For the apex atoms, the angles are $\theta \approx 1.09, \phi = 0, \frac{2\pi}{3}, \frac{4\pi}{3}$, and $\theta \approx 2.05, \phi = \frac{\pi}{3}, \pi, \frac{5\pi}{3}$. For the equator atom, $\frac{R_c}{R_a} \approx 1.1$.

One can calculate $\gamma_{4,3} = \gamma_{4,-3} = -3.11, \gamma_{4,0} = -0.858, \gamma_{2,0} = -0.399$.

Therefore, the matrix of crystal field is the following.

$$V_{CF} = \begin{bmatrix} C^4_{-2,-2}Y_{4,0} & 0 & 0 & C^4_{-2,1}Y_{4,-3} & 0 \\ 0 & C^4_{-1,-1}Y_{4,0} & 0 & 0 & C^4_{-1,2}Y_{4,-3} \\ 0 & 0 & C^4_{0,0}Y_{4,0} & 0 & 0 \\ C^4_{1,-2}Y_{4,3} & 0 & 0 & C^4_{1,1}Y_{4,0} & 0 \\ 0 & C^4_{2,-1}Y_{4,3} & 0 & 0 & C^4_{2,2}Y_{4,0} \end{bmatrix} U_{5,2,4} +$$

$$\begin{bmatrix} C^2_{-2,-2}Y_{2,0} & 0 & 0 & 0 & 0 \\ 0 & C^2_{-1,-1}Y_{2,0} & 0 & 0 & 0 \\ 0 & 0 & C^2_{0,0}Y_{2,0} & 0 & 0 \\ 0 & 0 & 0 & C^2_{1,1}Y_{2,0} & 0 \\ 0 & 0 & 0 & 0 & C^2_{2,2}Y_{2,0} \end{bmatrix} U_{5,2,2}.$$

One can calculate that $\frac{I_{5,2,4}}{I_{5,2,2}} = \frac{625}{2}$.

Since $R = 2.30$ Å, $\left(\frac{a_0}{RZ_{eff}}\right)^2 = \frac{0.0511}{Z_{eff}^2}$. So $\frac{U_{5,2,4}}{U_{5,2,2}} = \frac{15.97}{Z_{eff}^2}$.

If we let $\beta' = \frac{U_{5,2,4}}{U_{5,2,2}}$ and plug in the numbers for the Gaunt coefficients and the structure factors:

$$V_{CF} = U_{5,2,2} \begin{bmatrix} 0.11 - 0.04\beta' & 0 & 0 & 0.88\beta' & 0 \\ 0 & -0.06 + 0.16\beta' & 0 & 0 & 0.88\beta' \\ 0 & 0 & -0.11 - 0.24\beta' & 0 & 0 \\ 0.88\beta' & 0 & 0 & -0.06 + 0.16\beta' & 0 \\ 0 & 0.88\beta' & 0 & 0 & 0.11 - 0.04\beta' \end{bmatrix}$$

If we assume $Z_{eff} = 8$, $\beta' \approx 0.25$.

$$V_{CF} = U_{5,2,2} \begin{bmatrix} 0.10 & 0 & 0 & 0.22 & 0 \\ 0 & -0.02 & 0 & 0 & 0.22 \\ 0 & 0 & -0.17 & 0 & 0 \\ 0.22 & 0 & 0 & -0.02 & 0 \\ 0 & 0.22 & 0 & 0 & 0.10 \end{bmatrix}.$$

Obviously, the off-diagonal terms are the largest. So the 5d levels will be split into the following 3 levels, in the order of descending energy:

$(x^2 - y^2 + 2\lambda xz, xy + \lambda yz), z^2, (x^2 - y^2 - 2\lambda xz, xy - \lambda yz),$

where $\lambda > 0$ is a mixing factor that can be found from diagonalizing $V_{CF}$. When the off diagonal terms dominate, $\lambda \approx 1$.

With the analysis of spherical harmonic expansion, we can distinguish the $2e$ in the group theory analysis. In fact, we can name them as $e^\sigma = (x^2 - y^2 + 2\lambda xz, xy + \lambda yz)$ and $e^\pi = (x^2 - y^2 - 2\lambda xz, xy - \lambda yz)$, where the energy of $e^\sigma$ is the highest, and the energy of $e^\pi$ is the lowest.

## S7. Single-ion anisotropy of $Fe^{3+}$ in h-LuFeO$_3$ and o-LuFeO$_3$

In principle, the $3d^5$ configuration of $Fe^{3+}$ gives half-filled $3d$ orbitals, which corresponds to zero orbital angular moments. On the other hand, the high coercivity of LuFeO$_3$ [5,6] suggests significant magnetic crystalline anisotropy energy, which requires non-zero orbital angular momentum.

As discussed in Section 1 and 8, the hybridization between the Fe-$3d$ and O-$2p$ orbital makes the $3d^5$ configuration an approximation. In other words, there should be non-zero occupancy of the spin-minority bands, which may generate a certain orbital angular momentum. Below, we attempt to estimate the magnetic anisotropy energy. Since both hexagonal and orthorhombic LuFeO$_3$ are only weakly ferromagnetic with aniferromagnetic orders, single-ion anisotropy is actually concerned.

It is well known that magnetic anisotropy are closely related to the symmetry of the lattice structure. [7] We show that the local distortion from D$_{3h}$ to C$_s$ is the key for the anisotropy in h-LuFeO$_3$; the local distortion from O$_h$ to D$_{2h}$ is the key for the anisotropy in o-LuFeO$_3$.

In order to calculate the single-ion anisotropy, we consider the following interactions: crystal field interaction, exchange splitting between the spin up and spin down electrons, and the spin-orbit couplings. We assume that the exchange splitting has an energy scale of $E_{ex}$ which is larger than the energy scale of the crystal field energy (on the order of 1 eV) and the spin-orbit coupling (on the order of 50 meV for 3d transition metal atoms).

The single-ion anisotropy energy is calculated by comparing the energy of individual states when the spin is along $z$ and $x$ directions respectively. The spin up and down states are represented as $|\uparrow\rangle$ and $|\downarrow\rangle$ respectively; the spin along $x$ and $-x$ directions are represented as $\frac{1}{\sqrt{2}}(|\uparrow\rangle + |\downarrow\rangle)$ and $\frac{1}{\sqrt{2}}(|\uparrow\rangle - |\downarrow\rangle)$ respectively; the spin along $y$ and $-y$ directions are represented as $\frac{1}{\sqrt{2}}(|\uparrow\rangle + i|\downarrow\rangle)$ and $\frac{1}{\sqrt{2}}(|\uparrow\rangle - i|\downarrow\rangle)$ respectively.

In the calculation, we write down the Hamitonian that represent all three interactions and diagonalize to find the energy of the eigenstates. While the crystal field energy depends on the detailed local symmetry, the exchange splitting basically means all the spin-minority levels has an energy shift $E_{ex}$. Since the spin orbit interaction is much weaker than the other two interaction, the energy of the eigenstates are mostly determined by the exchange splitting and crystal field interactions; the spin-orbit interactions act as perturbation.

### S7.1 Hexagonal LuFeO$_3$

#### S7.1.1 D$_{3h}$ symmetry

*Spin along the z direction.*

In this case, we choose the following crystal field levels as the basis:

$$\phi_{1z} = |zz\uparrow\rangle, \phi_{2z} = |xy\uparrow\rangle, \phi_{3z} = |x^2-y^2\uparrow\rangle, \phi_{4z} = |xz\uparrow\rangle, \phi_{5z} = |yz\uparrow\rangle,$$
$$\phi_{6z} = |zz\downarrow\rangle, \phi_{7z} = |xy\downarrow\rangle, \phi_{8z} = |x^2-y^2\downarrow\rangle, \phi_{9z} = |xz\downarrow\rangle, \phi_{10z} = |yz\downarrow\rangle.$$

The Hamiltonian for the exchange splitting and the crystal field interaction is:

$$H_{CFEx}^{D3h} = \begin{bmatrix} E_{ex} & 0 & 0 & 0 & 0 & 0 & 0 & 0 & 0 & 0 \\ 0 & E_{ex}-a & 0 & 0 & 0 & 0 & 0 & 0 & 0 & 0 \\ 0 & 0 & E_{ex}-a & 0 & 0 & 0 & 0 & 0 & 0 & 0 \\ 0 & 0 & 0 & E_{ex}-b & 0 & 0 & 0 & 0 & 0 & 0 \\ 0 & 0 & 0 & 0 & E_{ex}-b & 0 & 0 & 0 & 0 & 0 \\ 0 & 0 & 0 & 0 & 0 & 0 & 0 & 0 & 0 & 0 \\ 0 & 0 & 0 & 0 & 0 & 0 & -a & 0 & 0 & 0 \\ 0 & 0 & 0 & 0 & 0 & 0 & 0 & -a & 0 & 0 \\ 0 & 0 & 0 & 0 & 0 & 0 & 0 & 0 & -b & 0 \\ 0 & 0 & 0 & 0 & 0 & 0 & 0 & 0 & 0 & -b \end{bmatrix}.$$

Note that due to the basis we choose, the Hamiltonian $H_{CFEx}^{D3h}$ is already diagonalized.

The Hamiltonian for the spin orbit interaction can be derived as

$$H_{SO}^z = \xi \vec{S} \cdot \vec{L} = \xi \begin{bmatrix} 0 & 0 & 0 & 0 & 0 & 0 & 0 & 0 & \frac{\sqrt{3}}{2} & \frac{\sqrt{3}}{2} \\ 0 & 0 & 1 & 0 & 0 & 0 & 0 & 0 & -\frac{1}{2} & \frac{1}{2} \\ 0 & 1 & 0 & 0 & 0 & 0 & 0 & 0 & \frac{1}{2} & -\frac{1}{2} \\ 0 & 0 & 0 & 0 & \frac{1}{2} & \frac{\sqrt{3}}{2} & \frac{1}{2} & \frac{1}{2} & 0 & 0 \\ 0 & 0 & 0 & \frac{1}{2} & 0 & -\frac{\sqrt{3}}{2} & \frac{1}{2} & \frac{1}{2} & 0 & 0 \\ 0 & 0 & 0 & \frac{\sqrt{3}}{2} & -\frac{\sqrt{3}}{2} & 0 & 0 & 0 & 0 & 0 \\ 0 & 0 & 0 & \frac{1}{2} & \frac{1}{2} & 0 & 0 & 0 & -1 & 0 \\ 0 & 0 & 0 & \frac{1}{2} & \frac{1}{2} & 0 & 0 & -1 & 0 & 0 \\ \frac{\sqrt{3}}{2} & -\frac{1}{2} & \frac{1}{2} & 0 & 0 & 0 & 0 & 0 & 0 & -\frac{1}{2} \\ \frac{\sqrt{3}}{2} & \frac{1}{2} & -\frac{1}{2} & 0 & 0 & 0 & 0 & 0 & -\frac{1}{2} & 0 \end{bmatrix}.$$

The Hamiltonian $H_{CFEx}^{D3h} + H_{SO}^z$ can be diagonalized using the perturbation theory. The results are

$$\phi'_{1z} = \phi_{1z} + \frac{\sqrt{3}}{2}\frac{\xi}{E_{ex}+b}(\phi_{9z}+\phi_{10z}), \quad E'_{1z} = E_{ex} + \frac{3}{2}\frac{\xi^2}{E_{ex}+b}$$

$$\phi'_{2z} = \frac{1}{\sqrt{2}}(\phi_{2z}+\phi_{3z}), \quad E'_{2z} = E_{ex} - a + \xi$$

$$\phi'_{3z} = \frac{1}{\sqrt{2}}(\phi_{3z}-\phi_{2z}) + \frac{1}{\sqrt{2}}\frac{\xi}{E_{ex}+b-a}(\phi_{9z}-\phi_{10z}), \quad E'_{3z} = E_{ex} - a - \xi + \frac{\xi^2}{E_{ex}-a+b}$$

$$\phi'_{4z} = \frac{1}{\sqrt{2}}(\phi_{4z}+\phi_{5z}) + \frac{1}{\sqrt{2}}\frac{\xi}{E_{ex}+b-a}(\phi_{7z}+\phi_{8z}), \quad E'_{4z} = E_{ex} - b + \frac{\xi}{2} + \frac{\xi^2}{E_{ex}+a-b}$$

$$\phi'_{5z} = \frac{1}{\sqrt{2}}(\phi_{4z}-\phi_{5z}) + \frac{\sqrt{6}}{2}\frac{\xi}{E_{ex}-b}\phi_{6z}, \quad E'_{5z} = E_{ex} - b - \frac{\xi}{2} + \frac{3}{2}\frac{\xi^2}{E_{ex}-b}$$

$$\phi'_{6z} = \phi_{6z} - \frac{\sqrt{3}}{2}\frac{\xi}{E_{ex}-b}(\phi_{4z}-\phi_{5z}), \quad E'_{6z} = -\frac{3}{2}\frac{\xi^2}{E_{ex}-b}$$

$$\phi'_{7z} = \frac{1}{\sqrt{2}}(\phi_{7z}+\phi_{8z}) - \frac{1}{\sqrt{2}}\frac{\xi}{E_{ex}+b-a}(\phi_{4z}+\phi_{5z}), \quad E'_{7z} = -a - \xi - \frac{\xi^2}{E_{ex}+a-b}$$

$$\phi'_{8z} = \frac{1}{\sqrt{2}}(\phi_{8z}-\phi_{7z}), \quad E'_{8z} = -a + \xi$$

$$\phi'_{9z} = \frac{1}{\sqrt{2}}(\phi_{9z} + \phi_{10z}) - \frac{\sqrt{6}}{2}\frac{\xi}{E_{ex}+b}\phi_{1z}, E'_{9z} = -b - \frac{\xi}{2} - \frac{3}{2}\frac{\xi^2}{E_{ex}+b}$$

$$\phi'_{10z} = \frac{1}{\sqrt{2}}(\phi_{9z} - \phi_{10z}) - \frac{1}{\sqrt{2}}\frac{\xi}{E_{ex}-a+b}(\phi_{3z} - \phi_{2z}), E'_{10z} = -b + \frac{\xi}{2} - \frac{\xi^2}{E_{ex}-a+b}.$$

*Spin along the x and y direction*

In this case, in order to use the same $H_{CFEx}^{D3h}$, we need to choose different basis and the corresponding spin-orbit Hamiltonian.

When the spins are along the $x$ axis, the basis of the spin-orbit interaction Hamiltonians are:

$$\phi_{1x} = \frac{1}{\sqrt{2}}(|zz \uparrow\rangle + |zz \downarrow\rangle), \phi_{2x} = \frac{1}{\sqrt{2}}(|xy \uparrow\rangle + |xy \downarrow\rangle),$$

$$\phi_{3x} = \frac{1}{\sqrt{2}}(|x^2-y^2 \uparrow\rangle + |x^2-y^2 \downarrow\rangle), \phi_{4x} = \frac{1}{\sqrt{2}}(|xz \uparrow\rangle + |xz \downarrow\rangle), \phi_{5x} = \frac{1}{\sqrt{2}}(|yz \uparrow\rangle + |yz \downarrow\rangle),$$

$$\phi_{6x} = \frac{1}{\sqrt{2}}(|zz \uparrow\rangle - |zz \downarrow\rangle), \phi_{7x} = \frac{1}{\sqrt{2}}(|xy \uparrow\rangle - |xy \downarrow\rangle),$$

$$\phi_{8x} = \frac{1}{\sqrt{2}}(|x^2-y^2 \uparrow\rangle - |x^2-y^2 \downarrow\rangle), \phi_{9x} = \frac{1}{\sqrt{2}}(|xz \uparrow\rangle - |xz \downarrow\rangle), \phi_{10x} = \frac{1}{\sqrt{2}}(|yz \uparrow\rangle - |yz \downarrow\rangle).$$

The Hamiltonian for the spin-orbit interaction is:

$$H_{SO}^x = \xi \begin{bmatrix} 0 & 0 & 0 & \frac{\sqrt{3}}{2} & 0 & 0 & 0 & 0 & 0 & -\frac{\sqrt{3}}{2} \\ 0 & 0 & 0 & 0 & \frac{1}{2} & 0 & 0 & 1 & \frac{1}{2} & 0 \\ 0 & 0 & 0 & \frac{1}{2} & 0 & 0 & 1 & 0 & 0 & \frac{1}{2} \\ \frac{\sqrt{3}}{2} & 0 & \frac{1}{2} & 0 & 0 & 0 & -\frac{1}{2} & 0 & 0 & \frac{1}{2} \\ 0 & \frac{1}{2} & 0 & 0 & 0 & \frac{\sqrt{3}}{2} & 0 & -\frac{1}{2} & \frac{1}{2} & 0 \\ 0 & 0 & 0 & 0 & \frac{\sqrt{3}}{2} & 0 & 0 & 0 & -\frac{\sqrt{3}}{2} & 0 \\ 0 & 0 & 1 & -\frac{1}{2} & 0 & 0 & 0 & 0 & 0 & -\frac{1}{2} \\ 0 & 1 & 0 & 0 & -\frac{1}{2} & 0 & 0 & 0 & -\frac{1}{2} & 0 \\ 0 & \frac{1}{2} & 0 & 0 & \frac{1}{2} & -\frac{\sqrt{3}}{2} & 0 & -\frac{1}{2} & 0 & 0 \\ -\frac{\sqrt{3}}{2} & 0 & \frac{1}{2} & \frac{1}{2} & 0 & 0 & -\frac{1}{2} & 0 & 0 & 0 \end{bmatrix}.$$

When the spins are along the y axis, the basis are

$$\phi_{1y} = \frac{1}{\sqrt{2}}(|zz \uparrow\rangle + i|zz \downarrow\rangle), \phi_{2y} = \frac{1}{\sqrt{2}}(|xy \uparrow\rangle + i|xy \downarrow\rangle),$$

$$\phi_{3y} = \frac{1}{\sqrt{2}}(|x^2-y^2 \uparrow\rangle + i|x^2-y^2 \downarrow\rangle), \phi_{4y} = \frac{1}{\sqrt{2}}(|xz \uparrow\rangle + i|xz \downarrow\rangle), \phi_{5y} = \frac{1}{\sqrt{2}}(|yz \uparrow\rangle + i|yz \downarrow\rangle),$$

$$\phi_{6y} = \frac{1}{\sqrt{2}}(|zz \uparrow\rangle - i|zz \downarrow\rangle), \phi_{7y} = \frac{1}{\sqrt{2}}(|xy \uparrow\rangle - i|xy \downarrow\rangle),$$

$$\phi_{8y} = \frac{1}{\sqrt{2}}(|x^2-y^2 \uparrow\rangle - i|x^2-y^2 \downarrow\rangle), \phi_{9y} = \frac{1}{\sqrt{2}}(|xz \uparrow\rangle - i|xz \downarrow\rangle), \phi_{10y} = \frac{1}{\sqrt{2}}(|yz \uparrow\rangle - i|yz \downarrow\rangle).$$

The Hamiltonian for the spin-orbit interaction is:

$$H_{SO}^{y} = \xi \begin{bmatrix} 0 & 0 & 0 & 0 & \frac{-i\sqrt{3}}{2} & 0 & 0 & 0 & \frac{i\sqrt{3}}{2} & 0 \\ 0 & 0 & 0 & \frac{i}{2} & 0 & 0 & 0 & 1 & 0 & \frac{i}{2} \\ 0 & 0 & 0 & 0 & \frac{i}{2} & 0 & 1 & 0 & \frac{i}{2} & 0 \\ 0 & \frac{-i}{2} & 0 & 0 & 0 & \frac{i\sqrt{3}}{2} & 0 & \frac{i}{2} & 0 & \frac{1}{2} \\ \frac{i\sqrt{3}}{2} & 0 & \frac{-i}{2} & 0 & 0 & 0 & \frac{i}{2} & 0 & \frac{1}{2} & 0 \\ 0 & 0 & 0 & \frac{-i\sqrt{3}}{2} & 0 & 0 & 0 & 0 & 0 & \frac{i\sqrt{3}}{2} \\ 0 & 0 & 1 & 0 & \frac{-i}{2} & 0 & 0 & 0 & \frac{-i}{2} & 0 \\ 0 & 1 & 0 & \frac{-i}{2} & 0 & 0 & 0 & 0 & 0 & \frac{-i}{2} \\ \frac{-i\sqrt{3}}{2} & 0 & \frac{-i}{2} & 0 & \frac{1}{2} & 0 & \frac{i}{2} & 0 & 0 & 0 \\ 0 & \frac{-i}{2} & 0 & \frac{1}{2} & 0 & \frac{-i\sqrt{3}}{2} & 0 & \frac{i}{2} & 0 & 0 \end{bmatrix}.$$

These Hamiltonians have the following features: 1) The diagonal terms are all zero. 2) The degenerate states are not directly coupled. 3) The degenerate states are not coupled by a third states. So the perturbation energy are all second order:

$$\phi_i' = \phi_i + \sum_i \frac{\langle \phi_i | H_{SO} | \phi_j \rangle}{E_i - E_j} \phi_j, E_i' = E_i + \sum_i \frac{|\langle \phi_i | H_{SO} | \phi_j \rangle|^2}{E_i - E_j}.$$

For example, when the spins are along the $x$ axis, the results of the energies are:

$$E_{1x}' = E_{ex} + \frac{3}{4}\frac{\xi^2}{b} + \frac{3}{4}\frac{\xi^2}{E_{ex}+b}$$

$$E_{2x}' = E_{ex} - a + \frac{1}{4}\frac{\xi^2}{b-a} + \frac{\xi^2}{E_{ex}} + \frac{1}{4}\frac{\xi^2}{E_{ex}+b-a}$$

$$E_{3x}' = E_{ex} - a + \frac{1}{4}\frac{\xi^2}{b-a} + \frac{\xi^2}{E_{ex}} + \frac{1}{4}\frac{\xi^2}{E_{ex}+b-a}$$

$$E_{4x}' = E_{ex} - b - \frac{3}{4}\frac{\xi^2}{b} - \frac{1}{4}\frac{\xi^2}{b-a} + \frac{1}{4}\frac{\xi^2}{E_{ex}+a-b} + \frac{1}{4}\frac{\xi^2}{E_{ex}}$$

$$E_{5x}' = E_{ex} - b - \frac{1}{4}\frac{\xi^2}{b-a} + \frac{3}{4}\frac{\xi^2}{E_{ex}-b} + \frac{1}{4}\frac{\xi^2}{E_{ex}+a-b} + \frac{1}{4}\frac{\xi^2}{E_{ex}}$$

$$E_{6x}' = -\frac{3}{4}\frac{\xi^2}{b} - \frac{3}{4}\frac{\xi^2}{E_{ex}-b}$$

$$E_{7x}' = -a + \frac{1}{4}\frac{\xi^2}{b-a} - \frac{\xi^2}{E_x} - \frac{1}{4}\frac{\xi^2}{E_x+a-b}$$

$$E_{8x}' = -a + \frac{1}{4}\frac{\xi^2}{b-a} - \frac{\xi^2}{E_x} - \frac{1}{4}\frac{\xi^2}{E_x+a-b}$$

$$E_{9x}' = -b - \frac{3}{4}\frac{\xi^2}{b} - \frac{1}{4}\frac{\xi^2}{b-a} - \frac{1}{4}\frac{\xi^2}{E_x+b-a} - \frac{1}{4}\frac{\xi^2}{E_x}$$

$$E_{10x}' = -b - \frac{1}{4}\frac{\xi^2}{b-a} - \frac{3}{4}\frac{\xi^2}{E_x+b} - \frac{1}{4}\frac{\xi^2}{E_x+b-a} - \frac{1}{4}\frac{\xi^2}{E_x}.$$

All the perturbations are in second order, which is small compared to the modification to energy generated by the spin in the $z$ direction.

Figure S7.1 depicts the energy diagram of 3d electron levels in D$_{3h}$ local symmetry. With the spin orbit coupling, the energy of a certain orbital actually depend on the spin direction. We define the single-ion anisotropy energy as

$$E_i^{an} = E_{iz} - E_{ix}.$$

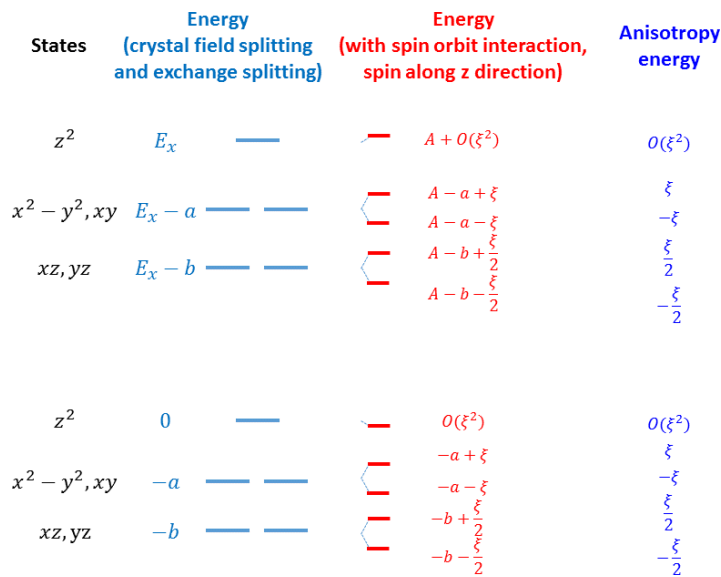

**Figure S7.1** Single-ion anisotropy energy of 3d electrons in a D$_{3h}$ local symmetry. The spin-orbit splitting when the spin is along the $x$ direction are all on the order of $\xi^2$.

Therefore, in case of D$_{3h}$ local symmetry, the modification of energy when spin is along the $z$ direction is dominant.

Using this model, one can analyze the single-ion anisotropy energy in the D$_{3h}$ symmetry.

For $Fe^{3+}$, if the electron occupation follows the nominal valence as $3d^5$ in the majority band, the total spin orbit coupling energy is canceled. So there will be no anisotropy, which is why the anisotropy is in general small (but non-zero) for $Fe^{3+}$, even if the local symmetry is anisotropic. The non-zero magnetic anisotropy is then expected to be related to the partially filled minority states due to the Fe 3d-O 2p hybridization. As shown in Section 8, the hybridization between the $x^2 - y^2$ and the O 2p states are the same as that between the $xy$ state and the O 2p states. This is also true for $xz$ and $yz$ states. So if we calculate the total energy using $\sum_i E_i n_i$, where $i$ is the index for states, and $n_i$ is the population of the $ith$ state, the first order modification to the energies in terms of $\xi$ is canceled. The result of the sum $\sum_i E_i n_i$ depends on the second order terms. In principle, if the hybridization level of all states are the same, the sum $\sum_i E_i n_i$ vanishes. On the other hand, due to D$_{3h}$ symmetry, the hybridization of the $z^2$ state is the strongest, followed by the $x^2 - y^2$ and $xy$ states, and the $xz$ and $yz$ states (see Section 8). This difference in hybridization creates an imbalance in $n_i$, causing a non-zero $\sum_i E_i n_i$. In this case, when the spins are along the $x$ axis, the minority states can interact with other minority states, pushing the $z^2$, $x^2 - y^2$, and $xy$ up and $xz$ and $yz$ states down. Combined with the population imbalance, the result is $\sum_i E_{ix} n_{ix} > \sum_i E_{iz} n_{iz}$. This means that the $z$ axis is the easy axis for the single-ion anisotropy. The fact that the spins are in the $x - y$ plane suggests that the triangular lattice plays an important role in the spin orientations.

In the $D_{3h}$ symmetry, the $x$ and $y$ directions are equivalent. So one finds $\sum_i E_{ix} n_{ix} = \sum_i E_{iy} n_{iy}$. To discuss the in-plane anisotropy, one needs to introduce the $C_S$ distortion, as shown below.

Note that, using this model, we can also analyze the single-ion anisotropy of Fe in $LuFe_2O_4$. For $Fe^{2+}$, the additional electron on $xz, yz$ orbital generates an anisotropy energy $-\frac{\xi}{2}$, which favors spin along the $z$ direction. This explains why the coercive field of $LuFe_2O_4$ is very large (close to 10 T at 4 K): the local environment is the anisotropic $D_{3h}$, and the $3d^6$ electronic configuration generates a single ion anisotropic energy on the order $-\frac{\xi}{2}$. [1]

### S7.1.2 $C_S$ symmetry

In order to understand the single-ion anisotropy in h-$LuFeO_3$, one needs to consider the structural distortion from $D_{3h}$ symmetry and calculate the anisotropy in the $C_S$ symmetry.

As shown in the Section 6, the distortion to $C_S$ symmetry will split the $e'$ and $e''$ levels. To simplify the calculation, we use a single parameter $d$ to represent the $D_{3h}$ to $C_S$ distortion, i.e. only consider the splitting between the $yz$ and $xz$ states because that is the largest. Using the basis $\{z^2, xy, x^2 - y^2, xz, yz\}$, the Hamiltonian is

$$H^{D3h}_{CFEx} = \begin{bmatrix} E_{ex} & 0 & 2d & 0 & 0 & 0 & 0 & 0 & 0 & 0 \\ 0 & E_{ex}-a & 0 & 0 & 0 & 0 & 0 & 0 & 0 & 0 \\ 2d & 0 & E_{ex}-a & 0 & 0 & 0 & 0 & 0 & 0 & 0 \\ 0 & 0 & 0 & E_{ex}-b+d & 0 & 0 & 0 & 0 & 0 & 0 \\ 0 & 0 & 0 & 0 & E_{ex}-b-d & 0 & 0 & 0 & 0 & 0 \\ 0 & 0 & 0 & 0 & 0 & 0 & 0 & 2d & 0 & 0 \\ 0 & 0 & 0 & 0 & 0 & 0 & -a & 0 & 0 & 0 \\ 0 & 0 & 0 & 0 & 0 & 2d & 0 & -a & 0 & 0 \\ 0 & 0 & 0 & 0 & 0 & 0 & 0 & 0 & -b+d & 0 \\ 0 & 0 & 0 & 0 & 0 & 0 & 0 & 0 & 0 & -b-d \end{bmatrix}.$$

By diagonalizing the total Hamiltonian numerically and comparing with the results when the spin is along the $y$ axis, one can discuss the anisotropy energy in the $x - y$ plane. As shown in Fig. S7.2, when $d = 0$ ($D_{3h}$ symmetry), within the $x - y$ plane, the spin orientation is isotropic, i.e. $E_y - E_x = 0$. When $d < 0$, the minority (unoccuopied) $yz$ orbit has a higher energy because of the $C_S$ distortion; its hybridization is also stronger. At the same time, the anisotropy energy $E_y - E_x > 0$, suggesting that the $x$ axis is an easy axis. Therefore, the spin is more likely to point toward $x$ direction, which is observed in hexagonal $LuFeO_3$. When $d > 0$, minority (unoccuopied) $yz$ orbit has higher energy and higher hybridization. Since the anisotropy energy of $xz$ orbit is $E_x - E_y > 0$, the $y$ axis becomes the easy axis. More details of the calculation is shown below.

When the spins are in the $x$ or $y$ direction, one can estimate the eigenenergies using the perturbation theory. For the minority states

$$E'_{1,x} = E_{ex} + \frac{2d^2}{a} + \frac{3}{4}\frac{\xi^2}{b-d} + \frac{3}{4}\frac{\xi^2}{E_x+b+d}$$

$$E'_{2,x} = E_{ex} - a + \frac{1}{4}\frac{\xi^2}{b-a+d} + \frac{\xi^2}{E_x} + \frac{1}{4}\frac{\xi^2}{E_x+b-a-d}$$

$$E'_{3,x} = E_{ex} - a - \frac{2d^2}{a} + \frac{1}{4}\frac{\xi^2}{b-a-d} + \frac{\xi^2}{E_x} + \frac{1}{4}\frac{\xi^2}{E_x+b-a+d}$$

$$E'_{4,x} = E_{ex} - b + d - \frac{3}{4}\frac{\xi^2}{b-d} - \frac{1}{4}\frac{\xi^2}{b-a-d} + \frac{1}{4}\frac{\xi^2}{E_x+a-b+d} + \frac{1}{4}\frac{\xi^2}{E_x+2d}$$

$$E'_{5,x} = E_{ex} - b - d - \frac{1}{4}\frac{\xi^2}{b-a-d} + \frac{3}{4}\frac{\xi^2}{E_x - b - d} + \frac{1}{4}\frac{\xi^2}{E_x + a - b - d} + \frac{1}{4}\frac{\xi^2}{E_x - 2d}$$

$$E'_{1,y} = E_{ex} + \frac{2d^2}{a} + \frac{3}{4}\frac{\xi^2}{b+d} + \frac{3}{4}\frac{\xi^2}{E_x + b - d}$$

$$E'_{2,y} = E_{ex} - a + \frac{1}{4}\frac{\xi^2}{b-a-d} + \frac{\xi^2}{E_x} + \frac{1}{4}\frac{\xi^2}{E_x + b - a + d}$$

$$E'_{3,y} = E_{ex} - a - \frac{2d^2}{a} + \frac{1}{4}\frac{\xi^2}{b-a+d} + \frac{\xi^2}{E_x} + \frac{1}{4}\frac{\xi^2}{E_x + b - a - d}$$

$$E'_{4,y} = E_{ex} - b + d - \frac{1}{4}\frac{\xi^2}{b-a-d} + \frac{3}{4}\frac{\xi^2}{E_x - b + d} + \frac{1}{4}\frac{\xi^2}{E_x + a - b + d} + \frac{1}{4}\frac{\xi^2}{E_x + 2d}$$

$$E'_{5,y} = E_{ex} - b - d - \frac{3}{4}\frac{\xi^2}{b+d} - \frac{1}{4}\frac{\xi^2}{b-a+d} + \frac{1}{4}\frac{\xi^2}{E_x + a - b - d} + \frac{1}{4}\frac{\xi^2}{E_x - 2d}$$

Then one can calculate anisotropy

$$E'_{1,y} - E'_{1,x} = \frac{3}{4}\frac{\xi^2}{b+d} + \frac{3}{4}\frac{\xi^2}{E_{ex}+b-d} - \frac{3}{4}\frac{\xi^2}{b-d} - \frac{3}{4}\frac{\xi^2}{E_{ex}+b+d}$$

$$E'_{2,y} - E'_{2,x} = \frac{1}{4}\frac{\xi^2}{b-a-d} + \frac{1}{4}\frac{\xi^2}{E_{ex}+b-a+d} - \frac{1}{4}\frac{\xi^2}{b-a+d} - \frac{1}{4}\frac{\xi^2}{E_{ex}+b-a-d}$$

$$E'_{3,y} - E'_{3,x} = \frac{1}{4}\frac{\xi^2}{b-a+d} + \frac{1}{4}\frac{\xi^2}{E_{ex}+b-a-d} - \frac{1}{4}\frac{\xi^2}{b-a-d} - \frac{1}{4}\frac{\xi^2}{E_{ex}+b-a+d}$$

$$E'_{4,y} - E'_{4,x} = \frac{3}{4}\frac{\xi^2}{E_{ex}-b+d} + \frac{3}{4}\frac{\xi^2}{b-d}$$

$$E'_{5,y} - E'_{5,x} = -\frac{3}{4}\frac{\xi^2}{E_{ex}-b-d} - \frac{3}{4}\frac{\xi^2}{b+d}.$$

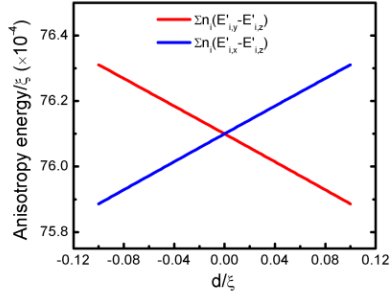

**Figure. S7.2** Anisotropy energy in the $x - y$ plane as a function of the C$_S$ distortion energy. The parameters assumed are $\xi = 50\ meV, a = 1.1\ eV, b = 1.4\ eV, E_{ex} = 4\ eV, n_1 = 0.18, n_2 = n_3 = 0.068, n_4 = n_5 = 0.046$.

The sum of the anisotropy energy $\sum(E'_{i,y} - E'_{i,x})$ is expected to be close to zero. On the other hand, since the hybridization is different for different states, the real anisotropy energy $\sum n_i(E'_{i,y} - E'_{i,x})$ is non-zero. In particular, since the hybridization of the $z^2$ state is the largest, the sign of $E'_{1,y} - E'_{1,x}$ determines the total anisotropy energy. One can expand the anisotropy energies assuming $d$ is small, the results are

$$E'_{1,y} - E'_{1,x} \approx -\frac{3}{2}\frac{\xi^2}{b^2}d + \frac{3}{2}\frac{\xi^2}{(E_x+b)^2}d$$

$$E'_{2,y} - E'_{2,x} \approx \frac{1}{2}\frac{\xi^2}{(b-a)^2}d - \frac{1}{2}\frac{\xi^2}{(E_x+b-a)^2}d$$

$$E'_{3,y} - E'_{3,x} \approx -\frac{1}{2}\frac{\xi^2}{(b-a)^2}d + \frac{1}{2}\frac{\xi^2}{(E_x+b-a)^2}d$$

$$E'_{4,y} - E'_{4,x} \approx -\frac{3}{4}\frac{\xi^2}{(E_x-b)^2}d + \frac{3}{4}\frac{\xi^2}{b^2}d$$

$$E'_{5,y} - E'_{5,x} \approx -\frac{3}{4}\frac{\xi^2}{(E_x-b)^2}d + \frac{3}{4}\frac{\xi^2}{b^2}d.$$

When $d < 0$ (the case for h-LFO at low temperature), the $E'_{1,y} - E'_{1,x} > 0$, suggesting that the spins are preferred to be along the $x$ direction, which is consistent with the experimental observations.

Figure S7.2 shows the numerical results of $\sum n_i(E'_{i,y} - E'_{i,z})$ and $\sum n_i(E'_{i,x} - E'_{i,z})$ as a function of C$_S$ distortion parameterized using variable $d$. The minority state population is chosen to be proportional to the hybridization (see Section 8), while the majority states are filled. Note that multiplying the population of all states does not change the results qualitatively. Indeed, when $d < 0$, $\sum n_i(E'_{i,x} - E'_{i,z}) < \sum n_i(E'_{i,y} - E'_{i,z})$, suggesting that the easy axis in the $x - y$ plane is the $x$ axis.

### S7.2 Orthorhombic LuFeO$_3$

In the orthorhombic LuFeO$_3$, the local environment of the Fe is the FeO$_6$ octahedral, which is approximately O$_h$ symmetry. In O$_h$ symmetry, the $x$, $y$, and $z$ directions are equivalent, making the anisotropy energy minimal. The distortion into D$_{2h}$ is expected to generate anisotropy.

We choose the same basis as we do in the analysis of D$_{3h}$ symmetry for the spin in the $x$, $y$ and $z$ directions. Then the spin-orbit coupling Hamiltonian are the same. The Hamiltonian for the crystal field and exchange splitting is

$$H^{C2h}_{CFEx} = \begin{bmatrix} E_{ex}+E_{10dq} & 0 & d & 0 & 0 & 0 & 0 & 0 & 0 & 0 \\ 0 & E_{ex} & 0 & 0 & 0 & 0 & 0 & 0 & 0 & 0 \\ d & 0 & E_{ex}+E_{10dq} & 0 & 0 & 0 & 0 & 0 & 0 & 0 \\ 0 & 0 & 0 & E_{ex}+2d & 0 & 0 & 0 & 0 & 0 & 0 \\ 0 & 0 & 0 & 0 & E_{ex}-2d & 0 & 0 & 0 & 0 & 0 \\ 0 & 0 & 0 & 0 & 0 & E_{10dq} & 0 & d & 0 & 0 \\ 0 & 0 & 0 & 0 & 0 & 0 & 0 & 0 & 0 & 0 \\ 0 & 0 & 0 & 0 & 0 & d & 0 & E_{10dq} & 0 & 0 \\ 0 & 0 & 0 & 0 & 0 & 0 & 0 & 0 & 2d & 0 \\ 0 & 0 & 0 & 0 & 0 & 0 & 0 & 0 & 0 & -2d \end{bmatrix},$$

where $E_{10dq}$ is the splitting between $t_{2g}$ and $e_g$ levels and $d$ is the parameter that represent the magnitude of the D$_{2h}$ distortion. Again, for simplicity, only a single parameter $d$ is used to represent the distortion; the tetragonal part of the distortion is ignored. When $d = 0$, the symmetry is O$_h$. Note that the sign of $d$ is the same as $b - a$, as shown in the Section 6.

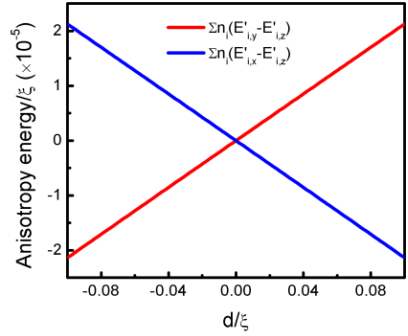

**Figure S7.3** Effect of lattice distortion from O$_h$ symmetry to D$_{2h}$ symmetry on the single ion anisotropy. The parameters assumed are $E_{ex} = 4\ eV$, $E_{10dq} = 1.4\ eV\ and\ \xi = 50\ meV, n_1 = 0, n_1 = n_3 = 0.07, n_2 = n_4 = n_5 = 0.02$.

Figure S7.3 displays the anisotropy energy of $\sum n_i(E'_{i,y} - E'_{i,z})$ and $\sum n_i(E'_{i,x} - E'_{i,z})$, calculated by diagonalizing the total Hamiltonian numerically. The minority state population is chosen to be proportional to the hybridization (see Section 8), while the majority states are filled. When distortion parameter $d > 0$, $\sum n_i(E'_{i,x} - E'_{i,z}) > \sum n_i(E'_{i,x} - E'_{i,z})$, suggesting that the $x$ axis is the easy axis, which is consistent with the experimental observations. [8]

Observed single-ion anisotropy

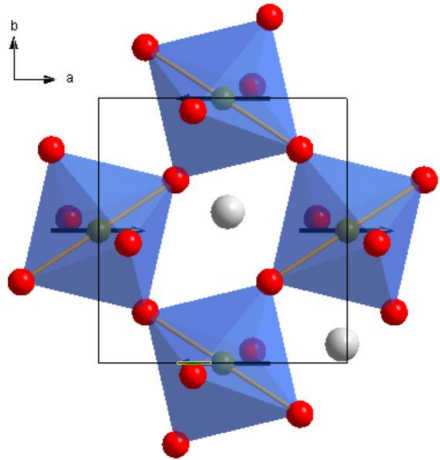

**Figure S7.4** Structure and spin orientation in orthorhombic LuFO$_3$. The arrows are the spin orientations. The rods indicate the shortest O-Fe-O path in the FeO$_6$; they are the local $x$ axis used in the analysis above.

As shown in Fig. S7.4, the observed spins are actually not pointing to the symmetry axis of the FeO$_6$ octahedra. Instead, they point toward the crystal $a$ axis, which is the shortest axis. This is an example in which the geometric arrangements of the magnetic ions play an important role. As shown in Fig. S7.4, the shortest axes (the $x$ axis) of the FeO$_6$ are indicated using solid rods. The angles between the solid rods are 52 degree. If the spins are all oriented according to the single-ion anisotropy, the antiparallel alignment required by the antiferromagnetism cannot be satisfied. Since the exchange interaction has a larger energy scale than that of the single-ion anisotropy, the spins are reoriented to form the antiferromagnetic order. To minimize the energy loss in the single-ion anisotropy, the spins are aligned along the $a$ axis to have a minimum common angle with the easy axis of the single-ion anisotropy. In other words, the observed magnetocrystalline anisotropy is consistent with the predicted easy axis from the single-ion anisotropy.

## S8. Hybridization of atomic orbitals

### S8.1 General concept

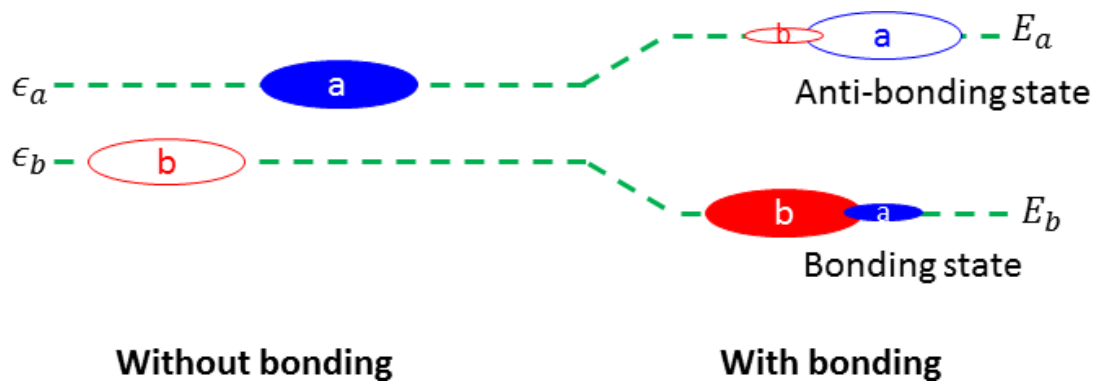

**Figure S8.1** Schematic illustration of hybridization

As shown in Fig. S7.1, when two atoms form an ionic bond, the new electronic (bonding state and anti-bonding) states are formed. The bonding (occupied) and anti-bonding (unoccupied) states are both superposition of the original atomic states; this is hybridization. Here we discuss the hybridization using a one-electron picture, to provide a simplest model.

Consider two bonding orbitals $|\phi_a>$ and $|\phi_b>$ (on atom $a$ and $b$ respectively), which are the eigenstates of Hamiltonian $H_a = \frac{\hbar^2}{2m}\nabla^2 + V_a$ and $H_b = \frac{\hbar^2}{2m}\nabla^2 + V_b$ respectively, where $V_a = -\frac{Z_a e^2}{|\vec{r}-\vec{r}_a|}$ and $V_b = -\frac{Z_b e^2}{|\vec{r}-\vec{r}_b|}$.

It follows that $H_a|\phi_a> = \epsilon_a|\phi_a>$ and $H_b|\phi_b> = \epsilon_b|\phi_b>$, where $\epsilon_a$ and $\epsilon_b$ are the eigenenergies.

In the bonded atoms, the Hamiltonian for the electron becomes $H = \frac{\hbar^2}{2m}\nabla^2 + V_a + V_b$.

Assuming the new eigenstates are $|\psi> = a|\phi_a> + b|\phi_b>$, if follows that
$$H\psi = \left(\frac{\hbar^2}{2m}\nabla^2 + V_a + V_b\right)(a|\phi_a> + b|\phi_b>) = a(\epsilon_a + V_b)|\phi_a> + b(\epsilon_b + V_a)|\phi_b>.$$

Consider the Schodinger equation $H|\psi> = E|\psi>$, one has
$$a(\epsilon_a + V_b)|\phi_a> + b(\epsilon_b + V_a)|\phi_b> = E(a|\phi_a> + b|\phi_b>).$$

Taking inner product of the above equation with $<\phi_a|$ and $<\phi_b|$, one reaches two linear equations:
$$a(\epsilon_a + <\phi_a|V_b|\phi_a>) + b<\phi_a|V_a|\phi_b> = aE,$$
$$a<\phi_b|V_b|\phi_a> + b(\epsilon_b + <\phi_b|V_a|\phi_b>) = aE.$$

If we define $V_{aba} \equiv <\phi_a|V_b|\phi_a>$, $V_{aab} \equiv <\phi_a|V_a|\phi_b>$, $V_{bba} \equiv <\phi_b|V_b|\phi_a>$, and $V_{bab} \equiv <\phi_b|V_a|\phi_b>$, the new states are
$$E_{a,b} = \frac{\epsilon_a + V_{aba} + \epsilon_b + V_{bab} \pm \sqrt{(\epsilon_a + V_{aba} - \epsilon_b - V_{bab})^2 + 4V_{aab}V_{bba}}}{2}$$
$$\psi_a = \phi_a - \frac{\epsilon_a + V_{aba} - \epsilon_b - V_{bab} - \sqrt{(\epsilon_a + V_{aba} - \epsilon_b - V_{bab})^2 + 4V_{aab}V_{bba}}}{2V_{aab}}\phi_b,$$

$$\psi_b = \phi_b + \frac{\epsilon_a + V_{aba} - \epsilon_b - V_{bab} - \sqrt{(\epsilon_a + V_{aba} - \epsilon_b - V_{bab})^2 + 4V_{aab}V_{bba}}}{2V_{bba}} \phi_a.$$

If $\epsilon_a + V_{aba} - \epsilon_b - V_{bab} \gg 4V_{aab}V_{bba}$, the solution are

$$E_a = \epsilon_a + V_{aba} + \frac{4V_{aab}V_{bba}}{2(\epsilon_a + V_{aba} - \epsilon_b - V_{bab})}$$

$$E_b = \epsilon_b + V_{bab} - \frac{4V_{aab}V_{bba}}{2(\epsilon_a + V_{aba} - \epsilon_b - V_{bab})}$$

$$\psi_a = \phi_a + \frac{V_{bba}}{2(\epsilon_a + V_{aba} - \epsilon_b - V_{bab})} \phi_b$$

$$\psi_b = \phi_b - \frac{V_{aab}}{2(\epsilon_a + V_{aba} - \epsilon_b - V_{bab})} \phi_a.$$

Furthermore, $V_{aba}$ and $V_{bab}$ are small compared with the $\epsilon_a - \epsilon_b$, so the solutions can be reduced to

$$E_a = \epsilon_a + \frac{4V_{aab}V_{bba}}{2(\epsilon_a - \epsilon_b)}$$

$$E_b = \epsilon_b - \frac{4V_{aab}V_{bba}}{2(\epsilon_a - \epsilon_b)}$$

$$\psi_a = \phi_a + \frac{V_{bba}}{2(\epsilon_a - \epsilon_b)} \phi_b$$

$$\psi_b = \phi_b - \frac{V_{aab}}{2(\epsilon_a - \epsilon_b)} \phi_a.$$

Therefore, the degree of hybridization can be estimated using the parameters $V_{bba}$ or $V_{aab}$. Note that $V_{bba} = V_{aab}$. For example, if an Fe-O bond is concerned, the contribution of oxygen orbital to the unoccupied Fe orbital is related to $V_{OOFe} = <\phi_O|V_O|\phi_{Fe}>$.

S8.2 Hybridization of between metal (Fe and Lu) and oxygen in h-LuFeO$_3$

One can see from the simple model of hybridization that the matrix element $V^i_{m_1,m_2} = \left\langle \psi^{m_1}_{Fe-3d} \middle| V_{Fe} \middle| \phi^{m_2}_{O-2p,i} \right\rangle$ is important, where $m_1, m_2$ are the magnetic quantum numbers, and $i$ is the index for the oxygen neighbor. In order to calculate the matrix, we define the $z$ axis along the vector that connects Fe and O atoms. In that case, if we write down the wave function of Fe and O using spherical harmonic function $\Psi^m_l$, the matrix element are only non-zero when $m_1 = m_2$, because $V_{Fe}$ does not depend on the azimuthal angle $\phi$. Therefore, we can rewrite the matrix element as $V^i_m = V^i_{m_1,m_2}$, where $m = m_1 = m_2$. When $m = 0, 1, 2$ the matrix element corresponds to a so called $\sigma, \pi, \delta$ bond, respectively. Since we are dealing with the interaction between Fe-3d and O-2p, the matrix elements are also written as $V^i_{pd\sigma}$ and $V^i_{pd\pi}$. According to Harrison, [9] the matrix elements are proportional to $d^{-3.5}$, where $d$ is the distance between the two atoms for a $pd$ bond.

Therefore, the calculation of the matrix elements $V^i_m$ comes down to transforming the atomic orbitals into the coordinate system mentioned above, i.e. $z$ axis along the vector that connects Fe and O atoms. Below, we calculate the individual $V^i_m$, starting from D$_{3h}$ symmetry and introduce C$_S$ symmetry later.

### S8.2.1 D$_{3h}$ symmetry

For the hybridization between the Fe and apex oxygen atoms, the coordinate system is automatically set for calculation. So the results are readily obtained:

$$V^{i=1}_{m=0} = V^{i=2}_{m=0} = V_{pd\sigma}R_c^{-3.5}$$

$$V^{i=1}_{m=1} = V^{i=2}_{m=1} = V^{i=1}_{m=-1} = V^{i=2}_{m=-1} = V_{pd\pi}R_c^{-3.5}.$$

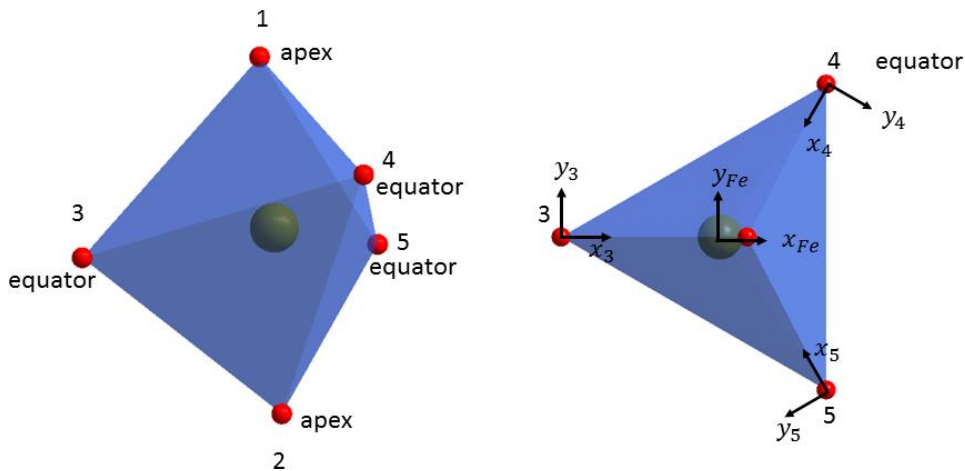

**Figure S8.2** The coordinate systems and oxygen indices used for calculating the hybridization. Here we use two coordinate systems. The $\{x_{oi}, y_{oi}z_{oi}\}$ $\{x_{Fe}, y_{Fe}, z_{Fe}\}$ system is shown in the right panel in which all the z axis are aligned to the crystalline c axis. When the hybridization is calculated, the atomic orbitals are transformed to the $\{x'_{oi}, y'_{oi}z'_{oi}\}$ $\{x'_{Fe}, y'_{Fe}, z'_{Fe}\}$ systems. In the $\{x'_{oi}, y'_{oi}z'_{oi}\}$ $\{x'_{Fe}, y'_{Fe}, z'_{Fe}\}$, the z' axis of the O and Fe sites are along the vector that connects Fe and O atoms. The axes that are not shown can be found using the cross products of the displayed axes.

The $m = \pm 2$ states of an Fe atom has no hybridization with the apex O atoms.

If we use the wave functions $\{z^2, xy, x^2 - y^2, xz, yz\}$ for an Fe site and $\{x, y, z\}$ for an O site, the matrix elements are $V^i_{\mu\nu}$

$$V^1_{z^2,z} = V^2_{z^2,z} = V_{pd\sigma}R_c^{-3.5}$$

$$V^1_{xz,x} = V^1_{yz,y} = V^2_{xz,x} = V^2_{yz,y} = V_{pd\pi}R_c^{-3.5},$$

where $\mu \in \{z^2, x^2 - y^2, xy, xz, yz\}$ and $\nu \in \{x, y, z\}$.

Putting these equations together, one gets the results in Table S8.1

Table S8.1

| $V^i_{\mu\nu} R_c^{3.5}$ | x | y | z |
|---|---|---|---|
| $2z^2 - x^2 - y^2$ | 0 | 0 | $V_{pd\sigma}$ |
| $\sqrt{3}(x^2 - y^2)$ | 0 | 0 | 0 |
| $2\sqrt{3}xy$ | 0 | 0 | 0 |
| $2\sqrt{3}xz$ | $V_{pd\pi}$ | 0 | 0 |
| $2\sqrt{3}yz$ | 0 | $V_{pd\pi}$ | 0 |

The $\{xy, x^2 - y^2\}$ states of Fe atom has no hybridization with the apex O atoms.

For the hybridization between Fe and equator oxygen atoms, the coordinate systems need to be transformed for the calculation.

As shown in Fig. S8,2, we choose z axis for both Fe and O sites to be along the crystal $c$ axis.

For the hybridization between Fe and 3-5 O (see Fig. S8.2), we can make the transformation by rotating the axis.

The transformation is $\begin{bmatrix} x \\ y \\ z \end{bmatrix}_{Fe} = \begin{bmatrix} \cos(\phi) & -\sin(\phi) & 0 \\ \sin(\phi) & \cos(\phi) & 0 \\ 0 & 0 & 1 \end{bmatrix} \begin{bmatrix} 0 & 0 & 1 \\ 0 & 1 & 0 \\ -1 & 0 & 0 \end{bmatrix} \begin{bmatrix} x' \\ y' \\ z' \end{bmatrix}_{Fe} =$

$\begin{bmatrix} 0 & -\sin(\phi) & \cos(\phi) \\ 0 & \cos(\phi) & \sin(\phi) \\ -1 & 0 & 0 \end{bmatrix} \begin{bmatrix} x' \\ y' \\ z' \end{bmatrix}_{Fe}$, where $\phi$ is $0$, $\frac{2\pi}{3}$, and $\frac{4\pi}{3}$, for 3, 4, and 5 oxygen atoms respectively.

The transformation is $\begin{bmatrix} x \\ y \\ z \end{bmatrix}_{Oi} = \begin{bmatrix} 0 & 0 & 1 \\ 0 & 1 & 0 \\ -1 & 0 & 0 \end{bmatrix} \begin{bmatrix} x' \\ y' \\ z' \end{bmatrix}_{Oi}$ for the $i$th oxygen atom.

Using the transformation, the relation between the wave functions are:

$$|2z^2 - x^2 - y^2\rangle = |2x'^2 - y'^2 - z'^2\rangle = -\frac{1}{2}|2z'^2 - x'^2 - y'^2\rangle + \frac{\sqrt{3}}{2}|\sqrt{3}(x'^2 - y'^2)\rangle$$

$$|\sqrt{3}(x^2 - y^2)\rangle = \frac{\sqrt{3}}{2}\cos(2\phi)|2z'^2 - x'^2 - y'^2\rangle + \frac{1}{2}\cos(2\phi)|\sqrt{3}(x'^2 - y'^2)\rangle - \sin(2\phi)|2\sqrt{3}y'z'\rangle$$

$$|2\sqrt{3}xy\rangle = \frac{\sqrt{3}}{2}\sin(2\phi)|2z'^2 - x'^2 - y'^2\rangle + \frac{1}{2}\sin(2\phi)|\sqrt{3}(x'^2 - y'^2)\rangle + \cos(2\phi)|2\sqrt{3}y'z'\rangle$$

$$|2\sqrt{3}xz\rangle = \sin(\phi)|2\sqrt{3}x'y'\rangle - \cos(\phi)|2\sqrt{3}x'z'\rangle$$

$$|2\sqrt{3}yz\rangle = -\cos(\phi)|2\sqrt{3}x'y'\rangle - \sin(\phi)|2\sqrt{3}x'z'\rangle$$

Table S8.2 shows the hybridization using the $(x', y', z')$ coordinate system of oxygen.

Table S8.2

| $V^i_{\mu\nu} R^{3.5}_a$ | $x_i'$ | $y_i'$ | $z_i'$ |
|---|---|---|---|
| $2z^2 - x^2 - y^2$ | | | $-\frac{1}{2}V_{pd\sigma}$ |
| $\sqrt{3}(x^2 - y^2)$ | | $-\sin(2\phi_i) V_{pd\pi}$ | $\dfrac{\sqrt{3}}{2}\cos(2\phi_i) V_{pd\sigma}$ |
| $2\sqrt{3}xy$ | | $\cos(2\phi_i) V_{pd\pi}$ | $\dfrac{\sqrt{3}}{2}\sin(2\phi_i) V_{pd\sigma}$ |
| $2\sqrt{3}xz$ | $-\cos(\phi_i) V_{pd\pi}$ | | |
| $2\sqrt{3}yz$ | $-\sin(\phi_i) V_{pd\pi}$ | | |

After transforming to the $(x, y, z)$ coordinate system of oxygen, the hybridization matrix elements are shown in the Table S8.3.

Table S8.3

| $V^i_{\mu\nu} R^{3.5}_a$ | $x_i$ | $y_i$ | $z_i$ |
|---|---|---|---|
| $2z^2 - x^2 - y^2$ | $-\dfrac{1}{2} V_{pd\sigma}$ | | |
| $\sqrt{3}(x^2 - y^2)$ | $\dfrac{\sqrt{3}}{2}\cos(2\phi_i) V_{pd\sigma}$ | $\sin(2\phi_i) V_{pd\pi}$ | |
| $2\sqrt{3}xy$ | $\dfrac{\sqrt{3}}{2}\sin(2\phi_i) V_{pd\sigma}$ | $\cos(2\phi_i) V_{pd\pi}$ | |
| $2\sqrt{3}xz$ | | | $\cos(\phi_i) V_{pd\pi}$ |
| $2\sqrt{3}yz$ | | | $\sin(\phi_i) V_{pd\pi}$ |

Adding all three sites (3-5) together, one gets the hybridizations strength in Table S8.4.

Table S8.4

| $\sum_i |V^i_{\mu\nu}|^2 R^7_a$ | $x_i$ | $y_i$ | $z_i$ |
|---|---|---|---|
| $2z^2 - x^2 - y^2$ | $\dfrac{3}{4}|V_{pd\sigma}|^2$ | | |
| $\sqrt{3}(x^2 - y^2)$ | $\dfrac{9}{8}|V_{pd\sigma}|^2$ | $\dfrac{3}{2}|V_{pd\pi}|^2$ | |
| $2\sqrt{3}xy$ | $\dfrac{9}{8}|V_{pd\sigma}|^2$ | $\dfrac{3}{2}|V_{pd\pi}|^2$ | |

| | | | | |
|---|---|---|---|---|
| $2\sqrt{3}xz$ | | | | $\frac{3}{2}|V_{pd\pi}|^2$ |
| $2\sqrt{3}yz$ | | | | $\frac{3}{2}|V_{pd\pi}|^2$ |

Arranging the hybridization in an in-plane ($p$) and out-of-plane ($s$) fashion, we get the Table S8.5.

Table S8.5

| $\sum_i |V_{\mu\nu}^i|^2$ | Apex-$p$ | Apex-$s$ | Equator-$p$ | Equator-$s$ |
|---|---|---|---|---|
| $2z^2 - x^2 - y^2$ | | $2|V_{pd\sigma}|^2 R_c^{-7}$ | $\frac{3}{4}|V_{pd\sigma}|^2 R_a^{-7}$ | |
| $\sqrt{3}(x^2 - y^2)$ | | | $\left(\frac{3}{2}|V_{pd\pi}|^2 + \frac{9}{8}|V_{pd\sigma}|^2\right) R_a^{-7}$ | |
| $2\sqrt{3}xy$ | | | $\left(\frac{3}{2}|V_{pd\pi}|^2 + \frac{9}{8}|V_{pd\sigma}|^2\right) R_a^{-7}$ | |
| $2\sqrt{3}xz$ | $2|V_{pd\pi}|^2 R_c^{-7}$ | | | $\frac{3}{2}|V_{pd\pi}|^2 R_a^{-7}$ |
| $2\sqrt{3}yz$ | $2|V_{pd\pi}|^2 R_c^{-7}$ | | | $\frac{3}{2}|V_{pd\pi}|^2 R_a^{-7}$ |

If we take the Harrison's assumption [9] $\frac{V_{pd\sigma}}{V_{pd\pi}} = -2.17$ and $\frac{R_c}{R_a} = 0.94$ in h-LFO, we can estimate the relative values for the hybridization, as shown in the Table S8.6.

Table S8.6

| $\sum_i |V_{\mu\nu}^i|^2 / |V_{pd\pi}|^2 R_a^{-7}$ | Apex-$p$ | Apex-$s$ | Equator-$p$ | Equator-$s$ |
|---|---|---|---|---|
| $2z^2 - x^2 - y^2$ | | 14.5 | 3.5 | |
| $\sqrt{3}(x^2 - y^2)$ | | | 6.8 | |
| $2\sqrt{3}xy$ | | | 6.8 | |
| $2\sqrt{3}xz$ | 3.08 | | | 1.5 |
| $2\sqrt{3}yz$ | 3.08 | | | 1.5 |

Or graphically, one can illustrate the hybridization as shown in the Figure S8.3

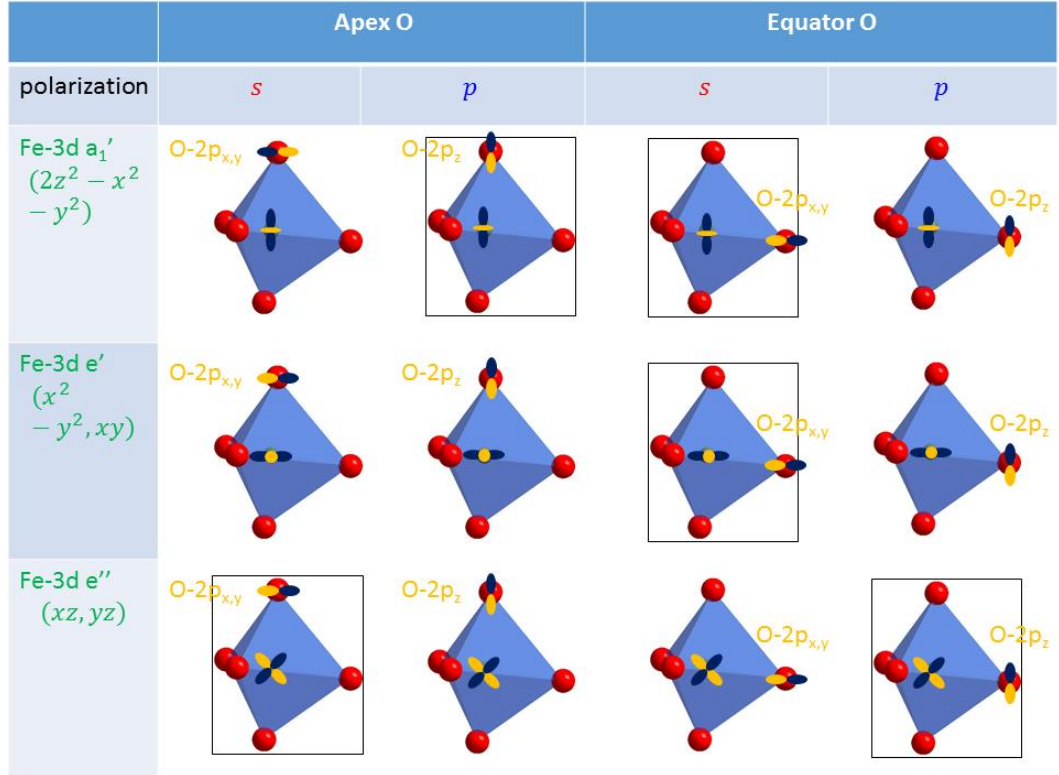

**Figure S8.3** Schematic illustration of relative geometric configuration between Fe-3d and O-2p orbitals. The boxes indicate significant hybridizations.

As shown in Fig. S8.3, the relative geometric configuration of different Fe-3d and O-2p orbitals are illustrated. There are 12 combinations of hybridizations coming from 3 Fe-3d orbitals, 2 O-2p orbital, and 2 types of O (apex and equator) atomic positions. From the relative geometric configuration of the orbitals, one can qualitatively estimate the configurations that have significant hybridizations; this results in 5 cases which are indicated by the boxes in Fig. S8.3. These 5 case can be further divided into two groups according to the polarization of the x-ray when the O-K edge excitation is concerned.

### S8.2.2 $C_S$ symmetry
Again, we discuss the effect of $C_S$ distortion in terms of lattice distortion mode $\Gamma_2^-$ and $K_1$. Although the displacement also changes the bond angles, but due to the $R^{-7}$ dependence, the main effect on the hybridization is coming from the atomic distance.

### $\Gamma_2^-$ mode
The hybridization between the Fe-3d and the apex oxygen atoms are affected. If we assume that the distance between the Fe atom and the apex oxygen atom changes from $R_c$ to $R_c + \delta$ and $R_c - \delta$. So the $\sum_i |V_{\mu\nu}^i|^2$ that contains $R_c^{-7}$ will have an factor $f_{hyb} = \frac{1}{2}\left(1 + \frac{\delta}{R_c}\right)^{-7} + \frac{1}{2}\left(1 - \frac{\delta}{R_c}\right)^{-7} = 1 + 21\left(\frac{\delta}{R_c}\right)^2$.

In h-LuFeO$_3$, $\frac{\delta}{R_c} \approx 3.7\%$, which results in $f_{hyb} = 1.015$.

This uniformly shifts the hybridization between the $a_1'$ and $e''$ states and the apex oxygen.

On the other hand, if one combines the effect of $\Gamma_2^-$ and $K_3$, a different effect on the hybridization of $xz$ and $yz$ states will occur. This effect comes from the hybridization between the $xz$ state of Fe-3d and the $z$ state of the apex oxygen because of the small angle $\theta_{g2}$.

Using the table derived for the Lu-5d and O-2p hybridization, this results in an additional term $\sum_i |V_{\mu\nu}^i|^2 = 2\left(|V_{pd\pi}|^2 + 3|V_{pd\sigma}|^2\right) R_c^{-7} \sin^2(\theta_{2g})$, which is second order in terms of $\sin(\theta_{2g})$

This additional hybridization is much smaller for the $yz$ state.

### $K_1$ mode

There are two parts of distortion in the $K_1$ mode, one is the displacement of the apex oxygen, the other is the displacement of the Fe within the triangular lattice, with respect to equator oxygen.

For the oxygen displacement, it generates a factor $f_{hyb} = \frac{1}{2}\left(1 + \frac{\delta}{R_c}\right)^{-7} + \frac{1}{2}\left(1 + \frac{\delta}{R_c}\right)^{-7} = 1 - 7\frac{\delta}{R_c}$. For h-LuFeO$_3$, $\frac{\delta}{R_c} \approx 0.018\%$, which corresponds to $f_{hyb} = 1 + 1.2 \times 10^{-3}$.

Again, this affects the $\sum_i |V_{\mu\nu}^i|^2$ that contains $R_c^{-7}$, which means the hybridizations between the $a_1'$ and $e''$ states and the apex oxygen.

For the Fe displacement, one needs to calculate

$$\sum_i |\sin(2\phi_i)|^2 R_{a,i}^{-7} = \sum_i |\sin(\phi_i)|^2 R_{a,i}^{-7} = \frac{3}{2} R_a^{-7}\left(1 - \frac{7}{2}\frac{\delta}{R_a}\right)$$

and

$$\sum_i |\cos(2\phi_i)|^2 R_{a,i}^{-7} = \sum_i |\cos(\phi_i)|^2 R_{a,i}^{-7} = \frac{3}{2} R_a^{-7}\left(1 + \frac{7}{2}\frac{\delta}{R_a}\right).$$

The resulting hybridization is summarized in the Table S7.7.

Table S8.7

| $\sum_i |V_{\mu\nu}^i|^2$ | Apex-p | Apex-s | Equator-p | Equator-s |
|---|---|---|---|---|
| $2z^2 - x^2 - y^2$ | | $2|V_{pd\sigma}|^2 R_c^{-7}$ | $\frac{3}{4}|V_{pd\sigma}|^2 R_a^{-7}$ | |
| $\sqrt{3}(x^2 - y^2)$ | | | $\left(\frac{3}{2}|V_{pd\pi}|^2 + \frac{9}{8}|V_{pd\sigma}|^2\right) R_a^{-7} + \frac{7}{2}\frac{\delta}{R_a}\left(-\frac{3}{2}|V_{pd\pi}|^2 + \frac{9}{8}|V_{pd\sigma}|^2\right) R_a^{-7}$ | |
| $2\sqrt{3}xy$ | | | $\left(\frac{3}{2}|V_{pd\pi}|^2 + \frac{9}{8}|V_{pd\sigma}|^2\right) R_a^{-7} - \frac{7}{2}\frac{\delta}{R_a}\left(-\frac{3}{2}|V_{pd\pi}|^2 + \frac{9}{8}|V_{pd\sigma}|^2\right) R_a^{-7}$ | |
| $2\sqrt{3}xz$ | $2|V_{pd\pi}|^2 R_c^{-7}$ | | | $\frac{3}{2}|V_{pd\pi}|^2 R_a^{-7}\left(1 + \frac{7}{2}\frac{\delta}{R_a}\right)$ |

| | | | |
|---|---|---|---|
| $2\sqrt{3}yz$ | $2|V_{pd\pi}|^2 R_c^{-7}$ | | $\frac{3}{2}|V_{pd\pi}|^2 R_a^{-7}(1 - \frac{7}{2}\frac{\delta}{R_a})$ |

It is clear that when $\frac{\delta}{R_a} > 0$, the hybridization of $\sqrt{3}(x^2 - y^2)$ will be stronger than that of $2\sqrt{3}xy$ state. Similarly, the hybridization of the $2\sqrt{3}xz$ state is stronger than that of the $2\sqrt{3}yz$ state. The relative change is $\frac{7}{2}\frac{\delta}{R_a}$ for $2\sqrt{3}xz$ and $2\sqrt{3}yz$ states but slightly smaller for the $\sqrt{3}(x^2 - y^2)$ and $2\sqrt{3}xy$ states. In h-LuFeO$_3$, $\frac{\delta}{R_a}$ can be as large as $6 \times 10^{-3}$ at low temperature.

### S8.3 Hybridization between Lu-5d and O-2p in h-LuFeO$_3$

In the C$_{3v}$ local environment, the Lu-5d orbitals are split into three energy levels by the crystal field: $e^\sigma = (x^2 - y^2 + 2\lambda xz, xy + \lambda yz)$, $a_1 = z^2$, $e^\pi = (x^2 - y^2 - 2\lambda xz, xy - \lambda yz)$ where $\lambda$ is a mixing factor that is on the order of 1. Significant hybridizations are expected in most cases between Lu-5d orbital and O-2p orbital.

The hybridization between the Lu-5d orbit and the equator oxygen can be calculated straightforwardly since their $x$, $y$ and $z$ axis can be readily aligned. The results are similar to the hybridization between Fe-3d and the apex oxygen atoms, which is shown again in the Table S8.8.

Table S8.8

| $V^i_{\mu\nu} R_c^{3.5}$ | x | y | z |
|---|---|---|---|
| $2z^2 - x^2 - y^2$ | 0 | 0 | $V_{pd\sigma}$ |
| $\sqrt{3}(x^2 - y^2)$ | 0 | 0 | 0 |
| $2\sqrt{3}xy$ | 0 | 0 | 0 |
| $2\sqrt{3}xz$ | $V_{pd\pi}$ | 0 | 0 |
| $2\sqrt{3}yz$ | 0 | $V_{pd\pi}$ | 0 |

For all the hybridization of Lu-5d with other 6 oxygen atoms, the calculation can be down using the coordinate system show in Fig. S8.4. In order to align the z axis of both Lu and O to the vector that connects the two atoms, one needs to make the following transformations:

$$\begin{bmatrix} x \\ y \\ z \end{bmatrix}_{Lu} = \begin{bmatrix} \cos(\phi) & -\sin(\phi) & 0 \\ \sin(\phi) & \cos(\phi) & 0 \\ 0 & 0 & 1 \end{bmatrix} \begin{bmatrix} \cos(\theta) & 0 & \sin(\theta) \\ 0 & 1 & 0 \\ -\sin(\theta) & 0 & \cos(\theta) \end{bmatrix} \begin{bmatrix} x' \\ y' \\ z' \end{bmatrix}_{Lu}$$

$$= \begin{bmatrix} \cos(\phi)\cos(\theta) & -\sin(\phi) & \cos(\phi)\sin(\theta) \\ \sin(\phi)\cos(\theta) & \cos(\phi) & \sin(\phi)\sin(\theta) \\ -\sin(\theta) & 0 & \cos(\theta) \end{bmatrix} \begin{bmatrix} x' \\ y' \\ z' \end{bmatrix}_{Lu}$$

for Lu, and

$$\begin{bmatrix} x \\ y \\ z \end{bmatrix}_{Oi} = \begin{bmatrix} \cos(\theta) & 0 & \sin(\theta) \\ 0 & 1 & 0 \\ -\sin(\theta) & 0 & \cos(\theta) \end{bmatrix} \begin{bmatrix} x' \\ y' \\ z' \end{bmatrix}_{Oi}$$ for $i$th oxygen atom, where $\phi$ is $0, \frac{\pi}{3}, \frac{2\pi}{3}, \pi, \frac{4\pi}{3}$ and $\frac{5\pi}{3}$, for 6, 4, 7, 2, 5, and 3 oxygen atoms respectively.

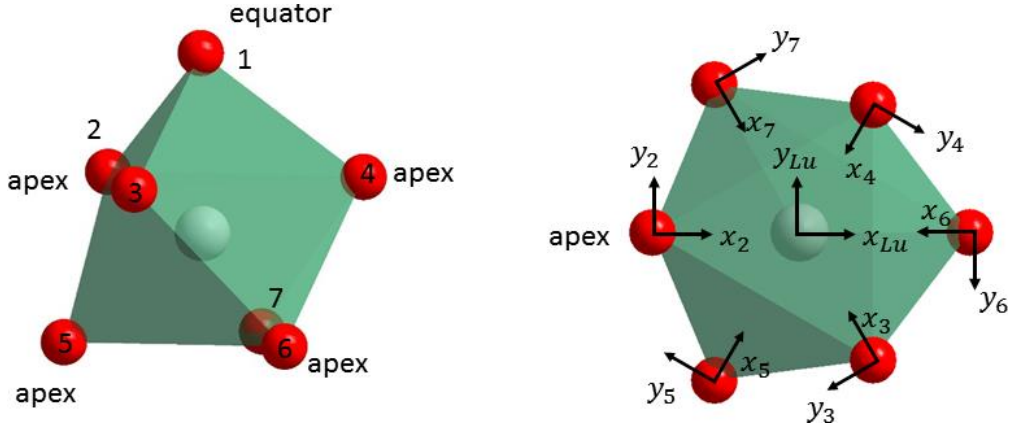

**Figure S8.4** The coordinate system and indices used to calculate the hybridization between Lu-5d and O-2p orbits.

The transformation of the Lu-5d wave function is

$$|2z^2 - x^2 - y^2\rangle$$
$$= \left(\frac{2 - 3\sin^2(\theta)}{2}\right)|2z'^2 - x'^2 - y'^2\rangle - \frac{\sqrt{3}}{2}\sin^2(\theta)|\sqrt{3}(x'^2 - y'^2)\rangle$$
$$- \frac{\sin(2\theta)}{2\sqrt{3}}|2\sqrt{3}x'z'\rangle$$

$$|\sqrt{3}(x^2 - y^2)\rangle = \frac{\sqrt{3}\cos(2\phi)\sin^2(\theta)}{2}|2z'^2 - x'^2 - y'^2\rangle - \cos(2\phi)\left(1 - \frac{\sin^2(\theta)}{2}\right)|\sqrt{3}(x'^2 - y'^2)\rangle$$
$$- \sin(2\phi)\cos(\theta)|2\sqrt{3}x'y'\rangle + \frac{1}{2}\cos(2\phi)\sin(2\theta)|2\sqrt{3}x'z'\rangle$$
$$- \sin(2\phi)\sin(\theta)|2\sqrt{3}y'z'\rangle$$

$$|2\sqrt{3}xy\rangle = \frac{\sqrt{3}\sin(2\phi)\sin^2(\theta)}{2}|2z'^2 - x'^2 - y'^2\rangle - \frac{\sin(2\phi)}{2}\left(1 - \frac{\sin^2(\theta)}{2}\right)|\sqrt{3}(x'^2 - y'^2)\rangle$$
$$+ \cos(2\phi)\cos(\theta)|2\sqrt{3}x'y'\rangle + \frac{1}{2}\sin(2\phi)\sin(2\theta)|2\sqrt{3}x'z'\rangle$$
$$+ \cos(2\phi)\sin(\theta)|2\sqrt{3}y'z'\rangle$$

$$|2\sqrt{3}xz\rangle = \frac{\sqrt{3}\cos(\phi)\sin(2\theta)}{2}|2z'^2 - x'^2 - y'^2\rangle - \frac{\cos(\phi)\sin(2\theta)}{2}|\sqrt{3}(x'^2 - y'^2)\rangle$$
$$+ \sin(\phi)\sin(\theta)|2\sqrt{3}x'y'\rangle + \cos(\phi)\cos(2\theta)|2\sqrt{3}x'z'\rangle - \sin(\phi)\cos(\theta)|2\sqrt{3}y'z'\rangle$$

$$|2\sqrt{3}yz\rangle = \frac{\sqrt{3}\sin(\phi)\sin(2\theta)}{2}|2z'^2 - x'^2 - y'^2\rangle - \frac{\sin(\phi)\sin(2\theta)}{2}|\sqrt{3}(x'^2 - y'^2)\rangle$$
$$- \cos(\phi)\sin(\theta)|2\sqrt{3}x'y'\rangle + \sin(\phi)\cos(2\theta)|2\sqrt{3}x'z'\rangle - \cos(\phi)\cos(\theta)|2\sqrt{3}y'z'\rangle.$$

With the transformed wave function, one can calculate the hybridization matrix, as shown in the Table S8.9.

Table S8.9

| $V^i_{\mu\nu} R_a^{3.5}$ | $x_i$ | $y_i$ | $z_i$ |
|---|---|---|---|
| $2z^2 - x^2 - y^2$ | $\left(\frac{2 - 3\sin^2(\theta)}{2}\right)\sin(\theta) V_{pd\sigma}$ $-\frac{\sin(2\theta)\cos(\theta)}{2\sqrt{3}} V_{pd\pi}$ | | $\left(\frac{2 - 3\sin^2(\theta)}{2}\right)\cos(\theta) V_{pd\sigma}$ $-\frac{\sin(2\theta)\sin(\theta)}{2\sqrt{3}} V_{pd\pi}$ |
| $\sqrt{3}(x^2 - y^2)$ | $\frac{\sqrt{3}\cos(2\phi)\sin^2(\theta)\sin(\theta)}{2} V_{pd\sigma}$ $+\frac{1}{2}\cos(2\phi)\sin(2\theta)\cos(\theta) V_{pd\pi}$ | $-\sin(2\phi)\sin(\theta) V_{pd\pi}$ | $\frac{\sqrt{3}\cos(2\phi)\sin^2(\theta)\cos(\theta)}{2} V_{pd\sigma}$ $-\frac{1}{2}\cos(2\phi)\sin(2\theta)\sin(\theta) V_{pd\pi}$ |
| $2\sqrt{3}xy$ | $\frac{\sin(2\phi)\sin^2(\theta)\sin(\theta)}{2} V_{pd\sigma}$ $+\frac{\sin(2\phi)\sin(2\theta)\cos(\theta)}{2} V_{pd\pi}$ | $\cos(2\phi)\sin(\theta) V_{pd\pi}$ | $\frac{\sqrt{3}\sin(2\phi)\sin^2(\theta)\cos(\theta)}{2} V_{pd\sigma}$ $+\frac{\sin(2\phi)\sin(2\theta)\sin(\theta)}{2} V_{pd\pi}$ |
| $2\sqrt{3}xz$ | $\frac{\sqrt{3}\cos(\phi)\sin(2\theta)\sin(\theta)}{2} V_{pd\sigma}$ $+\cos(\phi)\cos(2\theta)\cos(\theta) V_{pd\pi}$ | $-\sin(\phi)\cos(\theta) V_{pd\pi}$ | $\frac{\sqrt{3}\cos(\phi)\sin(2\theta)\cos(\theta)}{2} V_{pd\sigma}$ $+\cos(\phi)\cos(2\theta)\sin(\theta) V_{pd\pi}$ |
| $2\sqrt{3}yz$ | $\frac{\sqrt{3}\sin(\phi)\sin(2\theta)\sin(\theta)}{2} V_{pd\sigma}$ $+\sin(\phi)\cos(2\theta)\cos(\theta) V_{pd\pi}$ | $-\cos(\phi)\cos(\theta) V_{pd\pi}$ | $\frac{\sqrt{3}\sin(\phi)\sin(2\theta)\cos(\theta)}{2} V_{pd\sigma}$ $+\sin(\phi)\cos(2\theta)\sin(\theta) V_{pd\pi}$ |

As expected from the symmetry of the local environment, the hybridization between the Lu-5d and the apex oxygen are not anisotropic. That said, the hybridization between the Lu-5d and the equator oxygen is anisotropic. The Lu-5d $a_1$ hybridize more with the O-2p out-of-plane orbits ($s$ polarization); the Lu-5d $e^\sigma$, $e^\pi$ hybridize more with the O-2p in-plane ($p$ polarization) orbits.

In addition, the hybridization of the Lu-5d $a_1$ with the apex oxygen atoms are much less than that with the equator oxygen atoms, according to the calculated $V^i_{\mu\nu} R_a^{3.5}$ in the table above. If we take $\theta = 37$ degree as an approximation, the hybridization are $V^i_{a_1,x} R_a^{3.5} = 0.27 V_{pd\sigma} - 0.22 V_{pd\pi}$ and $V^i_{a_1,z} R_a^{3.5} = 0.36 V_{pd\sigma} - 0.57 V_{pd\pi}$. This also means that The Lu-5d $a_1$ hybridize more with the O-2p out-of-plane orbits ($s$ polarization). All of these are consistent with the experimental observation.

### S8.4 Hybridization of between Fe and oxygen in o-LuFeO$_3$

Since the local environment of Fe in o-LuFeO$_3$ is octahedral of O$_h$ symmetry, the hybridization is easier to calculate. The result for the O$_h$ symmetry is shown in Table S8.10, where the 1-6 oxygen atoms are located at $(R, 0, 0)$, $(-R, 0, 0)$, $(0, R, 0)$, $(0, -R, 0)$, $(0, 0, R)$, and $(0, 0, -R)$ respectively.

Table S8.10

| $V^i_{\mu\nu} R^{-3.5}$ | $x_{1,2}$ | $y_{1,2}$ | $z_{1,2}$ | $x_{4,6}$ | $y_{4,6}$ | $z_{4,6}$ | $x_{3,5}$ | $y_{3,5}$ | $z_{3,5}$ |
|---|---|---|---|---|---|---|---|---|---|
| $2z^2 - x^2 - y^2$ | 0 | 0 | $V_{pd\sigma}$ | 0 | $-\frac{1}{2}V_{pd\sigma}$ | 0 | $-\frac{1}{2}V_{pd\sigma}$ | 0 | 0 |
| $\sqrt{3}(x^2 - y^2)$ | 0 | 0 | 0 | 0 | $-\frac{\sqrt{3}}{2}V_{pd\sigma}$ | 0 | $\frac{\sqrt{3}}{2}V_{pd\sigma}$ | 0 | 0 |
| $2\sqrt{3}xy$ | 0 | 0 | 0 | $V_{pd\pi}$ | 0 | 0 | 0 | $V_{pd\pi}$ | 0 |
| $2\sqrt{3}xz$ | $V_{pd\pi}$ | 0 | 0 | 0 | 0 | 0 | 0 | 0 | $-V_{pd\pi}$ |
| $2\sqrt{3}yz$ | 0 | $V_{pd\pi}$ | 0 | 0 | 0 | $-V_{pd\pi}$ | 0 | 0 | 0 |

For the D$_{2h}$ distortion, $R_{12} = R, R_{35} = R - \delta, R_{46} = R + \delta$.

The relation value of hybridization $\sum_i |V^i_{\mu\nu}|^2 / |V_{pd\pi}|^2 R_a^{-7}$ are 7.1, 7.1, 2, 2, 2 for the $2z^2 - x^2 - y^2$, $\sqrt{3}(x^2 - y^2)$, $2\sqrt{3}xy$, $2\sqrt{3}xz$, and $2\sqrt{3}yz$ states respectively.

The total hybridization for the 3d orbitals is shown in Table S8.11.

Table S8.11

| | $|V^i_{\mu\nu}|^2 R^{-7}$ |
|---|---|
| $2z^2 - x^2 - y^2$ | $|V_{pd\sigma}|^2 \left[3 + 21\left(\frac{\delta}{R}\right)^2\right]$ |
| $\sqrt{3}(x^2 - y^2)$ | $3|V_{pd\sigma}|^2 \left[1 + 21\left(\frac{\delta}{R}\right)^2\right]$ |
| $2\sqrt{3}xy$ | $4|V_{pd\pi}|^2 \left[1 + 21\left(\frac{\delta}{R}\right)^2\right]$ |
| $2\sqrt{3}xz$ | $4|V_{pd\pi}|^2 (1 + \frac{7}{2}\frac{\delta}{R})$ |
| $2\sqrt{3}yz$ | $4|V_{pd\pi}|^2 (1 - \frac{7}{2}\frac{\delta}{R})$ |

Therefore, the states $2z^2 - x^2 - y^2$, $\sqrt{3}(x^2 - y^2)$, and $2\sqrt{3}xy$ are insensitive to the D$_{2h}$ distortion, while $2\sqrt{3}xz$ and $2\sqrt{3}yz$ are.

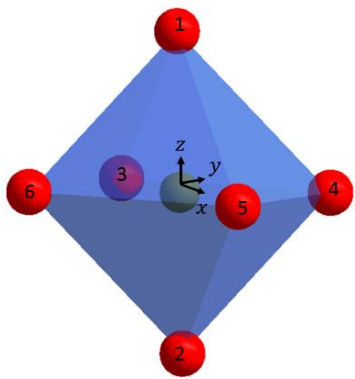

**Figure S8.5** The coordinate system and indices used in calculating the hybridization between Fe-3d and O-2p in an $O_h$ local environment.

## S9. Refinement of atomic positions in the h-LuFeO3 lattice structure at various temperatures

In the paramagnetic phase of h-LuFeO$_3$, the symmetry of the crystal structure can be described using space group P6$_3$mmc. In the ferroelectric phase (below 1050 K [10]), lattice distortion occurs, which changes the symmetry to P6$_3$cm. The distortions move the atoms away from high symmetric positions. The displacements of the atoms can be decomposed into 3 modes: $K_3$, $K_1$, and $\Gamma_2^-$. [10,11] The $K_3$ mode corresponds to a rotation of the FeO$_5$; the $\Gamma_2^-$ displaces the atoms along the $c$ axis which is expected to generate the electric polarization; the $K_1$ mode involves a displacement of Fe in the basal plane.

In order to measure the temperature dependence of the lattice distortion in the h-LuFeO$_3$ films, we carried out single-crystal x-ray diffraction measurements of 43 peaks (see the list of peaks below) at 7 temperatures. By fitting the measured peak intensities (areas), one can find the positions of the atoms. The distortions can be calculated from the displacement of the atoms from their high symmetry positions. Here we define the lattice distortion using 9 parameters (e.g. $dz_{Lu1}$), as shown in the Table S9.1.

Table S9.1 Definition of atomic displacements in the units of lattice constants.

| Site | Wyckoff position | x/a | y/b | z/c |
|---|---|---|---|---|
| Lu1 | 2a | 0 | 0 | $\frac{1}{4} + dz_{Lu1}$ |
| Lu2 | 4b | $\frac{1}{3}$ | $\frac{2}{3}$ | $\frac{1}{4} + dz_{Lu2}$ |
| Fe | 6c | $\frac{1}{3} + dx_{Fe}$ | 0 | 0 |
| O1 | 6c | $\frac{1}{3} + dx_{O1}$ | 0 | $dz_{O1}$ |
| O2 | 6c | $\frac{2}{3} - dx_{O2}$ | 0 | $dz_{O2}$ |
| O3 | 2a | 0 | 0 | $dz_{O3}$ |
| O4 | 4b | $\frac{1}{3}$ | $\frac{2}{3}$ | $dz_{O4}$ |

Temperature dependence structural distortion can then be represented using the 9 parameters in the table below. The last column is the data from Magome et al. [12]

Table S9.2 Measured atomic displacements at various temperatures.

| Variables ($\times 10^{-3}$) | 6 K | 100 K | 110 K | 130 K | 150 K | 200 K | 300 K | Error | 300 K (Magome) |
|---|---|---|---|---|---|---|---|---|---|
| $dz_{Lu1}$ | 28.0 | 27.7 | 27.3 | 26.7 | 25.8 | 25.3 | 24.1 | 0.7 | 22.1 |
| $dz_{Lu2}$ | -13.6 | -13.7 | -14.1 | -14.4 | -15.1 | -15.4 | -14.4 | 0.7 | -16.8 |
| $dx_{Fe}$ | 1.9 | 1.6 | 1.5 | 1.4 | 1.4 | 1.0 | 0.1 | 0.9 | 0 |
| $dx_{O1}$ | -41 | -43 | -44 | -44 | -43 | -43 | -21 | 7 | -30.3 |
| $dz_{O1}$ | 152 | 150 | 148 | 149 | 148 | 148 | 162 | 4 | 154.2 |
| $dx_{O2}$ | 7 | 5 | 4 | 3 | 3 | 2 | -5 | 5 | -17.7 |
| $dz_{O2}$ | 322 | 322 | 321 | 322 | 321 | 320 | 327 | 3 | 332 |
| $dz_{O3}$ | -60 | -50 | -50 | -60 | -60 | -60 | -50 | 10 | -28 |
| $dz_{O4}$ | 16 | 15 | 16 | 16 | 15 | 16 | 19 | 4 | 17 |

From the table above, one can see that the uncertainty for the positions of the oxygen atoms are much higher than those of the metal (Fe and Lu) atoms, which can be attributed to the small scattering factor of the oxygen atoms.

The next step is to decompose the displacement patterns into the three distortion modes. Due to the larger uncertainty of the position of the oxygen atoms, we choose to use the displacement of metal (Fe and Lu) atoms to represent the lattice distortions.

For $K_1$ mode, $dFe_{K_1} = dx_{Fe}$ is naturally chosen.

For $K_3$, we choose $dLu_{K_3} \equiv \frac{1}{2}(dz_{Lu1} - 2dz_{Lu2})$. Here $Lu1$ is the minority sites and $Lu2$ is the majority sites since the number of $Lu2$ is twice as much as that of $Lu1$.

For $\Gamma_2^-$, we choose $dLu_{\Gamma_2^-} = \frac{1}{2}(dz_{Lu1} + 2dz_{Lu2})$. One needs to be careful in explaining the values: it is not proportional to the electric polarization. This is because in the coordinate used here, Fe site is at the origin, which is not necessarily the center of the charge. One needs to know the oxygen positions to estimate the electric polarization. Unfortunately, the uncertainty of the oxygen positions is very high here.

Table S9.3 shows the temperature dependence of the parameters chosen to represent the three lattice distortions.

Table S9.3

| Variables ($\times 10^{-3}$) | 6 K | 100 K | 110 K | 130 K | 150 K | 200 K | 300 K | Error | 300 K ($Magma$) |
|---|---|---|---|---|---|---|---|---|---|
| $dLu_{K_3}$ | 27.6 | 27.5 | 27.8 | 27.8 | 28.1 | 28.0 | 26.4 | 0.8 | 27.9 |
| $dLu_{\Gamma_2^-}$ | 0.4 | 0.2 | -0.5 | -1.1 | -2.2 | -2.7 | -2.4 | 0.8 | -5.7 |
| $dFe_{K_1}$ | 1.9 | 1.6 | 1.5 | 1.4 | 1.4 | 1.0 | 0.1 | 0.9 | 0 |

The peaks we measured are:

(0, 1, 8), (1, 0, 8), (1, 0, 6), (1, 0, 4), (1, 1, 3), (1, 1, 4), (1, 1, 5), (1, 1, 6), (1, 1, 7), (1, 1, 8), (1, 1, 9), (1, 1, 10), (0, 2, 10), (0, 2, 8), (2, 0, 10), (2, 0, 8), (2, 0, 6), (2, 0, 4), (1, 2, 4), (1, 2, 6), (1, 2, 8), (1, 2, 10), (1, 2, 12), (2, 1, 12), (2, 1, 10), (2, 1, 8), (2, 1, 6), (2, 1, 4), (2, 2, 3), (2, 2, 4), (2, 2, 5), (2, 2, 6), (2, 2, 7), (2, 2, 8), (2, 2, 10), (2, 2, 11), (3, 0, 12), (3, 0, 10), (3, 0, 8), (3, 0, 6), (3, 0, 4), (0, 3, 8), (0, 3, 12 ).

## S10. Effect of the O-Fe-O bond angle on the density of states in a FeO$_5$ cluster

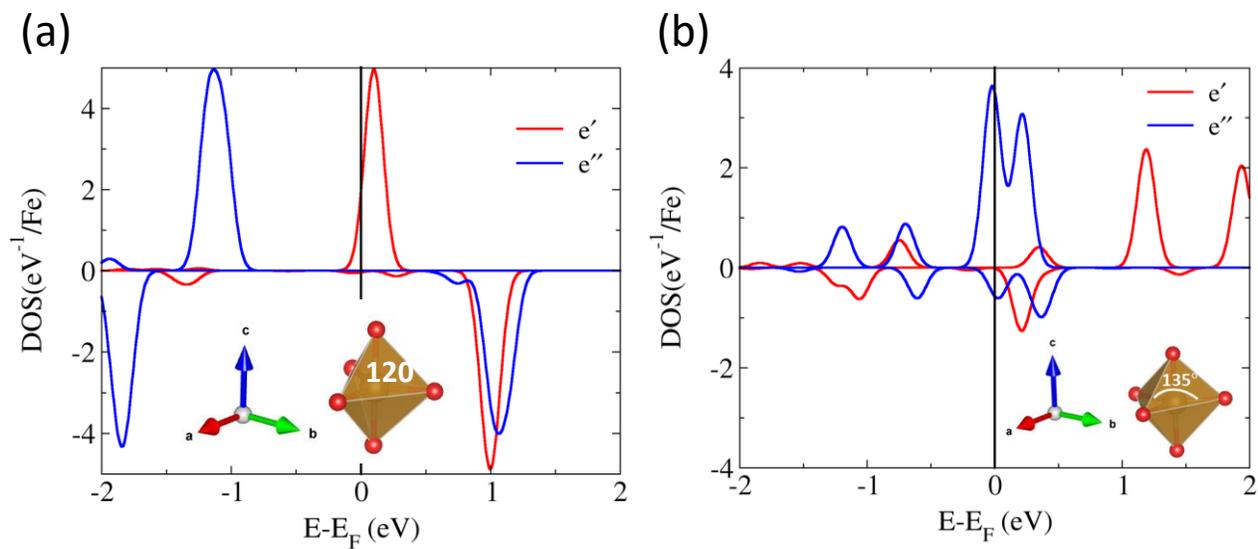

Figure S10: Density of states of FeO$_5$ cluster with ∠O-Fe-O =120° projected at Fe site and resolved into cubic harmonics according to the D$_{3d}$ symmetry (a) and those of FeO$_5$ cluster with ∠O-Fe-O =135° (b).

# References

[1]     K.-T. Ko, H.-J. Noh, J.-Y. Kim, B.-G. Park, J.-H. Park, A. Tanaka, S. B. Kim, C. L. Zhang, and S.-W. Cheong, Phys. Rev. Lett. **103**, 207202 (2009).

[2]     D.-Y. Cho, J.-Y. Kim, B.-G. Park, K.-J. Rho, J.-H. Park, H.-J. Noh, B. J. Kim, S.-J. Oh, H.-M. Park, J.-S. Ahn, H. Ishibashi, S.-W. Cheong, J. H. Lee, P. Murugavel, T. W. Noh, A. Tanaka, and T. Jo, Phys. Rev. Lett. **98**, 217601 (2007).

[3]     E. Pavarini, E. Koch, F. Anders, M. Jarrell, I. for A. Simulation, and A. S. C. Electrons, *Correlated Electrons: From Models to Materials Lecture Notes of the Autumn School Correlated Electrons 2012 at Forschungszentrum Jülich, 3 - 7 September 2012* (Forschungszentrum Jülich, Zentralbibliothek, Verl., Jülich, 2012).

[4]     M. S. Dresselhaus, G. Dresselhaus, and A. Jorio, *Group Theory Application to the Physics of Condensed Matter* (Springer-Verlag, Berlin, 2007).

[5]     J. a. Moyer, R. Misra, J. a. Mundy, C. M. Brooks, J. T. Heron, D. a. Muller, D. G. Schlom, and P. Schiffer, APL Mater. **2**, 012106 (2014).

[6]     W. Zhu, L. Pi, S. Tan, and Y. Zhang, Appl. Phys. Lett. **100**, (2012).

[7]     J. Stohr and H. C. Siegmann, *Magnetism from Fundamentals to Nanoscale Dynamics* (Springer, Berlin, 2006).

[8]     R. L. White, J. Appl. Phys. **40**, 1061 (1969).

[9]     W. A. Harrison, *Electronic Structure and the Properties of Solids : The Physics of the Chemical Bond* (Freeman, San Francisco, 1980).

[10]    W. Wang, J. Zhao, W. Wang, Z. Gai, N. Balke, M. Chi, H. N. Lee, W. Tian, L. Zhu, X. Cheng, D. J. Keavney, J. Yi, T. Z. Ward, P. C. Snijders, H. M. Christen, W. Wu, J. Shen, and X. Xu, Phys. Rev. Lett. **110**, 237601 (2013).

[11]    C. J. Fennie and K. M. Rabe, Phys. Rev. B **72**, 100103 (2005).

[12]    E. Magome, C. Moriyoshi, Y. Kuroiwa, A. Masuno, and H. Inoue, Jpn. J. Appl. Phys. **49**, 09ME06 (2010).